\documentclass[journal=jctcce,manuscript=article,chaptertitle=true]{achemso}

\usepackage[english]{babel}
\addto\captionsenglish{

}

\usepackage[sc]{mathpazo}        
\usepackage{units}               
\usepackage{xspace}
\usepackage{booktabs}            
\usepackage{dcolumn}             
\usepackage[table]{xcolor}
\usepackage{amsmath}
\usepackage{amssymb}             
\usepackage{multirow}
\usepackage{numprint}
\usepackage{datetime2}
\usepackage{caption}
\usepackage[normalem]{ulem}      
\usepackage{circledsteps}        
\usepackage[utf8]{inputenc}

\usepackage{listings}            
\definecolor{codegreen}{rgb}{0.00, 0.50, 0.00}
\definecolor{backcolor}{rgb}{0.93, 0.93, 0.93}

\lstdefinestyle{mdpStyle}{
    backgroundcolor=\color{backcolor},
    basicstyle=\ttfamily\small,
    captionpos=b,
    commentstyle=\color{codegreen},
    framexleftmargin=0.5em,
    keepspaces=true,
    morecomment=[l]{;},
    showspaces=false,
    showstringspaces=false,
}
\usepackage[numbers,super,sort&compress]{natbib}
\usepackage{graphicx}            
\usepackage{todonotes}
\usepackage{url}
\definecolor{LinkCol}{cmyk}{1.00 0.00 0.57 0.30} 
\usepackage{nameref}
\usepackage[section]{placeins}
\usepackage[%
colorlinks=true,
linkcolor=LinkCol,
urlcolor=LinkCol,
citecolor=LinkCol,
pdftitle={Constant pH Simulation with FMM Electrostatics in GROMACS}  
]{hyperref}

\newcommand{\gromacs}{GROMACS\xspace}
\newcommand{\xtc}{\texttt{.xtc}\xspace}
\newcommand{\trr}{\texttt{.trr}\xspace}
\newcommand{\mdp}{\texttt{.mdp}\xspace}

\newcommand{\lambdadyn}{$\lambda$-dynamics\xspace}
\newcommand{\Lambdadyn}{$\lambda$-Dynamics\xspace}

\newcommand{\pKa}   {\mbox{p$K_a$}\xspace}
\newcommand{\pKaRef}{\mbox{p$K_{a\text{,ref}}$}\xspace}
\newcommand{\pH}    {\mbox{pH}\xspace}  

\newcommand{\DGchem}{\mbox{$\Delta G_\text{chem}$}\xspace}
\newcommand{\DGmm}  {\mbox{$\Delta G_\text{MM}$}\xspace}

\newcommand{\Vmm}  {\mbox{$V_\text{MM}$}\xspace}
\newcommand{\VpH}  {\mbox{$V_\text{pH}$}\xspace}
\newcommand{\Vdw}  {\mbox{$V_\text{dw}$}\xspace}

\newcommand{\CPHParam}{(\textbullet)\xspace}

\newcommand{\ffCharmmUsed}{CHARMM36m\xspace}
\newcommand{\ffAmberUsed} {Amber99sb*-ILDN\xspace}
\newcommand{\Charmm}{CHARMM\xspace}
\newcommand{\Amber}{Amber\xspace}

\newcommand{\manuscriptTitleEliane}{Constant pH Simulation with FMM Electrostatics in GROMACS. \\
(A) Design and Applications }
\newcommand{\MPInat}{Theoretical and Computational Biophysics, 
Max Planck Institute for Multidisciplinary Sciences,
Am Fassberg 11,
37077 G\"ottingen,
Germany}

\author{Eliane Briand}
\affiliation{\MPInat}

\author{Bartosz Kohnke}
\affiliation{\MPInat}

\author{Carsten Kutzner}
\affiliation{\MPInat}

\author{Helmut Grubm\"uller}
\affiliation{\MPInat}
\email{hgrubmu@mpinat.mpg.de}

\title{\manuscriptTitleEliane}

\usepackage{datetime}

\begin{document}

\begin{abstract}
\noindent
The structural dynamics of biological macromolecules, such as proteins, DNA/RNA, or complexes thereof,
are strongly influenced by protonation changes of their typically many titratable groups, 
which explains their sensitivity to pH changes.
Conversely, conformational and environmental changes of the biomolecule affect the protonation state of these groups.
With few exceptions, conventional force field-based molecular dynamics (MD) simulations do not account for these effects, 
nor do they allow for coupling to a pH buffer.

Here we present a \gromacs implementation of a rigorous Hamiltonian interpolation \lambdadyn constant pH method, which rests on GPU-accelerated Fast Multipole Method (FMM) electrostatics. 
Our implementation supports both \ffCharmmUsed and \ffAmberUsed force fields and is largely automated to enable seamless switching from regular MD to constant pH MD, involving minimal changes to the input files. 
Here, the first of two companion papers describes the underlying constant pH protocol and sample applications to several prototypical benchmark systems such as cardiotoxin V, lysozyme, and staphylococcal nuclease. 
Enhanced convergence is achieved through a new dynamic barrier height optimization method, 
and high \pKa accuracy is demonstrated. 
We use Functional Mode Analysis and Mutual Information to explore the complex intra- and intermolecular couplings between the protonation states of titratable groups as well as those between protonation states and conformational dynamics. 
We identify striking conformation-dependent \pKa variations and unexpected inter-residue couplings. 
Conformation-protonation coupling is identified as a primary cause of the slow protonation convergence notorious to constant pH simulations involving multiple titratable groups, suggesting enhanced sampling methods to accelerate convergence. 
\end{abstract}


\section{Introduction}

The activity of hydrogen ions (pH) 
is one of the most important solution conditions for biomolecular systems, alongside temperature and ionic strength.
This importance stems from its direct influence on titratable residues, like Histidine (His), Aspartic acid (Asp), or Glutamic Acid (Glu), which can switch between protonated and deprotonated forms, thereby profoundly affecting the electrostatics of the biomolecule.
Alongside pH, other factors contribute to residue protonation: first, the intrinsic \pKa of the moiety (i.e., the free energy of deprotonation linked to the chemical bond breaking); second, and most relevantly in proteins, the local electrostatic and steric environment, which is time and conformation dependent. 
In turn, the charge difference between protonated and deprotonated residues influences the dynamics of the system through electrostatic interactions, ultimately affecting the conformations adopted by the protein.

A close interplay therefore exists between protonation state and conformational ensemble in proteins, which is relevant for biological phenomena such as (i) the stabilization of protein structure, 
either through polarization effects or direct salt bridge interactions, 
as found in the quaternary structure of hemoglobin;\cite{Bettati1998}
(ii) protein folding, as exemplified by the pH-dependent folding behavior of the model protein villin headpiece;\cite{tang2006} or 
(iii) the catalytic activities of enzymes, as in the acid-base-nucleophile catalytic triad central to hydrolases and transferases.

However, most molecular dynamics (MD) simulations today use fixed protonation states, which are typically determined heuristically based on the initial structure.
A fixed protonation state, regardless of its accuracy, cannot capture the aforementioned interplay between conformation and protonation, and therefore in many cases will not lead to accurate electrostatics and conformational dynamics.
Furthermore, fixed protonation states hinder simulations at specific pH values. 
To address these limitations, constant pH simulations (CPH-MD) have been developed
\cite{baptista1997,lee2004constant,Dlugosz2004,Mongan2004,khandogin2005constant,Meng2010,huang2018,
mertz1994molecular,borjesson2001,Buergi2002,Baptista2002,Donnini2011,wallace2012charge,
goh2014constant,swails2014,huang2016all,Radak2017,Wallace2011,Harris2022}
which enable dynamic protonation changes in response to pH and the electrostatic environment of titratable groups throughout the simulation.
The development of such CPH-MD methods presents various challenges, which we describe before presenting our solutions, as implemented in our \gromacs-based CPH-MD code.


Ideally, describing protonation in MD simulations by explicit hydronium ions and reactive titratable groups
would yield a constant pH method which accurately reproduces both thermodynamic and kinetic properties (depending  on the accuracy of the reaction model).
For the aim of obtaining an accurate conformational and titration ensemble, e.g., for calculating titration curves, this approach is inefficient for several reasons.
First, protonation changes within the protein core can be very slow, requiring tens of microseconds per protonation change event.\cite{Sham1999}
Second, the diffusion of explicit hydronium ions between protonatable residues adds another hurdle for a protonation event to occur.
Finally, physiological pH cannot be easily realized in typically sized simulation boxes with explicit, integer-charged hydronium ions.\cite{Donnini2011} 
For instance, a single ion in an 8~nm cubic box suitable for a hen egg-white lysozyme  simulation yields a pH of 2.5, five orders of magnitude more acidic than physiological conditions.
In the context of MD, an already sampling limited technique, alternative methods without those  impediments to protonation change are necessary.

As a result, a variety of constant pH methods have been developed that introduce virtual titration coordinates instead of explicit, diffusing protons,\cite{mongan2005review,Chen2014}
see de Oliveira et al.\cite{deOliveira2022} for a recent and concise review.
They can be broadly categorized by (a) how they treat the solvent --- implicitly\cite{baptista1997,lee2004constant,Dlugosz2004,Mongan2004,khandogin2005constant,Meng2010,huang2018} 
versus explicitly\cite{mertz1994molecular,borjesson2001,Buergi2002,Baptista2002,
Donnini2011,wallace2012charge,goh2014constant,swails2014,huang2016all,Radak2017,Wallace2011,Harris2022}--- and (b)
by the type of the titration coordinate --- discrete\cite{Buergi2002,Baptista2002,Walczak2002,Mongan2004,Dlugosz2004,Meng2010,itoh2011ph,swails2012,Radak2017} 
versus continuous.\cite{mertz1994molecular,baptista1997,borjesson2001,lee2004constant,khandogin2005constant,Donnini2011,
wallace2012charge,Wallace2011,goh2012,goh2014constant,huang2016all,huang2018,Harris2022,harris2019, harris2020} 

Discrete titration coordinate methods employ Monte Carlo (MC) steps at regular intervals to switch between two states for each titratable residue: protonated and deprotonated.
As the energy difference between these states in explicit solvent tends to be quite large due to the water orientation effect, earlier MC protocols used an implicit solvent auxiliary simulation based on Generalized Born\cite{Mongan2004,Meng2010,itoh2011ph} or Poisson-Boltzmann electrostatics,\cite{Dlugosz2004,Baptista2002,Walczak2002}
the result of which was used in the acceptance criteria to update the main simulation protonation state.
Several weaknesses of these approaches were identified:\cite{khandogin2005constant,lee2004constant}
the use of an auxiliary solver causes a discontinuity in forces and energy at the time of switching, 
resulting in instabilities; adapting a single site per MC step slows convergence in protonation space;
and the overall increase in computational effort as more titratable residues are considered renders them less applicable for larger proteins with tens to hundreds of such residues.
Recent MC protocols have partially mitigated these issues by substituting implicit solvent evaluations for brief non-equilibrium MD simulations\cite{Stern2007,Chen2015,Radak2017}
or thermodynamic integration,\cite{Buergi2002,vanGunsteren1993TI}
as well as efficiently choosing candidate residues for protonation state change,\cite{Radak2017}
showing that discrete coordinate methods are still under active development in the current constant pH method landscape.

In contrast to discrete titration methods, continuous titration methods, initially developed by Brooks et al.,\cite{Kong1996,lee2004constant,khandogin2005constant} use a continuous titration coordinate $\lambda$ to interpolate gradually between two Hamiltonians representing a protonated and a deprotonated state, respectively.\cite{tidor1993simulated}
This approach enables a gradual switching between protonated and deprotonated states via a continuum of non-physical intermediate states.
This avoids the large instantaneous energy difference of discrete methods by allowing a progressive rearrangement of the local environment, enabling both explicit solvation and more frequent protonation state changes.

However, without further adjustments, such a continuous titration coordinate leads
to a large fraction of simulation time spent in unphysical intermediate states.
This is mitigated through $\lambda$-coordinate transformation\citep{lee2004constant,Abrams2006,Vegt1998,Donnini2011,hayes2021blade} 
or a bias potential that favors the physical end states.\cite{Donnini2016}
Within this family of methods, two distinct but conceptually close approaches have been proposed for the interpolation between protonation states: Hamiltonian interpolation and charge interpolation\cite{aho2022constph,buslaev2022best} (more generally, parameter interpolation).
Although Hamiltonian interpolation is the physically canonical approach,
 early implementations faced computational challenges. 
The need for multiple electrostatic interaction evaluations -- one for each protonation state -- led to a linear slowdown as the number of protonatable residues increased.\cite{Donnini2011}
Charge interpolation offers a solution here, achieving better simulation performances that are independent of the number of titratable sites.\cite{aho2022constph, buslaev2022best, huang2016all}
We will discuss the differences between these interpolation schemes in our companion publication, including their effect on constant pH simulations.\cite{Kohnke2023}

The representation of tautomeric forms of protonatable residues, such as the $\delta$ and $\epsilon$ forms of singly protonated Histidine, poses an additional challenge in constant pH simulations.
This necessitates representing multiple protonated states for a given residue --- typically two --- rather than a simple binary switch between unique protonated and deprotonated forms, resulting in a three-state model.
Continuous titration methods often address this challenge by introducing a dual $\lambda$ coordinate system: one $\lambda$ governs protonation, while the other determines the tautomeric form.

Another challenge, independent of the titration coordinate choice, arises from the time-dependent protonation states, which result in a varying total charge of the simulation box.
Non-neutral, periodic simulation boxes cause simulation artifacts,\cite{hunenberger1999ewald,kastenholz2004influence,hub2014quantifying}
and transitions between differently charged systems under periodic boundary conditions fail to accurately represent real macroscopic systems, which are charge neutral.
Consequently, a non-neutral box raises questions about the validity of the resulting free energies.
These effects are particularly pronounced in inhomogeneous systems and when considering free energies,\cite{bignucolo2022galvani}
which is unfortunately the case in constant pH MD, given its energy-based protonation switch criteria.
This issue has been addressed in three main ways: 1) attempting an analytical correction\cite{hunenberger1999ewald,rocklin2013calculating} of the aforementioned effects,\cite{huang2016all}
2) computing an approximate correction specific to the system of interest with a short pre-simulation,\cite{Radak2017}
or 3) using buffer particles in bulk solvent that take on the necessary countercharge to maintain neutrality. \cite{wallace2012charge,chen2013introducing,Donnini2016}
The latter option, when combined with a sufficiently large simulation box, most closely mimics a real macroscopic system and is thus our chosen approach.

Finally, the correct treatment of Coulomb forces remains a key challenge in biomolecular simulations due to their long-range nature at the length scale of proteins, stemming from the slow $\nicefrac{1}{r}$ falloff of the electrostatic potential.
Historically favored plain cutoff methods resulted in stark simulation artifacts,\cite{schreiber1992cutoff}
leading to their replacement by periodic lattice-sum methods such as Particle Mesh Ewald (PME),\cite{Essmann:1995vj}
now the \textit{de facto} standard for  electrostatics in MD.
PME, based on the Fast Fourier Transform (FFT) of a discretized charge grid, is extremely efficient at low to moderate parallelization, especially when implemented for GPUs.\cite{ganesan2011fenzi,gpuAmber2013,pall2020}
Recently, this efficient GPU-accelerated PME approach has been successfully extended to constant pH simulations.\cite{hayes2021blade,aho2022constph,Harris2022}
However, at high parallelization, the global communication steps involved in FFTs become a scaling bottleneck,\cite{Dongarra2021}
with the number of messages increasing quadratically with the number of ranks.\cite{Hess:2008tf}
This limitation becomes particularly pressing for exascale computing, necessitating the development of Coulomb solvers with superior scaling properties.

We have therefore decided to implement the Fast Multipole Method (FMM)\cite{greengard1987} electrostatic solver.
FMM shows improved asymptotic scaling through astute use of hierarchical decomposition. It also avoids unfavorable scaling when combined with Hamiltonian interpolation $\lambda$-dynamics.
Moreover, FMM surpasses PME in terms of parallelization potential.\cite{arnold2013comparison,kohnke2020,bkohnke2021}
For instance, in the high parallelization regime, the MODYLAS MD simulation code has successfully deployed FMM electrostatics for a 10 M atoms MD system over 524,288 CPU cores.\cite{modylas2013}
Though our FMM implementation is not currently massively parallel,
and thus a typical biomolecule in solution
runs about three times slower compared to PME, 
it has already shown performance advantages for large sparse systems (e.g.\ droplets or vapour), 
where it outperforms PME on a single GPU.\cite{kohnke2020}

We have made the following choices for our GROMACS\cite{pall2020} constant pH MD implementation:

First, we used established techniques, including Hamiltonian interpolation \lambdadyn\cite{Kong1996} with a three-state model for protonatable residue tautomers\cite{khandogin2005constant} and charge buffers for electroneutrality.\cite{Donnini2016}

However, to further the conversation on optimal constant pH MD methodologies, we deviated from PME-based charge interpolation \lambdadyn, which is used in virtually all recent implementations.
Instead, we revisited the original formulation, Hamiltonian interpolation \lambdadyn.
Previously thought to be a dead-end due to performance concerns, we demonstrate that Hamiltonian interpolation can achieve computational efficiency through a novel calculation scheme within our FMM electrostatics solver.
Our companion publication\cite{Kohnke2023} provides a detailed explanation of this scheme and analysis of its computational performance. 
Here, we focus on describing our method and demonstrating its accuracy.

Second, we developed Dynamic Barrier Optimization (DBO), a sampling enhancement technique. DBO controls protonation/deprotonation transition rates by dynamically regulating free energy barrier heights.
This approach optimizes computational resource utilization.
We demonstrate that DBO achieves comparable \pKa accuracy while accelerating convergence in constant pH simulations.

Third, to assess the accuracy of our constant-pH implementation, we titrated benchmark proteins, which have NMR-determined residue \pKa values, namely cardiotoxin V, lysozyme and staphylococcal nuclease.
Additionally, we compared the performance of our code with other recent implementations using these benchmark systems.

Fourth, we developed novel post-simulation analysis tools to uncover the underlying drivers of protonation.
Using Mutual Information,\cite{thomas2006elements,shannon1948mathematical} we uncovered interactions between titratable residues in a model-free manner.
Using Functional Mode Analysis,\cite{hub2010detection,Krivobokova2012} we characterized the couplings between protein conformation and protonation states.
This approach yielded both qualitative insights into the coupling mechanisms and quantitative data on conformation-induced \pKa shifts.
We applied these techniques to the aforementioned benchmark systems, illustrating these protonation coupling effects. 
Our analysis also explains why specific residues pose challenges for accurate titration in constant pH simulations in general.
 Furthermore, we present methods to monitor convergence for these problematic residues using these techniques.

Finally, we posit that methodological complexities and setup difficulties impede the widespread adoption of constant pH methodology. 
Improved software design can mitigate many of these challenges.
Therefore, to leverage the existing expertise of MD users, we integrated our code into the \gromacs\cite{pall2020} software suite, adhering closely to its established usage and settings conventions.
Through carefully selected and assessed default parameters, and extensive setup automation, we demonstrate  that an acquainted \gromacs user can simulate a protein using our constant pH code with very little additional effort.

\section{Methods}

\subsection{Constant pH Simulations Based on \Lambdadyn}
\label{sec:lambdadyn_cph_intro}
%
%

The \lambdadyn method described here builds upon previous work.\cite{Kong1996,Donnini2011,dobrev2017,dobrev2020}
The present section summarizes the main concepts of \lambdadyn-based constant pH simulations,
while subsequent sections, starting from~\ref{sec:pfc}, introduce novel features of our FMM-based implementation.

In \lambdadyn, titratable \emph{sites} --- typically protein residues that change their total charge by $\pm 1$ upon (de)protonation ---
are described by a $\lambda$-dependent Hamiltonian $\mathcal{H}(\lambda)$\cite{tidor1993simulated}
which is a weighted sum of the Hamiltonians representing the protonated ($\mathcal{H}_\text{A}$) 
and the deprotonated forms ($\mathcal{H}_\text{B}$) of the site.
The variable $\lambda$ linearly interpolates between these Hamiltonians, with each chemically distinct state of a site referred to as a \emph{form}.
Sites can encompass not only protonatable residues but also peptide termini, ionizable lipids, and small molecules such as ligands or drug-like compounds.

In contrast to free energy methods like thermodynamic integration\cite{vanGunsteren1993TI} (TI)
which use $\lambda$ as an input parameter to drive the system Hamiltonian from the A to the B state, here, $\lambda$ is treated as an additional, dynamic degree of freedom --- a pseudo-particle with mass, velocity, and consequently, kinetic energy $\frac{1}{2}m\dot{\lambda}^2$.\cite{tidor1993simulated,Kong1996}
Forces acting on the $\lambda$-pseudo-particle, given by $F_\lambda = -\partial\mathcal{H} / \partial\lambda$,
derive from the following extended Hamiltonian
\begin{equation}
\mathcal{H}(\lambda) = (1-\lambda) \mathcal{H}_\text{A} + \lambda \mathcal{H}_\text{B} + \frac{1}{2}m\dot{\lambda}^2 + V(\lambda).
\label{eq:H_basic}
\end{equation}
            
To properly control the dynamics of the $\lambda$-particle,
an additional potential $V(\lambda)$ is introduced,
which

\begin{itemize}
\item [(R1)] Accounts for the chemical energy difference between protonated and deprotonated states not captured by the molecular mechanics (MM) force field,
\item [(R2)] Incorporates the chosen pH by establishing a reference free energy of protonation $\Delta G_{0}$ for each titratable group,
\item [(R3)] Biases the $\lambda$-coordinate towards values near 0 or 1, representing fully protonated or deprotonated states, respectively, while disfavoring intermediate, unphysical states,
\item [(R4)] Establishes a barrier between protonated and deprotonated state 
that allows controlling the transition kinetics.
\end{itemize}

A two-well potential,\cite{Donnini2011,Donnini2016,sppexaFinalReport2020} detailed in section~\ref{sec:biasPotential}, has been implemented to meet all four requirements (R1-R4).
To control protonation and de-protonation rates, achieve high transition rates while avoiding unphysical intermediate states, we dynamically adjust the height of the central barrier as described below  (R4).

As in previous work,\cite{dobrev2017} a second $\lambda$-coordinate is used to alternate between protonation tautomers
(e.g., histidine $\delta$ and $\epsilon$ forms), 
interpolating between all forms with the following expression ($\bar{\lambda} := 1 - \lambda$),
\begin{equation}
\mathcal{H}(\lambda_p, \lambda_t) =  \bar{\lambda}_p \left (  \bar{\lambda}_t \mathcal{H}_\text{A} + \lambda_t \mathcal{H}_\text{B} \right) \\
 + \lambda_p  \left ( \bar{\lambda}_t \mathcal{H}_\text{C} + \lambda_t \mathcal{H}_\text{D} \right ) + \frac{m}{2} \left(\dot{\lambda}_p^2 + \dot{\lambda}_t^2 \right ) + V(\lambda_p,\lambda_t).
\label{eq:H_tautomers}
\end{equation}
This scheme can be extended to any number of forms with additional $\lambda$-coordinates
(which is already supported by our FMM electrostatics solver),
although two are sufficient for a constant pH treatment of protein residues and termini.
Our current implementation applies the full Hamiltonian (eq~\ref{eq:H_tautomers}) to all titratable sites. For sites lacking tautomers, we set $\mathcal{H}_A = \mathcal{H}_B$ and $\mathcal{H}_C = \mathcal{H}_D$.
Chemically equivalent tautomeric states have the same microscopic \pKa for the transition toward one or the other tautomer form.
This phenomenon is exemplified by the two protonated forms of Glutamate (Glu), which differ only in which oxygen atom of the carboxyl group bears the proton.
Conversely, Histidine (His) has one double protonated form, but two chemically distinct single protonated forms ($\delta$ and $\epsilon$ tautomers) with different microscopic \pKa values.

In principle, any chemical moiety or residue can be made dynamically protonatable through this scheme.
However, because MM force fields are not parameterized to give correct (de)protonation energies, a calibration process, described below, is necessary for each protonatable residue.
This computationally intensive procedure is required once for titratable residue types (e.g., Glu, His ...) in the used force field.
This calibration is pre-calculated here for most relevant residues, with a \pKa close to physiological pH, 
namely His, Glu, and aspartic acid (Asp) (c.f.\ Supplementary Information).

Depending on the particular force field, 
the protonated and deprotonated form of a titratable residue ($\mathcal{H}_\text{A}$ and $\mathcal{H}_\text{B}$)
can differ in ways other than charge, 
e.g., bond and Lennard-Jones interaction parameters.
In our current implementation, we posit that these differences do not affect the dynamics as much as electrostatics.
Consequently, we will limit ourselves here to the change of charges of the atoms comprising the sites.
All other interaction parameters reflect those of the fully protonated form.

To offer both ease of use and flexibility,  
most constant pH-related parameters described below are user-controllable
 via \gromacs input file directives, indicated by the symbol \CPHParam.
We have defined default settings for protein simulations to mitigate the potential configuration burden arising from this flexibility. 
These parameters, also employed in our simulations, facilitate reproduction of our settings for other systems, as detailed in the \nameref{PractAspectSubSection} subsection.

\subsection{Calibration of Protonation Dynamics}
\label{sec:biasPotential}

The $V(\lambda)$ potential term in the Hamiltonian (eq~\ref{eq:H_basic}) ensures 
that the protonation behavior produced by the MD simulation at a given pH 
satisfies the four requirements (R1-R4) listed above.
This is achieved by
calibrating the chemical free energy of protonation against experiments (R1),
by coupling the protonation behavior to an external pH bath (R2),
concentrating the $\lambda$ values on physical states (R3), and
controlling the (de)protonation rates by a central barrier (R4).
Accordingly, we break down $V(\lambda)$ into separate contributions
\begin{equation}
V(\lambda) = \Vmm(\lambda) + \VpH(\lambda) + \Vdw(\lambda) \ .
\label{eq:Vmm_VpH_Vdw}
\end{equation}
\Vmm and \VpH together provide the calibration (R1) and (R2), whereas \Vdw addresses both (R3) and (R4).
These three individual contributions will be explained in the following sections.

\subsubsection*{General approach}

The free energy of protonation derived from an MM simulation, \DGmm,
does not generally match  the actual chemical free energy of protonation, \DGchem, as the potential function defined by the force field does not take into account the energy of bond cleavage and formation resulting from chemical reactions.
To correct this mismatch, \DGmm is calibrated to be equal to \DGchem for reference compounds, namely single  residues in water, using the $\Vmm(\lambda)$ potential, such that constant pH simulation of these reference compounds yields the same \pKa as experimental titrations.\cite{Donnini2011}

Therefore, prior to the constant pH simulation,
Thermodynamic Integration (TI) simulations of the individual amino acid in solution were performed
to determine the free energy landscape along $\lambda$ over the entire $\lambda$ interval for the force field used.
Defining $\Vmm(\lambda) := -\DGmm(\lambda)$ establishes an energy landscape that is perfectly flat for the reference residue at $\pH = \pKaRef$.
Subsequently, when the residue is placed in an environment different from the calibration conditions --- e.g., within a protein --- the local environment results in a deviation from this flat potential, corresponding to a shift in \pKa, which is the goal of constant pH MD simulations.

Simulations of the reference compound in such a flat potential landscape would yield uniformly distributed $\lambda$ values, including the unphysical intermediate states.
Therefore, to concentrate the distribution on physically meaningful end states around $\lambda = 0$ and $\lambda = 1$,
a double-well potential \Vdw with a central barrier is used.
The barrier is chosen high enough to minimize the time spent in unphysical semi-protonated states, 
but low enough to allow for enough transitions between states to obtain a chosen sampling efficiency.

Finally, while a flat potential is adequate for $\pH = \pKaRef$, the population ratio for the protonated and deprotonated forms should be different when $\pH \neq  \pKaRef$, which is achieved by using \VpH, a pH-dependent potential setting the relative height of the two wells according to the free energy of deprotonation at a given pH, of the form

\begin{equation}
\DGchem = \VpH(\lambda=1) - \VpH(\lambda=0) = (\text{ln } 10) R T (\pKaRef - \pH) \ .
\label{eq:V_lambda}
\end{equation}

\subsubsection*{Details of the TI protocol}

We collected the derivatives $\left\langle \partial \mathcal{H}_\text{ref} / \partial \lambda \right\rangle_\lambda$ at discrete points along the $\lambda$ coordinate.
Tautomerism was included by considering a two-dimensional free energy landscape $\Vmm(\lambda) = \Vmm(\lambda_p, \lambda_t)$ (see eq \ref{eq:H_tautomers}).
Accordingly, the derivatives were collected on a 2D grid, with either $\lambda := \lambda_p$ or $\lambda := \lambda_t$, computed using
\begin{equation}
\frac{\partial \Vmm}{\partial \lambda}
= \left\langle \frac{\partial \mathcal{H}_\text{ref}(\lambda)}{\partial \lambda} \right\rangle_{\lambda_p, \lambda_t}
= \left\langle \frac{\partial V_{\text{ref}}^\text{Coul}}{\partial \lambda} \right\rangle_{\lambda_p, \lambda_t}
\end{equation}

for combination of values of $\lambda_p$ and $\lambda_t$.
The last relation, equating $\mathcal{H}_\text{ref}$ (the full Hamiltonian) to $V_{\text{ref}}^\text{Coul}$ (the electrostatic potential energy) holds here because the protonated and deprotonated form of a residue differ only by charge.

%
%
%
%
%
%

In our implementation, a polynomial is fitted to $\partial \Vmm / \partial \lambda$ 
as collected on the ($\lambda_p$, $\lambda_t$) grid, with each $\lambda$ varying
from $-0.1$ to 1.1 with a step of 0.2, and a finer step of 0.05 around 0 and 1 (see Supplementary Information section 1 for the list of values), which we empirically found to be sufficient granularity for protein residues.
At every grid point, $\partial V / \partial \lambda_p$ and $\partial V / \partial \lambda_t$ 
were collected for 60~ns to thoroughly average over all conformations of the residue. 

\begin{sloppypar}
Calibration is specific to each residue and to each supported force field (here, \ffCharmmUsed and \ffAmberUsed), 
as well as to the distance between the individual residue and its charge buffer site 
(though the distance is fixed to 3 nm here, it is possible to calibrate for any distance), 
and the ionic strength of the solution.
The 2D grids are fitted to polynomials of degree 5 in both dimensions \CPHParam, 
including all mixing terms (e.g., $c_{2,3}\cdot \lambda_{p}^{2}\cdot\lambda_{t}^{3}$), which we empirically found in test calculations to provide sufficient accuracy for proteins.
\end{sloppypar}

The time required for convergence was determined by repeating the TI and the fit procedure 
for three fully independent replicas and comparing the resulting fit surfaces.
For the final calibration data provided as Supplementary Material, all three replicas were used to fit the polynomial.

Note that, as a consequence of calibrating against a reference which is a single amino acid in solution (eq~\ref{eq:V_lambda}),
constant pH simulations produce $\Delta$\pKa to the reference compound, not absolute \pKa.
We nonetheless report absolute \pKa for ease of interpretation, in line with practices in the field.
 
To allow for a flexible control of the height of the central barrier
and the position, shape, and depth of the two wells independently,
we represent the potential $\Vdw(\lambda)$ using cubic Hermite splines.

\subsection{Partition Function Correction}
\label{sec:pfc}

As a side effect, the typically asymmetrical shape of the two potential wells of $\Vdw(\lambda)$ exerts an entropic bias on the protonated versus deprotonated population.
To compensate for this bias, we calculate the respective partition functions,

\begin{align}
Z_\text{prot}   &= \int_{\lambda=0}^{\lambda=0.5} e^{-\beta V(\lambda)} \\
Z_\text{deprot} &= \int_{\lambda=0.5}^{\lambda=1} e^{-\beta V(\lambda)} .
\end{align}

Subsequently, the well depths are iteratively refined until the resulting free energy $\Delta G = - \nicefrac{1}{\beta} \ln{(Z_\text{prot} / Z_\text{deprot})}$ agrees with the target free energy $\Delta G = k_B T \cdot \ln{10} \cdot (\pKaRef - \pH)$ within numerical accuracy.

If enabled \CPHParam, the Partition Function Correction is automatically applied when needed, 
i.e., before the first simulation step, and when adjusting the barrier height or well position (see next section).
We emphasize that this correction is not applied \textit{post hoc}; rather, it is computed at runtime, ensuring accurate protonation state ratios during the simulation.

\subsection{Dynamic Barrier and Well Optimization}
\label{sec:dbo}

The double well potential and Partition Function Correction described above satisfy requirements R1 and R2 (see Section~\ref{sec:lambdadyn_cph_intro}), while partially addressing R3 and R4. However, two issues remain to be resolved.

First, although the double well potential has wells centered at 0 and 1, the average $\lambda$ value within each well typically deviates from these ideal values, e.g.  when the local environment exerts strong forces on $\lambda$ through other terms of the Hamiltonian (eq~\ref{eq:H_basic}), causing the averaged protonated and deprotonated charges to diverge from their physically meaningful integer values.

Second, whereas the central barrier suppresses undesirable intermediate states, it also suppresses transitions between the protonated and deprotonated state, which are necessary for convergence in protonation space. 
Here, a practical and user-controllable trade-off between sampling efficiency and avoiding non-physical protonation states is desirable.

To address these concerns, we developed a Dynamic Barrier and Well Optimization (DBO) feature, 
which dynamically regulate the wells position and barrier height of the double well potential \Vdw.

The first algorithm fine-tunes the position of the well to ensure the desired average $\lambda$ value of 0 and 1, respectively, when in the well.
Our implementation automatically collects statistics on the $\lambda$ values when in the vicinity of each of the wells,
as well as the fraction of time spent there, in blocks of 40~ps duration \CPHParam.
The criteria for proximity to the well located at  $\lambda = 0$ and  $\lambda = 1$ is  $\lambda < 0.2$ and $\lambda > 0.8$, respectively.
At the end of each of these blocks, if $\lambda$ has remained more than 70\% of the time in the block near one endpoint, 
and if the difference between the average $\lambda$ and the ideal value 0 or 1 (depending on the considered well) is greater than a tolerance of 0.03 \CPHParam,
the well position is shifted laterally by 0.5 times this difference.
A maximum adjustment range ($\pm$0.08 \CPHParam) is used to avoid artifacts in highly coupled residues.

The second algorithm adjusts the barrier height for each residue independently (as well as for the protonation and tautomer coordinates within a residue if applicable).
It is inspired by the Adaptive Landscape Flattening technique typically used in ligand-binding \lambdadyn.\cite{raman2020automated,hayes2017adaptive}
Statistics on the number of frames spent in-transition, defined as satisfying $0.2 < \lambda < 0.8$, are collected in 1~ns blocks \CPHParam.
The central barrier \Vdw is adjusted up or down in 1.0~kJ/mol increments \CPHParam in a 1--20~kJ/mol interval \CPHParam 
until the fraction of in-transition frame reaches the given target value,
which is 25\% \CPHParam for this work, with a tolerance of 5\% \CPHParam. 
The starting barrier height is the same as for runs without barrier adjustment, namely 6 kJ/mol \CPHParam.

For the tautomer coordinate, two different barriers are used \CPHParam, depending on whether the site is protonated or deprotonated, 
as the force felt in the tautomerically degenerate state is close to zero, 
and the barrier should therefore be lowered to achieve the desired transition rate in that state.
The aim of this algorithm is to make the transition rates of all sites similar, 
taking into account additional barriers that the local environment of the respective titratable site may or may not induce, and without manual intervention by the user.

Following each adjustment, the $\lambda$-trajectory frames are tagged with a censoring flag for a duration of 10~ps \CPHParam.
These censored frames are not considered during post-simulation analyses to eliminate possible relaxation effects after the small but instantaneous change of $V(\lambda)$.
Both the barrier and the well adjustment can be enabled independently if desired \CPHParam.
In practice, DBO adjustments occur frequently during the first tens of nanoseconds of the simulation, after which
the regulated parameters typically stabilize until larger conformational changes occur.

\subsection{Sites Acting as Charge Buffers}

In the context of periodic boundary conditions and free energy calculations, 
 electroneutrality of the simulation box is crucial.\cite{hub2014quantifying,chen2014effects}
To ensure that this requirement is met despite the dynamically changing charges of titratable residues, 
buffer sites that take up a compensating charge\cite{wallace2012charge} have been used.
Water molecules were selected for this role, with the oxygen atom switching from its charge of $-0.834~e$ to $+0.166~e$ in the protonated buffer state, 
as described previously.\cite{dobrev2017}
An explicit negative buffer state was not necessary, as permanent negative ions (Cl$^-$) are automatically added 
to compensate for a positively charged initial state of the residues during system preparation. 
In the current implementation, each titratable residue is paired with a corresponding charge buffer site, 
and the total charge of each pair is kept constant.

To avoid artificial electrostatic interactions between a residue and its buffer site, 
the distance between the oxygen of the buffer molecule and the C$_\alpha$ of the titratable site is restrained to 3 nm \CPHParam ($> 3$ Debye lengths),
with a force constant of $k_\text{restr} = 50 $ kJ mol$^{-1}$nm$^{-1}$ \CPHParam. 
Although a distance of $3 $~nm was chosen here, 
longer distances are possible but require $\Vmm$ recalibration.

As the charge buffers should be located in bulk solvent, 
a minimum distance restraint was also added between each buffer molecule and every $\mathrm{C}_\alpha$ atom \CPHParam, 
as well as between the buffer themselves, ensuring a distance of at least 2~nm \CPHParam.
The group of atoms for which restraints are generated can be extended through a user setting to, for example, lipid headgroups, as necessary for the system of interest.
The one-sided restraint was achieved with an harmonic potential using the same force constant $k_\text{restr} = 50 $ kJ mol$^{-1}$nm$^{-1}$ \CPHParam, which applies force only when the distance is less than the threshold.
This reuses the existing \gromacs distance restraint code 
and is automatically introduced at the preprocessing stage.
The \gromacs \texttt{genion} tool, which places ions during system preparation, has been enhanced to add these buffers automatically, in accordance with the above distance conditions.

We are aware that for larger systems, 
the number of buffer sites may become so large as to require larger simulation boxes, which decreases simulation speed; 
a possible future solution is to adopt a strategy that exploits the fact that 'nearly all protonated' 
and 'nearly all deprotonated' states are rare in this case, 
and can therefore be neglected, thus markedly reducing the required number of buffer sites.\cite{Donnini2016}

\subsection{Force Field Modification}
The force fields used for this work --- and for which this implementation provides full support --- 
are \ffCharmmUsed\cite{huang2017} and \ffAmberUsed.\cite{best2009,Lindorff2010}
For the former, the recommended \Charmm-modified TIP3P water model\cite{huang2017} is used, 
whereas for the latter, unmodified TIP3P molecules are used.\cite{jorgensen1983comparison}

Glu and Asp residues each have two carboxylate oxygens susceptible to protonation.
For technical reasons, in many implementations of constant pH MD, including ours, both of these oxygens carry a hydrogen atom (proton), with the deprotonated states described by assigning zero charge to one or both of the protons.
However, because charge-neutralized hydrogens tend to adopt an undesirable \textit{anti} conformation 
that unphysically persists after the charge is restored,
we added a dihedral restraint to enforce correct stereochemistry.
This restraint is flat-bottomed and harmonic ($V(\phi) = \frac{1}{2}k(\phi - \phi_0)^2$ where $|\phi| > \phi_0 $), 
acting when the angle increases beyond 50 degrees \CPHParam toward the \textit{anti} conformation ($\phi_0 = 50^{\circ} = 0.87 \,\mathrm{rad}$), with a stiff force constant of $k=$ 30 kJ mol$^{-1}$rad$^{-2}$ \CPHParam.
The threshold $\phi_0$ was chosen such that $|\phi|$ is always lower in the protonated state, thus having no influence on the angle distribution in that state while effectively suppressing the \textit{anti} conformation in the deprotonated form.

In \ffAmberUsed, the charges of all atoms, including backbone atoms, differ between protonated and deprotonated forms of each protonatable residue.
As the backbone atoms have electrostatic interactions with the side chain atoms of the $N-1$ and $N+1$ residues (beyond the 1-3 exclusion), 
this would require a separate reference compound calibration for each possible pair of neighboring residues, which is impractical.
We have therefore chosen to follow previous work on this matter \cite{Mongan2004,dobrev2017} and deviate from the force field-specified backbone charges for ease of implementation.
The backbone atomic charges of the deprotonated form were thus reassigned to be the same as in the protonated form.
This backbone charge change caused a slight deviation from neutrality over the whole residue, which was compensated by distributing the opposite charge uniformly on the side chain atoms of the deprotonated form.
Note that this procedure does not need to be applied to \ffCharmmUsed due to unchanging backbone charges.

\subsection{Constant pH Simulations with FMM Electrostatics}
\label{sec:cph_with_fmm_estats}

Among all  Coulomb interactions between particle pairs in the simulation system, only a small fraction is affected by changes in $\lambda$ variables, even for systems with numerous titratable sites.
One should therefore expect minimal computational overhead for constant pH simulations relative to conventional fixed charge simulations.
However, a naive use of PME for constant pH necessitates computing electrostatics twice per $\lambda$-variable, one for each form of the titratable residue.
This duplication arises from the non-local nature of the Fourier transform underlying PME, requiring separate charge grids for each protonation state.
In previous implementations, the computational cost thus scaled with the number of sites,\cite{Donnini2011,dobrev2017} resulting in prohibitive slowdowns (e.g., 100-fold for a protein with 100 titratable sites) compared to conventional simulations.

We opted for rigorous Hamiltonian interpolation (eq~\ref{eq:H_basic}) over the charge scaling formulation, despite the good performance of the latter with PME.\cite{aho2022constph}
This choice required replacing the PME electrostatic solver with the Fast Multipole Method (FMM), whose locality maintains low computational overhead.
Our FMM implementation promises efficiency comparable to charge interpolation methods while enabling a correction that restores the proper behavior of Hamiltonian interpolation.\cite{Kohnke2023}

A detailed description of our FMM implementation, its integration with \lambdadyn, and comprehensive performance analysis is published separately.\cite{Kohnke2023}

\subsection{Simulations at Constant pH: Parameters and Setup}
%
%

All MD simulations were performed at a temperature of 300~K and a salt concentration of 150~mM NaCl.
Coulomb interactions were calculated using our GPU-accelerated FMM.\cite{Kohnke2023}
Simulations were conducted in the NVT ensemble (see also Supplementary Information 4.1), as our FMM implementation has currently not been validated for use with a barostat.
We employed the Bussi-Parrinello-Donadio thermostat\cite{bussi2007canonical} with a coupling time $\tau = 0.1$ ps and an integration step size of 2 fs.

Specific simulation settings are detailed in the template \mdp files provided in the Supplementary Information. 
We employed standard force field parameters, except for constant pH-specific settings.

\subsubsection*{Constant pH Parameters}

For the $\lambda$ particles, the Velocity Verlet integrator\citep{swope1982computer} with an integration step size of 2~fs was used.
The protonation state of the system was written to output files every 0.5~ps \CPHParam.

Test simulations suggested a mass of 60~u for the $\lambda$ particles, as an optimal trade-off between transition rates (affecting titration curve convergence) and integrator stability.
We note that \lambdadyn is rather insensitive to the particular choice of this mass, within an order of magnitude.

The temperature in the $\lambda$-subsystem was maintained, as in the MD simulations, 
by the Bussi-Parrinello-Donadio thermostat\cite{bussi2007canonical}
with a coupling constant of $\tau = 1$~ps \CPHParam.
Test simulations have shown that this particular thermostat is insensitive to the choice of coupling time for MD simulations in general,\cite{bussi2007canonical} within a 0.01 to 10~ps range, such that $\tau = 0.1$ ~ps could have been used as well.
We additionally verified that the choice of $\tau$ in that interval does not affect the \pKa in constant pH titrations.
Our implementation also supports the Andersen thermostat \CPHParam as an alternative.

A 50~ps $\lambda$-equilibration run, with position-restrained heavy atoms and a low $\lambda$-barrier, preceded all \lambdadyn simulations to ensure initial protonation states compatible with the conformation.

\subsection{Titration Simulations}
\label{sec:titration_simulation}

Constant pH MD enables computational titration of proteins. 
This involves simulating the protein at various pH values, using multiple replicas per pH point, to generate an analog of an experimental titration curve for each residue.
Such series of simulations are used to determine the \pKa of these residues 
as well as to study the pH-dependent behavior of the protein or residue of interest, including conformational change and coupling of protonation.
Table~\ref{tab:titrations} summarizes the parameters used for titration simulations for 5 test systems.
        
In the first group of titration simulations, the single residues Glu, Asp and His were titrated, each with both termini capped by methyl groups.
This methyl capping was realized by adding the ACE and NME pseudo-residues, 
which are acetyl and N-methyl moieties attached via peptide bonds, at the N-terminus and C-terminus respectively.
Parameters for ACE and NME are available in both the \ffCharmmUsed and \ffAmberUsed force fields.
Because these methyl-capped residues are the reference compounds that were used to calibrate the force field compensation potential $\Vmm(\lambda)$, this constituted a self-consistency check, as computational titration should reproduce the reference \pKa for these residues.

In the second set of titrations, the pentapeptides GEAEG, GEAHG, GHAEG and GHAHG were titrated, starting from an initial conformation of a straight chain (backbone in the same plane) 
with extended side chains, generated using the Avogadro software (version 1.2).\cite{hanwell2012avogadro}

\begin{table}[tbp]
\centering
\resizebox{\columnwidth}{!}{
\begin{tabular}{lclllcccc}
\toprule
                                    & PDB                                  & \multicolumn{3}{c}{reference \pKa}        & \# of titr. &\# of    & time per   & box  \\
Protein or residue                  & code                                 &    Asp    &    Glu    &    His            & residue &  replicas & repl. (ns) & (nm) \\ \midrule
Capped Asp, Glu, His             &                                      &    4.0    &    4.4    &    6.38           &  1 &  20    &   15       & 6    \\
Pentapeptides GXAXG                 &                                      &    N/A       &    4.08   &    6.54           &   2 & 30    &   50       & 6    \\
Cardiotoxin V                       & 1CVO\cite{singhal1993solution}      &    3.65   &    4.25   &    6.42$^\dagger$ &  4 &  40    &   100      & 8    \\
Hen lysozyme                        & 2LZT\cite{ramanadham1990refinement} &    3.65   &    4.25   &    6.42           &  10 &  40    &   75       & 8    \\
SNase mutant $\Delta \mathrm{PHS}$  & 3BDC\cite{Castaneda2009}            &    3.90   &    4.36   &    6.46           &  19 &  40    &   75       & 8    \\
\bottomrule
\end{tabular}
}
\caption{Specifications of titration simulations carried out with \ffCharmmUsed and \ffAmberUsed.
$^\dagger$Microscopic \pKa{}s 6.53 and 6.94.}
\label{tab:titrations}
\end{table}

The third group was the titration of snake cardiotoxin V from the Chinese cobra \emph{Naja naja atra}.\cite{singhal1993solution}
To facilitate a direct comparison with the recent PME-based constant pH implementation in \gromacs,\cite{aho2022constph,buslaev2022best} 
we chose simulation conditions for this protein that matched as closely as possible this work, including the pH range (from 1 to 8) and the simulation duration.\cite{buslaev2022best}

The fourth and fifth titration sets used the hen egg lysozyme and \textit{Staphylococcus} nuclease mutant $\Delta \mathrm{PHS}$, respectively.
Both of these proteins are typical benchmark systems for constant pH methods, thus enabling comparison with previous constant pH implementations.\cite{Harris2022,huang2016all,swails2014constant,aho2022constph,buslaev2022best} 

Titrations were performed over a \pH range encompassing all experimentally reported \pKa values, extending 1 pH unit beyond extremes in 0.5 unit steps.
Replica sets began with identical protein structures but different ion positions and initial velocities. 
To further decorrelate the individual trajectories, each replica was equilibrated for 1~ns prior to the  actual sampling run.
All titrations were performed both in \ffCharmmUsed and \ffAmberUsed to evaluate force field dependence, as well as with and without Dynamic Barrier Optimization to assess its impact on convergence and accuracy.
\textit{Staphylococcus} nuclease was an exception, simulated only with Dynamic Barrier Optimization enabled.

Comparing computational titration results to experimental Nuclear Magnetic Resonance (NMR) data presents challenges, particularly the choice of reference \pKa for each amino acid in water. 
Because directly measuring the resonance of the titrating proton is usually impossible due to the fast exchange kinetics, a nearby non-titrating atom ($^1$H, $^{13}$C, or $^{15}$N) is used as a reporter instead.
The choice of this reporter affects the measured \pKa values in the reference compound and in the protein.
Discrepancies of up to 1~\pH unit have been observed in multiple titrations of lysozyme, depending on the selected reporting atom.\cite{webb2011} 
We partially mitigated this issue by setting our reference \pKa values based on the reported values of the \pKa for single residues in the NMR titration work cited for each protein, ensuring direct \pKa comparability with experimental results.
These reference \pKa values are configurable in the input files \CPHParam .

\subsection{\pKa Calculation}
\label{sec:pka_caculation}

For each simulation frame in the trajectory, residues were assigned protonated ($\lambda_p \geq 0.5$) or deprotonated ($\lambda_p < 0.5$) states.
Because our Partition Function Correction ensures correct populations for both halves of the double well potential,  frames with $\lambda_p$ near 0.5 can be retained, contrary to common practice, thus improving sampling.

For each replica and for each pH value, the deprotonation fraction, which is the ratio $x = N_{p = 0}/N$ between the number $N_{p = 0}$ of deprotonated and the total number $N$ of frames was computed for each pH point and each replica, respectively. 
The resulting titration curve $\tilde{x}(\mathrm{pH})$ was fitted to a Henderson-Hasselbalch (H-H) equation, $\tilde{x}(\mathrm{pH}) = \frac{1}{10^{(\mathrm{pK}_a - \mathrm{pH})} + 1}$ to determine the \pKa, using SciPy's non-linear least squares routine.\cite{SciPy20200}
For pentapeptide titrations, we fitted the fractions $x$ and pH values to a Hill equation model, $\tilde{x}(\mathrm{pH}) = \frac{1}{10^{n(\mathrm{pK}_a - \mathrm{pH})} + 1}$, yielding both a \pKa and a cooperativity parameter $n$.

The two microscopic \pKa values of His (p$K_{a,\delta}$ and p$K_{a,\epsilon}$) were determined similarly with a H-H curve, with ratios  $x_{\mathrm{micro,}\delta} = N_{p = 0,t = 0}/(N_{p = 1} + N_{p = 0, t = 0})$ and $x_{\mathrm{micro,}\epsilon} = N_{p = 0,t = 1}/(N_{p = 1} + N_{p = 0, t = 1})$, respectively, 
with $N_{p = 0,t = 0}$ the number of deprotonated frames for the $\delta$ tautomer, and $N_{p = 0,t = 1}$ the number of deprotonated frames for the $\epsilon$ tautomer.

\label{BootstrappingError}

Error bars were estimated using bootstrapping. 
Specifically, for each pH point, we resampled with replacement $R$ data points from the set of deprotonation fraction $x$, where  $R$ is the number of replicas for that particular run (see Table~\ref{tab:titrations}).
The H-H curve fitting was done as described above, to obtain a \pKa and, where appropriate, a Hill coefficient $n$.
This procedure was repeated 5000 times, with reported confidence intervals corresponding to a 5\% significance level for both \pKa and $n$.

\subsection{Analysis of Direct Residue-Residue Coupling}
\label{sec:directCoupling}

The titration curves of monoprotic molecules  follow the well-known H-H sigmoid curve, but macromolecular polyprotic acids and bases, including proteins, 
exhibit more complex titration behavior.\cite{ullmann2003relations,Onufriev2001}
This behavior arises from protonation coupling, which falls into several qualitative categories, like residue-residue coupling\cite{wyman1990binding,ben2013cooperativity,hill2013cooperativity} or protonation-conformation coupling\cite{shi2012thermodynamic,machuqueiro2006constant,sarkar2020ph,di2012ph} (Section~\ref{ProtonationConformationCouplingMethodParagr}).
In this section, we address this first category, coupling between several protonatable residues, typically mediated by electrostatic interaction between them, which we refer to as \textit{direct residue-residue coupling}.

We first screen for interactions between titratable sites using Normalized Mutual Information (NMI) , a metric based on the information theoretic quantity Mutual Information (MI),\cite{thomas2006elements,shannon1948mathematical} normalized to conveniently adopt values between 0 and 1.
Here, NMI quantifies how much the protonation state of one site informs about the protonation state of the other, detecting correlations between the two in a symmetric, quantitative and model-free manner.
Once clusters of interacting residues are identified through NMI,
an effective model of this coupling can be fitted.

Here we will fit macroscopic titration curves\cite{ullmann2003relations,wallace2012charge} for such interacting clusters.
First, the macroscopic titration curve is extracted from the simulation by averaging the total number of protons bound to this cluster, $X$, with this average, $\left \langle X \right \rangle$, being computed for each pH value and each replica.
Subsequently, a suitable macroscopic titration equation, e.g., for two interacting residues

\begin{equation}
\left \langle X \right \rangle = \frac{10^{\mathrm{p}K_{a,2} - \mathrm{pH}} + 2\cdot 10^{\mathrm{p}K_{a,1} + \mathrm{p}K_{a,2} - 2\cdot\mathrm{pH}}  }{1 + 10^{\mathrm{p}K_{a,2} - \mathrm{pH}} + 10^{\mathrm{p}K_{a,1} + \mathrm{p}K_{a,2} - 2\cdot\mathrm{pH}} },
\end{equation}

is fitted. 
From this fit, the macroscopic \pKa values $\mathrm{p}K_{a,1} $ and $\mathrm{p}K_{a,2}$ are obtained, which describe the collective titration behavior.
No clusters involving more than 2 residues were identified in this work, though the described approach generalizes to any number of protonatable sites.

To calculate the NMI, we discretized the $\lambda_p$ trajectories into binary trajectories.
Each frame was classified as protonated or deprotonated based on whether its $\lambda_p$ value exceeded 0.5.
In our work, trajectory frames were saved every 0.5~ps, implicitly low-pass filtering the data to mitigate repeat boundary crossing effects.
For the trajectories of each replica, pH point and protonatable residue, separately, $\mathrm{NMI} = 2 \cdot \mathrm{I}(\mathrm{X};\mathrm{Y})/(\mathrm{H}(\mathrm{X}) + \mathrm{H}(\mathrm{Y}))$ was computed, where $\mathrm{I}$ is the mutual information and $\mathrm{H}$ is the information theoretic entropy of the trajectory.
This latter term is defined as $\mathrm{H} = \sum^{N}_{i=1} p_i \ln p_i$, with $p_i$ the probability of state $i$ (among $N$) in each frame, in our case the two protonated and deprotonated states.
Both $\mathrm{I}$ and $\mathrm{H}$ were computed with the C++ library InfoTheory.\cite{infotheory}
NMI values close to 1 correspond to a high correlation between protonation states, denoting coupling between residues, as opposed to values close to 0, which indicate statistical independence.

We found that the NMI metric can take on spuriously high values when both of the trajectories have very low, but not zero, entropy.
This can occur when residues are almost always protonated except for a few frames, sometimes leading to false positives for coupling.
To eliminate these, we considered the mean NMI and entropy $\mathrm{H}$ values over all replicas at a given pH point.

Specifically, we considered the conjunction of mean NMI and mean $\mathrm{H}$ values larger than an empirically determined threshold of 0.1, across replicas and at any pH point, to be indicative of coupling.

\subsection{Analysis of Protonation-Conformation Coupling}
\label{ProtonationConformationCouplingMethodParagr}

The second coupling, mentioned above, is \textit{protonation-conformation coupling}\cite{shi2012thermodynamic,machuqueiro2006constant,sarkar2020ph,di2012ph} and can also be studied using constant pH simulations as follows.
This coupling arises from changing protein conformations, which lead to varying local environments for titratable residues and ultimately to shifts in their \pKa. Conversely, changes in protonation state can trigger conformational rearrangements.

This type of coupling can be detected and quantified via Partial Least Squares-based (PLS) Functional Mode Analysis\cite{hub2010detection,Krivobokova2012} (FMA). While csonceptually similar to Principal Component Analysis (PCA) of protein trajectories, FMA identifies collective motions that correlate most with protonation state rather than structural variance.

For each titratable residue, we performed FMA using the protonation state ($\lambda_p$) as the target observable and the heavy atom positions in the protein as the structural ensemble.
To detect protonation-conformation coupling, representative structures of the protonated, deprotonated, and intermediate states should be observed in the input trajectory.
We achieved this by selecting trajectories from replicas whose pH was closest to the \pKa of the residue of interest. 
We kept 20\% of the replicas as a validation set to compute the explained variance in protonation. 
After correcting for periodic boundary effects, we combined the remaining 80\% of trajectories into a structural ensemble, fitting them to the average protein structure's backbone.
The FMA was performed on this ensemble using the first 20 PLS components.
The number of PLS components should be large enough to capture most of the variation in $\lambda_p$, 
and is optimally determined by cross-validation for each residue individually.
Across all analyzed residues, we found that the explained variance in protonation plateaued past 10 to 15 PLS components.
For consistency and simplicity, we therefore used 20 PLS components for all residues.

\begin{figure}[hbtp]
\includegraphics[width=0.90\textwidth]{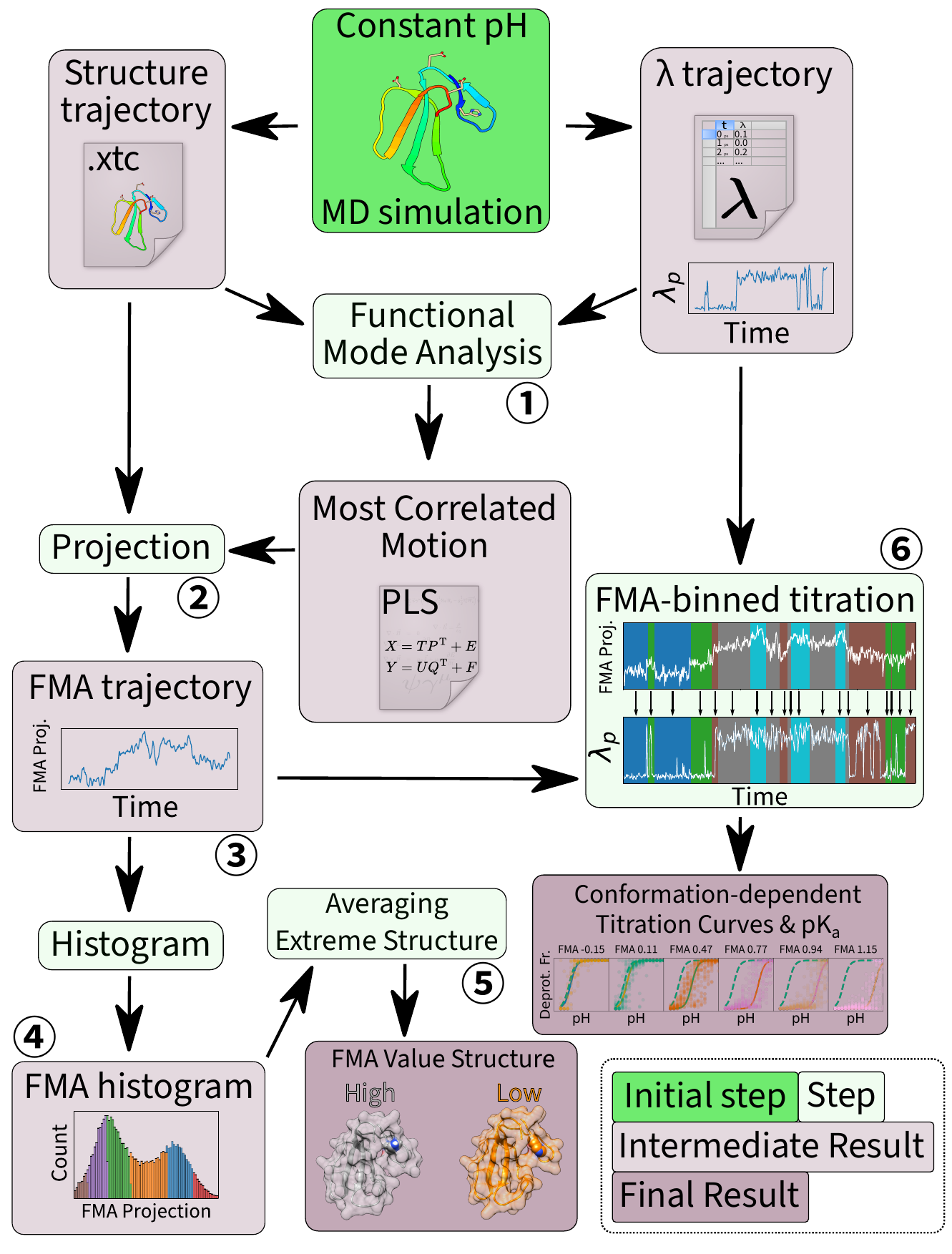} 
\caption{\textbf{Flowchart of FMA-related analysis}, starting with constant pH MD (green), followed by intermediate processing (light green), yielding intermediate (light purple) and final results (dark purple).
Circled numbers are referenced in the main text. 
Icons, plots and data in this figures are for illustration purposes.}
\label{fig:FMAFlowchart}
\end{figure}

Figure~\ref{fig:FMAFlowchart} illustrates several interdependent analyses we carried out based on the FMA \Circled{1}.
First, we projected \Circled{2} the structure trajectories of each replica at each pH value onto the most correlated motion. 
This yielded a one-dimensional projection of the high-dimensional conformational space onto the collective motion best predicting protonation/deprotonation, called an FMA trajectory \Circled{3}.
We then histogrammed these FMA trajectories \Circled{4} to create discrete bins based on percentiles ( $\leq$ 5th percentile, 5th to 25th, 25th to median, median to 75th, 75th to 95th, and finally $>$ 95th).

Selecting and averaging \Circled{5} structures from the highest and lowest bins yielded representative protonated and deprotonated state structures, called \emph{high} and \emph{low FMA value structures}, respectively.
These were used for subsequent analyses of the coupling mechanism.

In parallel to structure averaging, we quantified the \pKa shift associated with the conformation change.
To this end, we grouped all $\lambda$ trajectory frames according to their respective bins in the FMA histogram, whose boundaries were defined above.
For each group, we fitted a H-H titration curve to obtain \pKa values, as described in Section~\ref{sec:pka_caculation}. 
This process, called \emph{FMA-binned titration} \Circled{6}, yields a series of titration curves and \pKa values as a function of FMA values.

\subsection{Practical Aspects}
\label{PractAspectSubSection}

The constant pH functionality described above has been integrated into the \gromacs  MD simulation software suite,\cite{Hess:2008tf,Abraham2015}
so that users can perform constant pH MD simulations with minimal changes to their existing workflow.
Specifically, routine constant pH MD simulations require only a few additions to the MD parameter file (\mdp extension), 
as exemplified in Figure~\ref{fig:minimalMDP}.

\begin{figure}[h!]
\begin{lstlisting}
; Enable constant pH
lambda-updater-enable        = yes
constantph-enable            = yes

; Select lambda dynamics mode, with pH = 7.4
constantph-mode              = dynamics
constantph-fixed-ph          = 7.4
; Output lambda value every 1000 steps
constantph-nstout            = 1000

; Use the FMM electrostatic solver
; d = 2, p = 8 are appropriate settings for single precision
; in a 8x8x8 nm box
coulombtype                  = FMM       
fmm-override-multipole-order = 8
fmm-override-tree-depth      = 2
\end{lstlisting}
\caption{Minimal but fully functional \mdp parameters for constant pH simulations.}
\label{fig:minimalMDP}
\end{figure}

The parameters denoted by \CPHParam in the previous sections correspond either to additional directives that may be added to this \mdp file, to modify and adapt the default set-up; or to modifications in the topology and force field files.

Starting from any PDB file suitable for non-CPH MD,
the \texttt{pdb2gmx} tool now offers new command line switches (\texttt{-ldglu}, \texttt{-ldhis}, and \texttt{-ldasp}) to enable protonation for Glu, His, and Asp, respectively. 
This tool generates a topology file with the necessary CPH MD information, provided a suitably modified force field is used (see Supplementary information for the \ffCharmmUsed and \ffAmberUsed force field files used in this work)

After protein solvation, the \gromacs tool \texttt{genion} adds both the standard ions and the charge buffer sites for constant pH, 
automatically placing the latter at appropriate distances.

Following an unmodified energy minimization step, equilibration proceeds with new parameters to control the constant pH code.
These additional parameters, parsed as usual by the GROMACS \texttt{grompp} command, are listed in Figure~\ref{fig:minimalMDP}, for a constant pH simulation at pH 7.4.
While numerous other constant pH parameters are available, those shown in Figure~\ref{fig:minimalMDP} will likely suffice for most scenarios, as we have carefully set appropriate general-purpose defaults for the others.

Simulations are run using the GROMACS command \texttt{mdrun}, with the additional argument \texttt{-lambdaout} specifying the filename for storing $\lambda$ values for each residue.
As in non-CPH MD simulations, the protein trajectory is stored in \xtc or \trr files.

For detailed information on all constant pH parameters, \mdp templates, analysis protocols, and Fast Multipole Method (FMM) settings, we direct readers to the tutorials and documentation available at \url{https://www.mpinat.mpg.de/grubmueller/gromacs-fmm-constantph}.
The source code can be accessed at \url{https://gitlab.mpcdf.mpg.de/grubmueller/fmm}.

\section{Results and Discussion}
\label{sec:result_begin}
 
Our results are grouped according to three goals: (1) assessing the accuracy of our CPH method; (2) providing sample applications for our analysis tools for studying protonation coupling; and (3) investigating and addressing sampling issues.

To these ends, we examine test systems of increasing complexity, from single residues for consistency checks, via pentapeptides, cardiotoxin V, lysozyme, and ending with staphylococcal nuclease, a challenging CPH benchmark system.
For each system, we compare computed \pKa{} values with NMR titration results, then present the results of our analysis methods, focusing on certain protonation coupling relations  with interesting properties.
Where major differences occur between force fields, we comment on results for both \ffCharmmUsed and \ffAmberUsed; otherwise only for \ffCharmmUsed for brevity.
Thus, unless explicitly specified, figures illustrate the results of \ffCharmmUsed simulations.

We conclude by addressing three general topics: (1) sampling improvement via Dynamic Barrier Optimization; (2) the impact of protonation coupling on convergence, and proposed solutions; and (3) overall accuracy of our CPH code, considering force field influence and residue-specific performance (His, Asp, Glu).

Computational performance and implementation details of the Fast Multipole Method underlying our CPH implementation will be presented in our companion publication.\cite{Kohnke2023}

\subsection{Single Residue Titration}

We first performed computational titration on methyl-capped residues: His, Glu, and Asp.
As these molecules serve as reference compounds for which calibration data were collected,
 our titration calculations should recover their reference \pKa values, providing a self-consistency check of our implementation.

\begin{figure}[h!]
\includegraphics[width=0.95\textwidth]{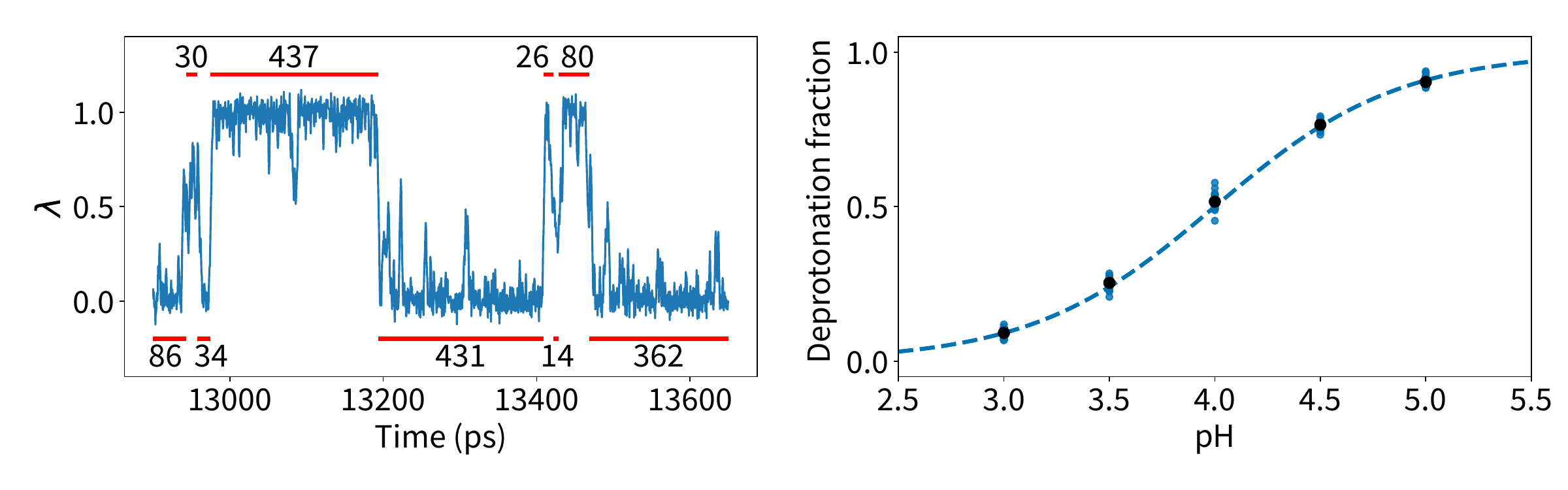} 
\caption{\textbf{Titration of an Asp residue}. 
Left: Excerpt of a trajectory of the $\lambda_{p}$ value at $\mathrm{pH} = 4$. 
Red intervals indicate the number of frames assigned to the protonated and deprotonated states.
Right: Resulting titration curve, with
blue dots showing the fraction of deprotonated frames for each replica at each pH point.
Black dots show the average of all replicas.
The dashed blue line is a Henderson-Hasselbalch (H-H) fit.}
\label{fig:ExampleAspTrajectoryTitrationCurve}
\end{figure}

The computational titration process, described in Methods sections \ref{sec:titration_simulation} and \ref{sec:pka_caculation}, is illustrated here in detail for the Asp residue in Figure~\ref{fig:ExampleAspTrajectoryTitrationCurve}.
The left panel shows an excerpt of a $\lambda_{p}$ trajectory, representing the time-dependent fractional protonation state. 
Frames are assigned to protonated or deprotonated states (red bars) based on their $\lambda_{p}$ values, with the number of frames in each state recorded for each replica and pH value.
The right panel shows the ratio of deprotonated frames to total frames ('deprotonation fraction') as a function of the pH at which each simulation was run.
Fitting a titration curve, such as the H-H sigmoidal curve (dashed line in the right panel), yields the \pKa value of the residue. 
The scatter of the deprotonation fraction for the individual trajectories allows assessment of convergence and \pKa uncertainty estimation. 
For these single residues, the small scatter indicates the absence of sampling issue.s

\begin{table}[h]
\centering
\begin{tabular}{lllll}
\toprule
Residue & Macro/Micro \pKa &  \ffCharmmUsed & \ffAmberUsed & Reference \\
 \midrule
His & Macro & $6.39 \pm 0.01$ & $6.36 \pm 0.02$ & $6.38$ \\
 & Micro $\delta$ & $6.53 \pm 0.01$ & $6.51 \pm 0.02$ & $ 6.53$ \\
 & Micro $\epsilon$ & $6.93 \pm 0.02$ & $6.88 \pm 0.03$& $6.92$ \\
 \midrule
Glu & Macro & $4.38 \pm 0.01$ & $4.36 \pm 0.02$ & $4.40$ \\
 \midrule
Asp  & Macro  & $3.98 \pm 0.01$ & $3.98 \pm 0.02$ & $4.00$ \\
\bottomrule
\end{tabular}
\caption{\textbf{Summary of \pKa values obtained from single residue titration for} for His, Glu and Asp, using two different force fields, with respective reference \pKa values.}
\label{tab:titrations_single_res}
\end{table}

Table~\ref{tab:titrations_single_res} presents the \pKa values from our simulated titrations, which exhibit very small uncertainties and align closely with their reference values, constituting a successful self-consistency test for both force fields used.
In particular, these results demonstrate that the calibration potential $\mathrm{V}_{\mathrm{MM}}$ for each residue was correctly implemented, using a polynomial of sufficient degree to accurately fit the free energy surface and with enough sampling for convergence of the calibration simulations, as well as for the computational titrations.

\FloatBarrier
\subsection{Pentapeptide Titration}
 
Unlike protonatable residues in proteins, the above single residues lack both the influence of neighboring protonatable groups and conformation-linked effects --- two phenomena of great interest to CPH MD practitioners.
To assess these effects while avoiding sampling issues, we selected a test system of appropriate, small size.

We therefore revisited the pentapeptides from our previous work.\cite{dobrev2020}
These peptides, with sequence GEAEG, GEAHG, GHAEG and GHAHG, were designed with two protonatable residues (His or Glu) on the same side of the peptide, opening the possibility of electrostatic interactions and (anti)cooperativity behavior, a form of residue-residue coupling.

\begin{figure}[h!]
\includegraphics[width=0.90\textwidth]{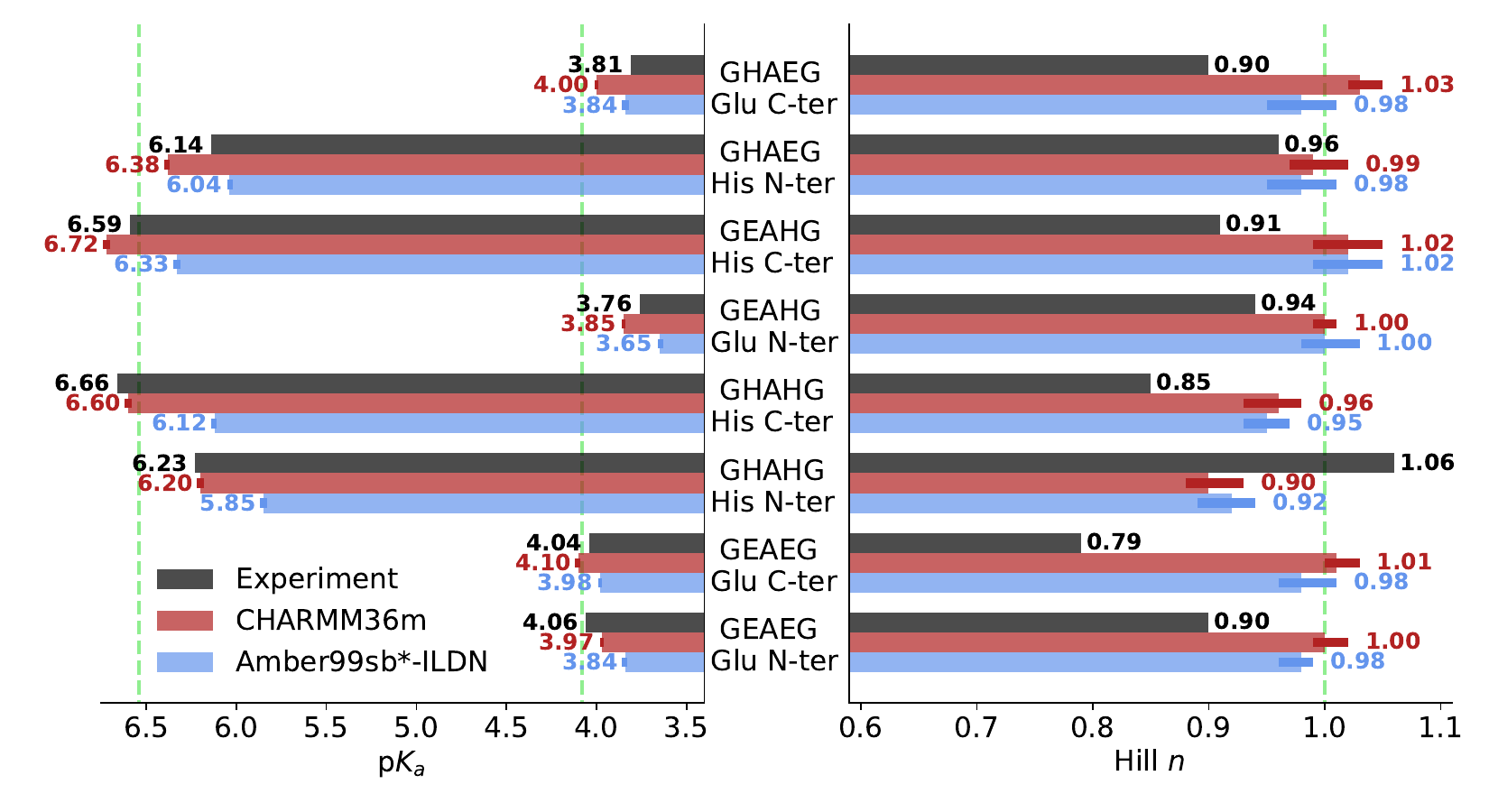} 
\caption{\textbf{ Computational vs.\ experimental titration of short peptides.} Shown are \pKa values (left) and Hill coefficients (right), calculated using \ffCharmmUsed (red) and \ffAmberUsed (blue), and measured by NMR (black).
Error bars indicate bootstrapped 95\% confidence intervals.
The reference \pKa values for single residues (Glu 4.08, His 6.54) as well as the unity Hill coefficient $n$ are shown as dashed green lines, representing the behavior of isolated solvated amino acids for comparison.}
\label{fig:PentapeptidePlotpKaN}
\end{figure}

We first examined whether the simulation duration used, 50~ns per replica, was sufficient to achieve protonation convergence.
As can be seen from Supplementary Information 2.1, all replicas yields similar protonation fractions with small scatter, and all titration curves are fitted well by the sigmoidal H-H curves.  We therefore concluded that the simulations were sufficiently converged.

Figure~\ref{fig:PentapeptidePlotpKaN} compares computational and NMR-measured \pKa values (left) and Hill coefficients $n$ (right).
Computational titration accurately reproduces all experimental \pKa{} values, capturing both the shift direction relative to isolated amino acids and the absolute values.
The overall \pKa{} Root Mean Squared Error (RMSE) is 0.13 for \ffCharmmUsed and 0.26 for \ffAmberUsed, with the latter moderately underestimating His \pKa.

In the simulations, all pentapeptides followed a H-H model, with Hill coefficients close to $n=1.0$ (Figure~\ref{fig:PentapeptidePlotpKaN}, right panel), meaning that the protonation state of each protonatable residue was largely independent of the protonation state of the second protonatable residue.
We also tested this finding using mutual information analysis.
Every protonatable residue pair has NMI values below 0.1, often even below 0.01.
This confirms that the protonation states of the two titratable residues are uncorrelated.

The simulations thus rule out residue-residue coupling by two separate methods.
However, experimental measurements (Figure~\ref{fig:PentapeptidePlotpKaN}, right panel) contrast with these results, showing Hill coefficients as low as 0.79 for GEAEG.
We discussed this discrepancy in our previous work,\cite{dobrev2020} which highlights that a straightforward interpretation of the measured Hill coefficient is challenging, complicating direct comparison to constant pH MD results.
Notably, this issue does not affect \pKa value comparisons.
Furthermore, including this study, a total of three simulation studies using two different constant pH MD implementations and three different force fields have now obtained similar unity Hill coefficients, in contradiction to experimental results.

\FloatBarrier
\subsection{Cardiotoxin Titration}
\label{sec:CardiotoxinTitr}
Next, we studied the small, 62-residue protein cardiotoxin V, which contains four titratable residues.
This protein is widely used as a benchmark system for CPH MD methods, as its small size allows for long simulations, keeping sampling issues under control.

\begin{figure}[h]
\includegraphics[width=0.90\textwidth]{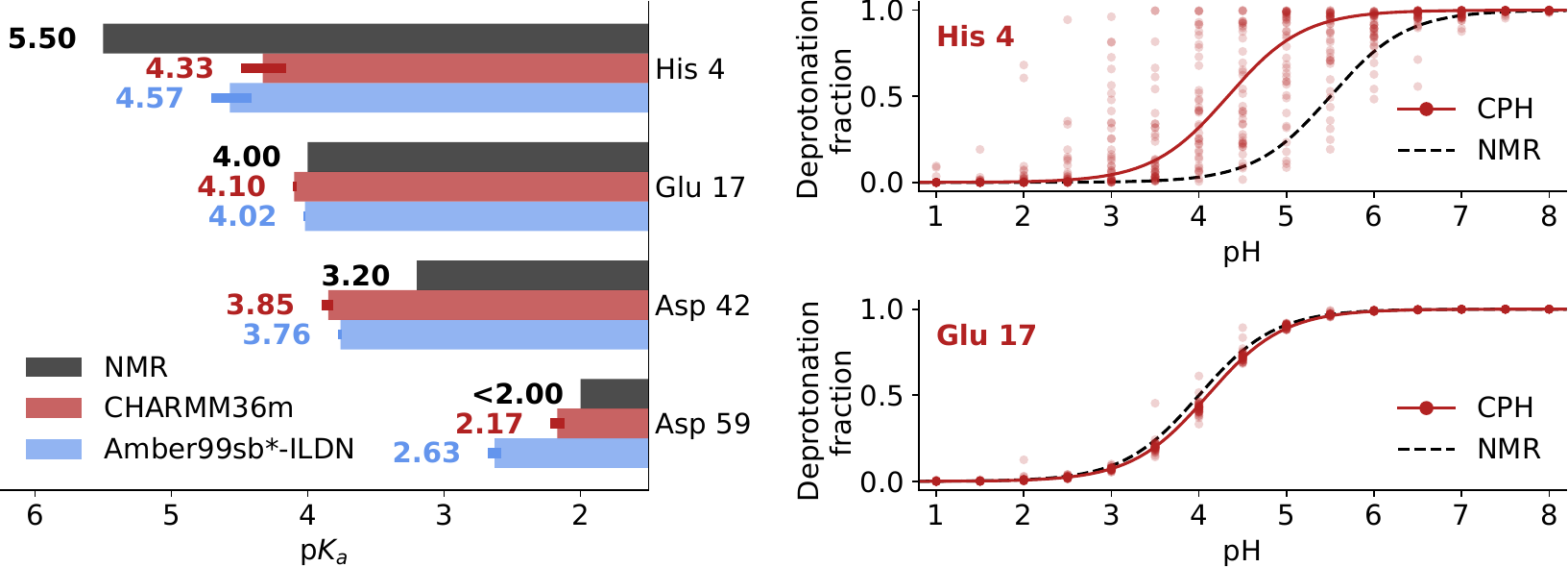} 
\caption{\textbf{ Computational vs.\ experimental titration of Cardiotoxin V.} 
Left: \pKa values calculated using \ffCharmmUsed (red) and \ffAmberUsed (blue), and measured by NMR (black).
Error bars indicate bootstrapped 95\% confidence intervals.
Right: Titration curve for His 4 (top) and Glu 17 (bottom), with fraction of deprotonated frames for each replica (point) and the H-H fit to simulation data as solid line.
The dashed line is a H-H curve based on the NMR \pKa values.}
\label{fig:CardiotoxinPlotpKaExampleTitrCurve}
\end{figure}

\subsubsection{Aggregate accuracy and inter-replica spread}

We first evaluated the overall accuracy of our constant pH simulations against \pKa values measured by NMR.
To this end, we conducted constant pH MD simulations using 40 replicas of 100~ns duration per pH value, spanning a pH range of 1 -- 8.
The overall \pKa  error in these simulations is small ($\mathrm{RMSE}<0.7$ for both force fields).
At the residue level, quite different degrees of convergence and, hence, accuracies are achieved.
The right panel of Figure~\ref{fig:CardiotoxinPlotpKaExampleTitrCurve} illustrates two contrasting examples.
For Glu~17 (bottom), we observed a very small inter-replica spread.
Every replica (transparent circle) at each pH point (x-axis) shows a similar deprotonation fraction (y-axis), indicating good overall convergence of this titration.
Its \pKa is also close to the measured value.
In contrast, His~4 (top) displays a wide spread in deprotonation fraction,
indicating that each replica experienced a very different free energy of protonation $\Delta G_{\mathrm{prot}}$, causing incomplete convergence of the overall titration.
This lack of convergence is associated here with a marked \pKa error.
We observed similar "spread-replica titrations" for certain residues in all other test systems described below.
Elucidating the underlying causes of this $\Delta G_{\mathrm{prot}}$ heterogeneity would help enhance convergence, or reduce the required simulation time to reach a desired accuracy.

To this end, we speculated that this behavior stems from protonation coupling -- either between protonation sites or between a titratable group and a slow process like conformational dynamics.
To test this idea, we used our analysis tools to detect these couplings, and
asked whether or not they correlate with observed spread-replica titration, as well as investigated the mechanism of these couplings.

\subsubsection{Analysis of protonation coupling}

We first searched for direct residue-residue coupling, where the protonation of one titratable residue influences that of another residue, mainly through electrostatic interactions.
To this end, we analyzed correlated protonation/deprotonation transitions for all pairs of residues in cardiotoxin V using NMI (Methods section~\ref{sec:directCoupling}).
This analysis yielded very small NMI values, below 0.1 at all pH, indicating no significant direct residue-residue coupling, an expected outcome given the large separation ($> 8 $ \AA) between all four titratable residues.

We therefore investigated a possible second coupling type, conformation-protonation coupling, where structural changes shift the protonation of a particular residue, or, vice versa, protonation changes induce structural changes.
To examine this coupling, we applied FMA to each titratable residue (Methods section~\ref{ProtonationConformationCouplingMethodParagr}).

\begin{figure}[htb]
\includegraphics[width=0.90\textwidth]{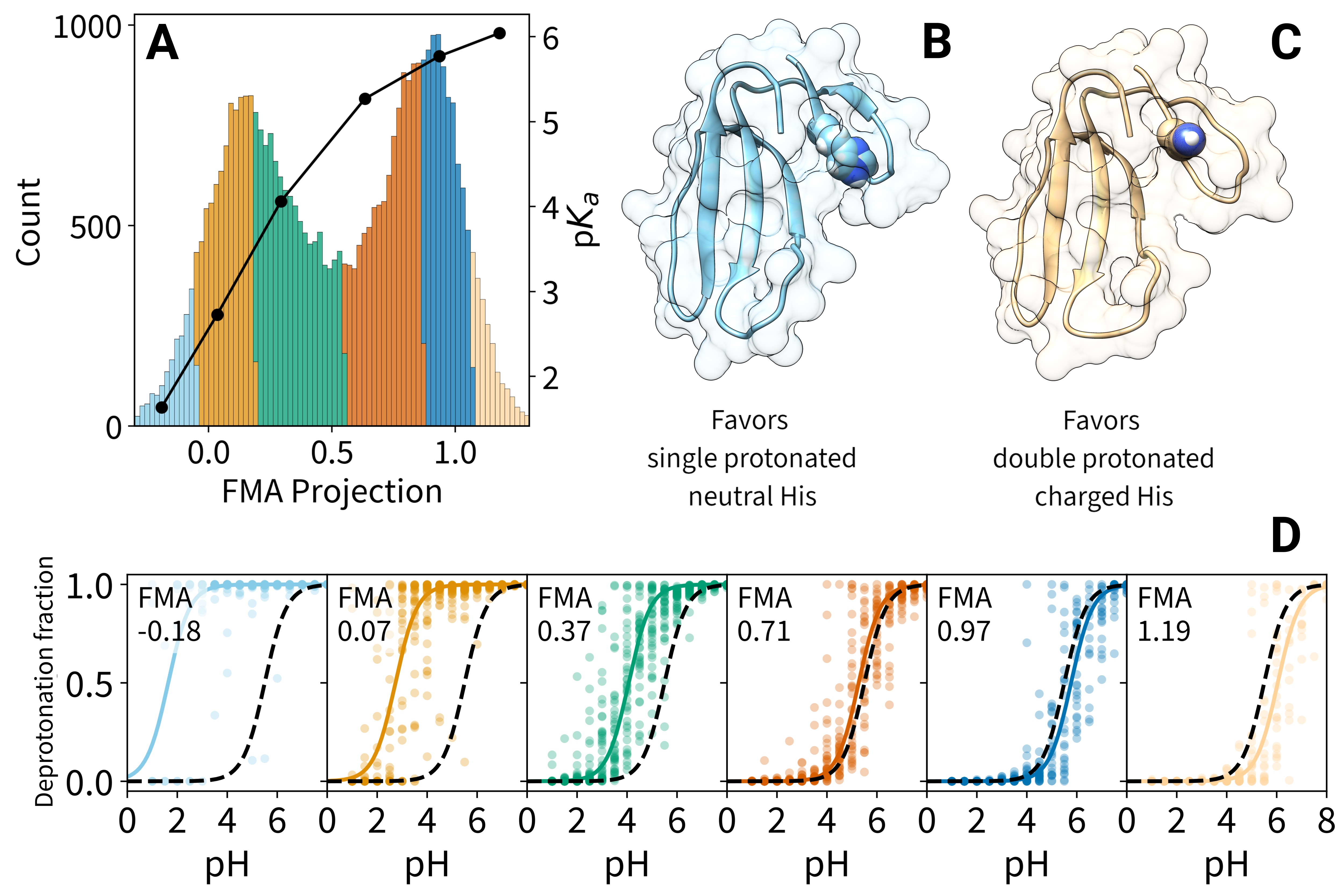} 
\caption{\textbf{FMA analysis for the His~4 residue of cardiotoxin V}.
(A): Histogram of FMA projection values for all replicas at pH = 4.5 (left axis) 
and \pKa as a function of the projection value (black curve, right axis).
(B), (C): Low and high FMA value structure, respectively.
(D): Separate titration curves for the six colored bins shown in A (colors correspond), 
with fraction of deprotonated frames for each replica (points), H-H-fits to these fractions (solid lines), and H-H-curves (black dashed lines) based on the measured \pKa value.
}
\label{fig:CardiotoxinHis4FMAPanel}
\end{figure}

Indeed, for \textbf{His 4}, FMA reports that 53\% of the protonation variance is explained by conformation coupling.
Such strong coupling may thus explain the spread-replica titration seen for this residue. 

To characterize this coupling, we examined the histogram of FMA projection values (Figure~\ref{fig:CardiotoxinHis4FMAPanel}A), which exhibited two broad peaks around 0.2 and 0.8.
These peaks represent two main conformations associated with predominantly single protonated and double protonated states of the residue, respectively (Figure~\ref{fig:CardiotoxinHis4FMAPanel}B and \ref{fig:CardiotoxinHis4FMAPanel}C).
To study the motion associated with the coupling, we contrasted protein structures representative of both ends of the FMA value range, obtained by averaging protein conformations within the two outermost FMA bins (Figure~\ref{fig:CardiotoxinHis4FMAPanel}A, blue and beige sections), yielding what we call the low and high FMA value structures (Figure~\ref{fig:CardiotoxinHis4FMAPanel}B and \ref{fig:CardiotoxinHis4FMAPanel}C).
Key structural differences are evident in the Cys 3 to Tyr 12 loop and the His 4 side chain orientation.
At low FMA values (Figure~\ref{fig:CardiotoxinHis4FMAPanel}B), His 4 is partially buried within the more hydrophobic loop backbone environment,  while at high FMA values (Figure~\ref{fig:CardiotoxinHis4FMAPanel}C), it becomes solvent-exposed.
In both conformations, side chains of the loop adapt to accommodate His 4, resulting in additional local structural changes.

To elucidate the relationship between conformational changes and observed protonation states, we computed the \pKa of His~4 separately for several conformations along the FMA coordinate. 
These conformations are indicated by different colors in Figure~\ref{fig:CardiotoxinHis4FMAPanel}A. The resulting titration curves, with corresponding colors, are shown in Figure~\ref{fig:CardiotoxinHis4FMAPanel}D.
We call this process FMA-binned titration (Methods section~\ref{ProtonationConformationCouplingMethodParagr}).
The resulting \pKa values as a function of the FMA coordinate are shown as a black line in Figure~\ref{fig:CardiotoxinHis4FMAPanel}A.
As can be seen, the solvent-exposed conformation (Figure~\ref{fig:CardiotoxinHis4FMAPanel}C) shows a \pKa of 5.7, closer to the solution \pKa of histidine (6.4), whereas the burying of the histidine within the more hydrophobic environment (Figure~\ref{fig:CardiotoxinHis4FMAPanel}B) drastically reduces the \pKa by more than 4 units.

\begin{figure}
\includegraphics[width=0.95\textwidth]{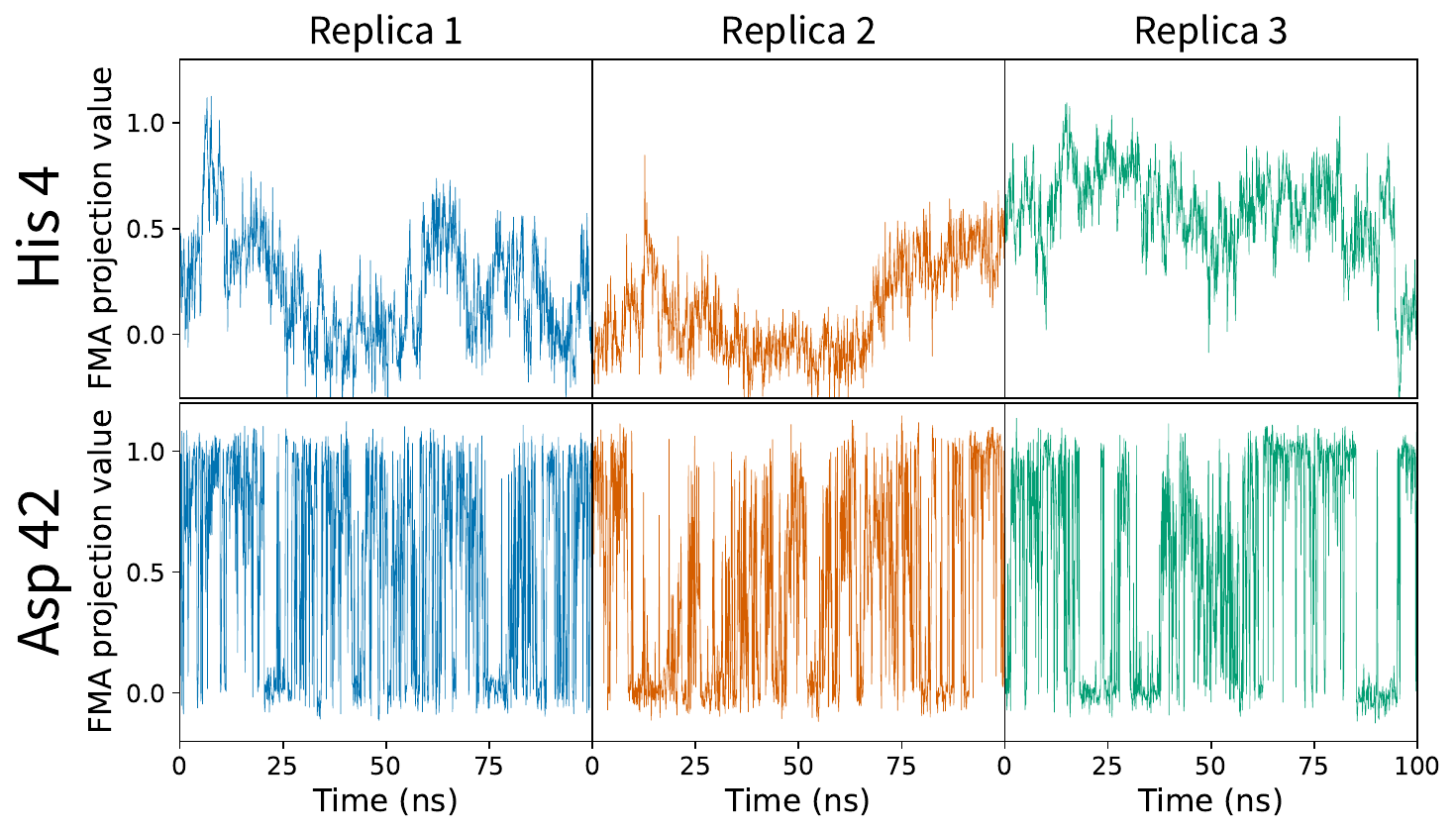} 
\caption{\textbf{Sample FMA trajectories for the His 4 and Asp 42 residues} of cardiotoxin V, for the three first replicas, at \pH $= 4.0$.}
\label{fig:FmaTrajCompareHis4Asp42}
\end{figure}

These results establish that the conformational state is a major determinant of the protonation of His~4.
Examination of the time evolution of the FMA coordinate (Figure~\ref{fig:FmaTrajCompareHis4Asp42}, top row) shows that the transition between these conformational states occurs very slowly, observed only once or twice within each 100~ns replica simulation.
Although we have not established strict causality, it is likely that the protonation fraction follows the conformational dynamics, explaining the slow convergence.
Each replica samples a partially distinct subset of conformations associated with this coupling (Figure~\ref{fig:FmaTrajCompareHis4Asp42}, top row), resulting in varying fractions of deprotonated states, thus explaining the observed large inter-replica spread.

It is instructive to also examine the other three protonatable residues of cardiotoxin V along similar lines.
As these residues do not exhibit spread-replica titration to the extent of His 4, they nicely illustrate other possible coupling behaviors.

\begin{figure}[htbp]
\includegraphics[width=0.95\textwidth]{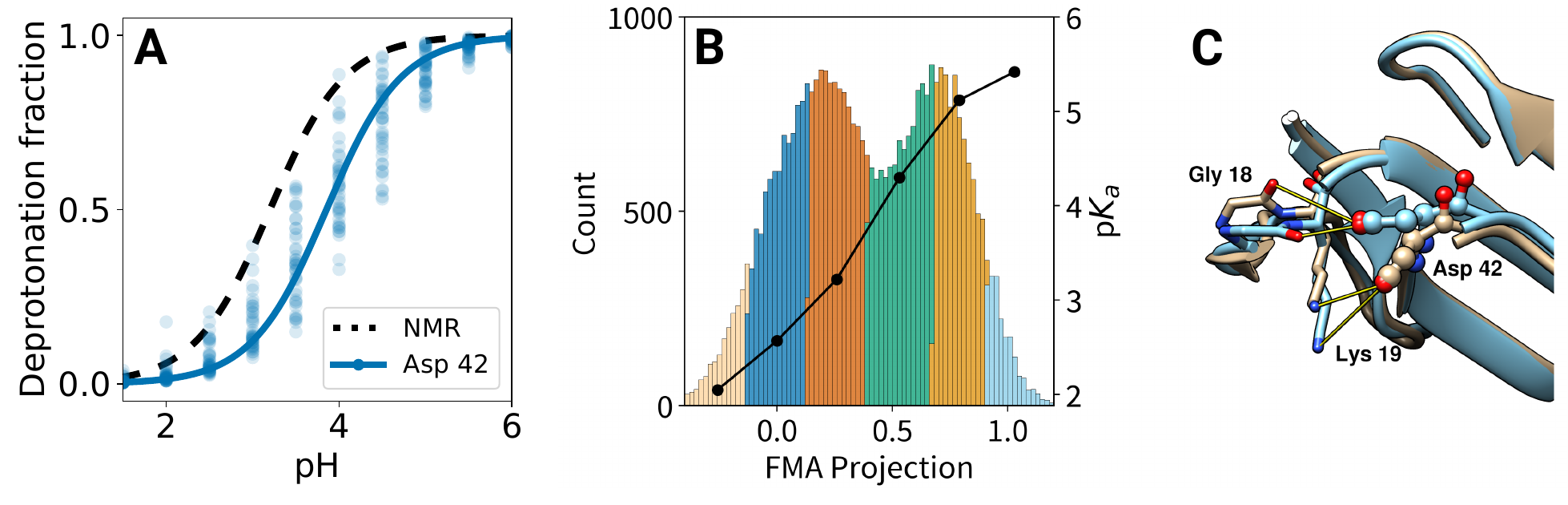} 
\caption{
\textbf{FMA analysis for the Asp~42 residue of cardiotoxin V}.
(A): Computational titration of Asp~42, with fraction of deprotonated frames for each replica (point) and the H-H fit to simulation data as solid line.
The dashed line is a H-H curve based on the NMR \pKa values.
(B): Histogram of FMA projection values for all replicas at pH = 4.0 (left axis) 
and \pKa as a function of the projection value (black curve, right axis).
(C): Low and high FMA value structure in beige and cyan, respectively, with pseudo-bond for relevant stabilizing interactions with Asp~42.}
\label{fig:CardiotoxinAsp42FMAPanel}
\end{figure}

For \textbf{Asp~42}, FMA explains 58\% of protonation variance, indicating substantial conformation-protonation coupling despite small inter-replica spread (Figure~\ref{fig:CardiotoxinAsp42FMAPanel}A), which warrants closer inspection.
Similar to His~4, the FMA analysis (Figure~\ref{fig:CardiotoxinAsp42FMAPanel}B) reveals two main conformations, differing primarily in the Pro~16 -- Asn~20 loop facing Asp~42 (Figure \ref{fig:CardiotoxinAsp42FMAPanel}C).
In the high FMA value conformation, Asp~42 forms a hydrogen bond with the Gly~18 backbone carbonyl (distance 2.7 \r{A}), thereby favoring protonation and shifting its \pKa to 5.8.
In contrast, in the low FMA value conformation, Asp~42 forms a salt-bridge to Lys 19 (distance 3.3 \r{A}), thus stabilizing the deprotonated form and lowering its \pKa to 3.0.

Compared to the rather large conformational change seen for His~4, the side chain reorientation correlating with Asp~42 protonation (Figure~\ref{fig:CardiotoxinAsp42FMAPanel}C) is small and local, suggesting faster kinetics for the Asp~42 coupling.
Indeed, more than 35 transitions are seen in the respective FMA trajectories (Figure~\ref{fig:FmaTrajCompareHis4Asp42}, bottom row),
 which in turn explains the faster convergence of individual replicas, leading to a smaller spread visible in Figure~\ref{fig:CardiotoxinAsp42FMAPanel}A.
These findings demonstrate that protonation-conformation coupling does not necessarily impede convergence by slowing protonation/deprotonation rates.

We note that otherwise identical constant pH simulations with \ffAmberUsed show much smaller coupling for Asp~42 (17\% variance explained), and more distant interactions with Gly~18 and Lys~19, underscoring considerable force field dependence.

In contrast to His 4 and Asp 42, FMA analysis of \textbf{Asp 59}, located near the C-terminus, explains 27\% of its protonation variance, indicating only weak coupling.
Correspondingly, the histogram of FMA values shows only one conformation, corresponding to a broad peak near 0.5.
The structural dynamics most correlated with the protonation state of Asp 59 are motions involving the C-terminal residues spanning Thr 58 to Asn 62, exhibiting an amplitude under 2.5~\r{A}.
There are also no marked changes of interactions to adjacent residues.
Despite these rather small changes in the Asp 59 environment, we observed unexpectedly large \pKa fluctuations, between 1 and 3 (data not shown).
This exemplifies that conformation-protonation coupling does not generally require two distinct conformations.
In this case, we similarly observed fast convergence and only very small inter-replica spread.

For the fourth titratable residue, \textbf{Glu 17}, the FMA projection explains less than 1\% of the protonation variance, thus ruling out any conformation-protonation coupling.
Similarly, very fast convergence results (Figure~\ref{fig:CardiotoxinPlotpKaExampleTitrCurve}).

We emphasize that, although suggestive, we have not yet established here causal connections between conformational changes and protonation states, only correlations. 
Within the present constant pH framework, this could be achieved by enforcing changes of the respective FMA coordinate, e.g., via Essential Dynamics\cite{deGroot1996towards} using FMA projection as a restrained variable, to see if the expected protonation changes follow. 
Conversely, enforced protonation changes resulting in corresponding structural changes would establish the reverse causality.
Although these approaches are feasible, they lie beyond the scope of this paper.

\FloatBarrier
\subsection{Lysozyme Titration}
\label{LysoTitr}

Next, we studied Hen Egg-White Lysozyme (HEWL), which is a widely used prototypical benchmark system for constant pH simulations, with several independent NMR titration measurements of its protonatable residues.\cite{bartik1994measurement,webb2011}
With ten such residues, HEWL enables us to explore more complex protonation couplings than the four-residue cardiotoxin V system.

\begin{figure}[h]
\includegraphics[scale=0.50]{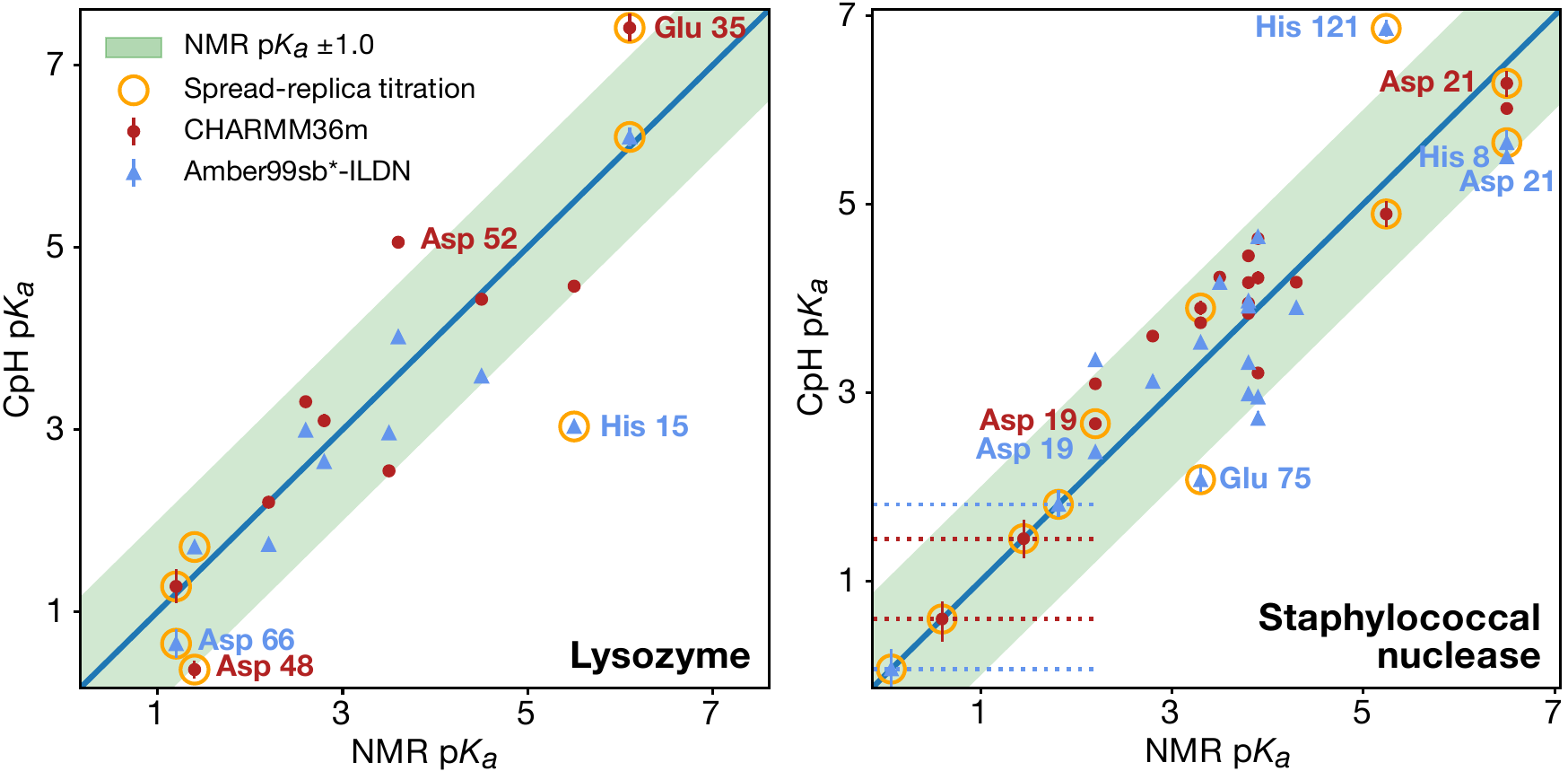} 
\caption{\textbf{Comparison of residue \pKa values from NMR titration ($x$ axis)\cite{webb2011,Castaneda2009} and CPH-MD simulation ($y$ axis)}, in HEWL (left) and staphylococcal nuclease (right),
for \ffCharmmUsed (red circles) and \ffAmberUsed (blue triangles), with bootstrapped 95\% confidence intervals as error bars. 
The green region indicates $\le 1$ pH point of difference between NMR and simulation.
Residues with spread-replica titration are highlighted by orange circles.
A dashed horizontal range indicates residues for which only a measured upper bound of 2.2 was reported, compatible with the respective \pKa values calculated from CPH simulations.
(\pKa table available in Supplementary Information section 2.3 and 2.4.)}
\label{fig:LysozymeSnasePkaChart}
\end{figure}

\subsubsection{Aggregate Accuracy and Outliers}

We first evaluated the overall accuracy of our implementation for this test system.
To this end, we carried out constant pH MD simulations of HEWL, with 40 replicas of 75~ns for each pH point over a range from $-1$ to $9$.
Figure~\ref{fig:LysozymeSnasePkaChart} (left) compares our constant pH simulation results to measured \pKa values for all titratable groups, resulting in a \pKa RMSE of 0.85 and 0.90 for \ffCharmmUsed and \ffAmberUsed, respectively.
Most \pKa values fall within 1 unit of their NMR reference (green region).
Our results align with other state-of-the-art CPH implementations, including GROMACS,\cite{aho2022constph} Amber,\cite{Harris2022} and \Charmm\cite{huang2016all} simulation codes (RMSE: 0.98, 0.83, and 0.92, respectively).

\begin{figure}[h]
\includegraphics[scale=0.7]{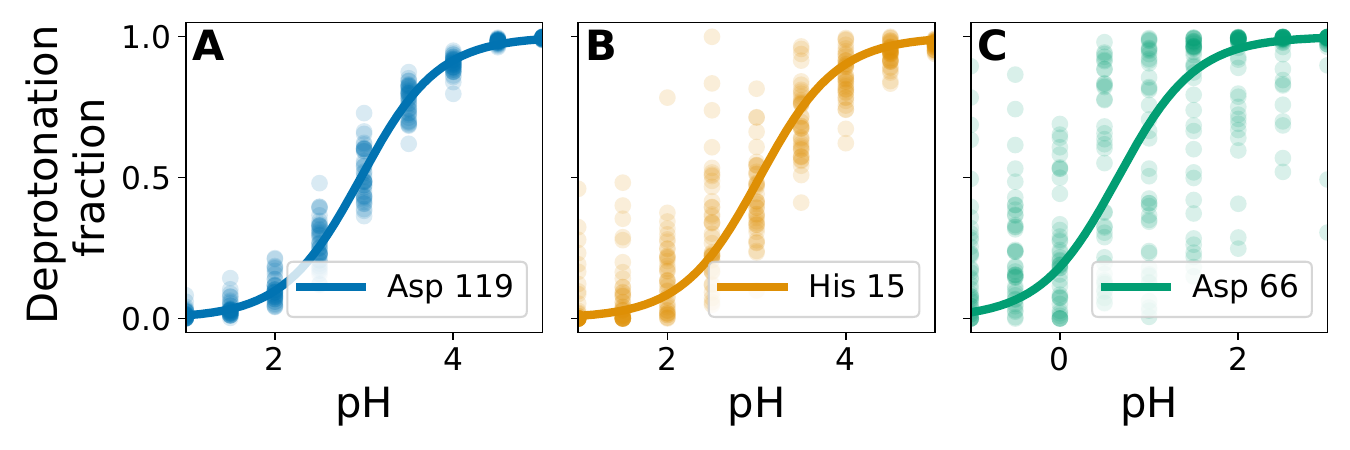} 
\caption{\textbf{Computational titration curves of Asp 119, His 15 and Asp 66} for HEWL, 
using \ffAmberUsed.
Transparent points show the deprotonation fraction computed for each replica.
The solid line corresponds to the H-H sigmoid fitted to all replicas.}
\label{fig:LysozymeShowUnconvAmber}
\end{figure}

Residues for which less accurate \pKa values are predicted by our simulations fall into two main categories.
The first group comprises those with large shifts from solution \pKa, such as Glu 35 and Asp 48.
These shifts are known to be generally challenging for CPH MD codes, because the strong electric fields involved are not accurately described  by fixed point charge, non-polarizable force fields.\cite{first2020beyond,lin2023amoeba}
The second category consists of residues with spread-replica titrations.
This effect, already detected for His~4 of cardiotoxin, is also seen to varying degrees in HEWL.
In contrast to relatively well converged residues such as Asp 119 (Figure~\ref{fig:LysozymeShowUnconvAmber}A), 
spread-replica titrations show  mild (Figure~\ref{fig:LysozymeShowUnconvAmber}B), to rather pronounced (Figure~\ref{fig:LysozymeShowUnconvAmber}C) spread in deprotonation fraction across replicas (transparent points),  leading to less accurate H-H fits (solid line) and, hence, generally larger \pKa deviations.
In total, seven out of 20 titrations show substantial spread-replica effect (Figure~\ref{fig:LysozymeSnasePkaChart} left, orange circles), including His 15 and Asp 48.

\subsubsection{Analysis of Protonation Coupling}

We examined residue-residue coupling using NMI analysis (Methods section~\ref{sec:directCoupling}) to further study the link between spread-replica titrations and protonation coupling.
Only the Asp~48 -- Asp~66 residue pair showed positive NMI results, with values up to 0.25 at pH~$=0$, in line with the above-mentioned link.

Given these results, the protonation fractions of this pair are best described by a macroscopic titration curve, which accounts for the coupling (Methods section~\ref{sec:directCoupling}).
Unfortunately, in the respective titration NMR study, the protonation fraction of the pair is only described by two independent H-H curves, such that the resulting \pKa values cannot be directly compared.
For a fair comparison, we therefore also used two independent H-H curves.
We refer the reader to the next system, staphylococcal nuclease, for two examples of macroscopic titration curve modeling.

To test whether the second type of coupling, conformation-protonation coupling, also occurs, we used FMA (Methods section~\ref{ProtonationConformationCouplingMethodParagr}), focusing on residues which show so far unseen effects.
The FMA analysis explains 77\% and 51\% of protonation variance for Asp 48 and Asp 66, respectively, demonstrating strong protonation coupling for both residues.
Together with their relatively large distance of more than 10 \r{A}, this result suggests that conformation-protonation coupling --- rather than direct electrostatic interactions --- mediate the previously detected residue-residue coupling.

\begin{figure}[h]
\includegraphics[scale=0.25]{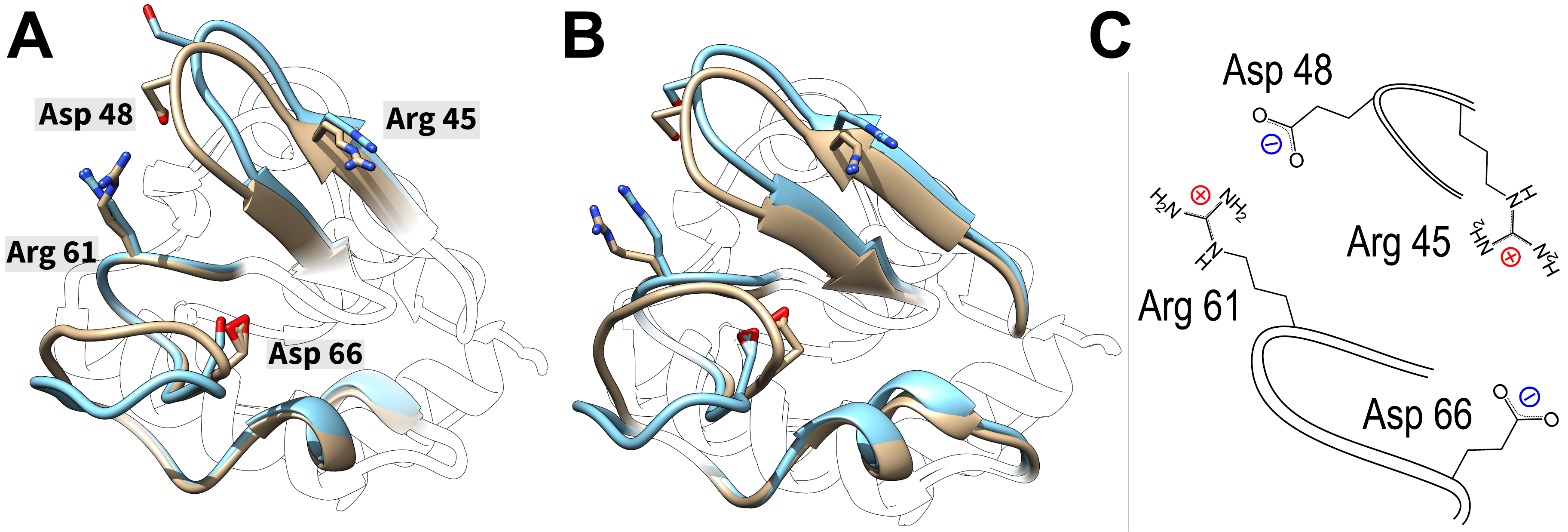} 
\caption{
\textbf{Conformation-protonation coupling in HEWL.} 
Superimposed are the structures corresponding to the FMA projection low (beige) and high (cyan) values for Asp 48 (A) and Asp 66 (B), respectively. Highlighted by color are only those regions which show marked structural differences. (C) Sketch of the proposed coupling mechanism discussed in the text..}
\label{fig:LysozymeAsp48Asp66StructreMCMSchem}
\end{figure}

Closer inspection of the involved structural changes (Figure~\ref{fig:LysozymeAsp48Asp66StructreMCMSchem}A-B, beige and blue) revealed by FMA provides further evidence for this scenario.
Indeed, the protonation of both residues Asp 48 (Panel A) and Asp 66 (Panel B) correlates with conformational changes of the same regions, involving a loop between Arg 45 and Thr 61, and another loop between Trp 63 and Leu 75.
As can be seen, the conformational changes primarily involve a concerted motion of these two loops either towards each other (low FMA values, beige), or apart (high FMA values, blue).
This motion suggests a coupling mechanism, sketched in Figure~\ref{fig:LysozymeAsp48Asp66StructreMCMSchem}C.
When Asp 48 and Arg 61 are close, the stronger electrostatic interaction with the positively charged Arg 61 favors the deprotonated form of Asp 48.
Similarly, when Asp 66 and Arg 45 are close, Arg 45 may fulfill a comparable role, though weaker due to its greater distance.
Because both Asp -- Arg pairs are located on the two loops, the concerted motion of the loops changes the distance within each pair cooperatively, such that their respective protonation/deprotonation dynamics occurs jointly.
This provides a nice example of protonation coupling mediated via concerted conformational motions. 
A straightforward generalization of the FMA-binned titration analysis, used previously for single residues, to the binning of protonation states along more than one FMA coordinate would provide a more quantitative picture, but is not implemented so far.

The remaining spread-replica titrations, Glu 35 and His 15, exhibited strong conformation-protonation coupling, akin to His 4 in cardiotoxin V (Section~\ref{sec:CardiotoxinTitr}), further corroborating our hypothesis linking coupling and spread-replica titration.
Of note, the His 15 coupling manifests solely in \ffAmberUsed simulations, similarly to Asp 42 in cardiotoxin V, where only \ffCharmmUsed showed a coupling, again underscoring that force field choice and comparisons are critical.

\FloatBarrier
\subsection{Staphylococcal Nuclease Titration}

Next, we studied staphylococcal nuclease $\Delta\mathrm{PHS}$ (\textbf{pH} \textbf{S}tabilized mutant), our final protein system.
This enzyme is, like HEWL, a widely used constant pH simulation benchmark.\cite{huang2016all,Harris2022,arthur2011predicting}
It constitutes a challenging test of our implementation, owing to its high density of titratable residues (twice that of lysozyme), many of which are buried and/or adjacent to other protonatable sites.

\subsubsection{Aggregate accuracy}

We first examine the overall \pKa{} accuracy of our constant pH implementation for this test system.
To this end, we carried out constant pH MD simulations, split in 40 replicas of 75~ns for each \pH value within a range from 1 to 8.
 Figure~\ref{fig:LysozymeSnasePkaChart} (right) compares our simulation results to measured \pKa{} values for all titratable groups, demonstrating that virtually all titrations fall within 1 pH unit of the NMR reference values (green region). 
Our simulations yield \pKa RMSEs of 0.53 and 0.83 for \ffCharmmUsed and \ffAmberUsed, respectively. 
Notably, the \ffCharmmUsed simulations demonstrate excellent accuracy for this challenging system, outperforming other state-of-the-art constant pH codes, including the \Amber code\cite{Harris2022} (RMSE 0.76) and the \Charmm code\cite{huang2016all} (RMSE 0.80), both of which use the CHARMM22 force field. 
Our \ffAmberUsed simulations, while less accurate than our \ffCharmmUsed simulations, remain competitive with these other codes.

We attribute the improved accuracy of our \ffCharmmUsed simulations to two factors: (1) the use of the more recent \ffCharmmUsed force field (versus CHARMM22) and, (2) much longer simulation times and, hence, better convergence.
We gathered a total of 3~microseconds of simulation per pH point across all replicas, compared to the 10 to 40~ns of previous studies reported here for comparison,\cite{Harris2022,huang2016all} corresponding to a hundredfold increase in sampling.
This enhancement was facilitated by FMM-driven performance increases,\cite{bkohnke2021,kohnke2020,Kohnke2023} and our use of 40 replicas (versus 3-5 previously).
This approach allows for parallel execution of relatively short simulations, markedly enhancing sampling at affordable wallclock time.
As discussed in section~\ref{ReplicaDurationAndConvergence}, some residues exhibit slow, coupled titration dynamics, a particularly salient issue in the staphylococcal nuclease system, which is ameliorated by increased sampling.

\begin{figure}
\includegraphics[scale=0.5]{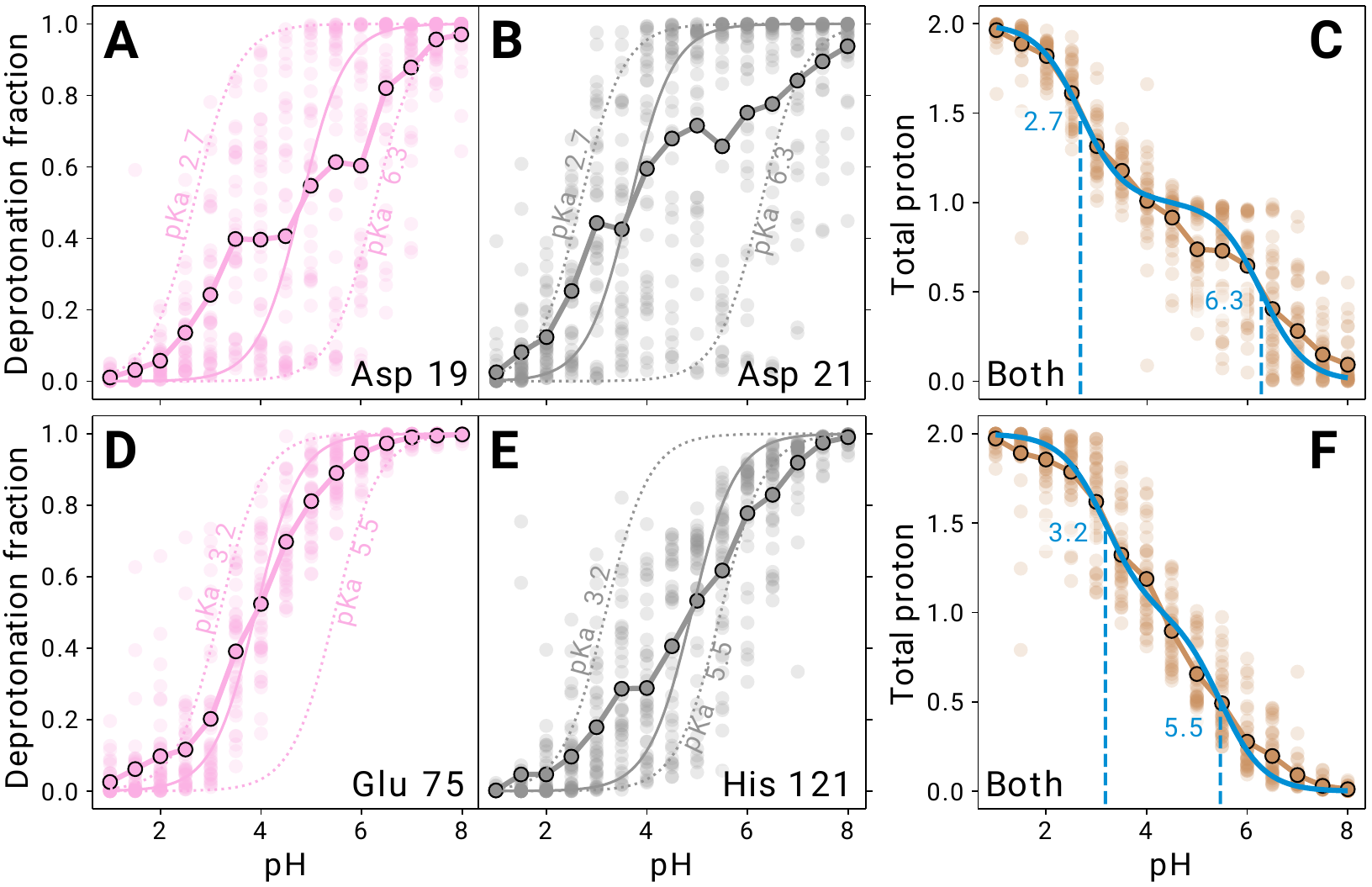} 
\caption{
\textbf{Residue-residue coupling in the Asp 19 (A) -- Asp 21 (B) and Glu 75 (D) -- His 121 (E) pairs of staphylococcal nuclease}.
(A,B,D,E): Computational titration curves for individual residues.
(C,F): Macroscopic titration curves for the pairs, i.e., number of protons bound to both residues.
Deprotonation fraction (A,B,D,E) or number of proton bound (C,F) shown for each replica (transparent circle) and averaged across replicas (black outlined circles).
Fit to both per-residue and macroscopic titration curves shown as solid lines.
Inflection points in the macroscopic titration curve (dashed vertical lines) mark the macroscopic \pKa values, with the corresponding H-H curves shown as dashed lines.
}
\label{fig:SNaseCoupledExplainChart}
\end{figure}

As observed in the HEWL system, errors in \pKa values primarily stem from inaccurate electrostatics for residues with large \pKa shifts, such as Asp 21 and Asp 77, as well as spread-replica titrations (Figure~\ref{fig:LysozymeSnasePkaChart} right, orange circles).
Following our earlier approach, we investigate couplings in this system to test the hypothesized correlation between protonation coupling and spread-replica titration.
In particular, the Asp 19 -- Asp 21 pair exhibits notable inter-replica spread in deprotonation fraction in our simulations (Figure~\ref{fig:SNaseCoupledExplainChart}A and \ref{fig:SNaseCoupledExplainChart}B). 
Castaneda et al. identified this pair as highly coupled in their NMR titration study.\cite{Castaneda2009}
We therefore explored this pair in more detail to test whether such strong coupling is also seen in our simulations, and whether quantitative agreement with experiment is achieved for this complex case.

\subsubsection{Asp 19 -- Asp 21 Pair}

To this end, we first examined the titration curves obtained from the simulations (Figure~\ref{fig:SNaseCoupledExplainChart}, top row).
As can be seen, the titration curves for Asp~19 and Asp~21 exhibit substantial inter-replica spread (transparent circles).
Interestingly, the average deprotonation fraction across all replicas (black outlined circles) shows marked deviation from a sigmoid titration curve, indicating pronounced residue-residue coupling.

Indeed, an NMI analysis yielded values of 0.20 and 0.17 for \ffCharmmUsed and \ffAmberUsed, respectively, underscoring strong residue-residue coupling.
In contrast to the Asp~48 -- Asp~66 pair in HEWL, the close proximity of Asp~19 and Asp~21 suggests that their coupling arises from direct electrostatic interactions, i.e., that 
the negative charge of a deprotonated Asp disfavors the deprotonation of its neighbor, resulting in protonation anticooperativity.

In further contrast to the Asp 48 -- Asp 66 pair, an NMR study using the appropriate macroscopic titration curve is available for Asp 19 -- Asp 21 (Methods section~\ref{sec:directCoupling}). 
This enabled a more direct comparison between simulation and experiment.
Indeed, the macroscopic titration curve provided a better fit to the data, as shown in Figure~\ref{fig:SNaseCoupledExplainChart} (A and B versus C), thus demonstrating quantitative agreement.
With this model, the two residues Asp 19 and Asp 21 show macroscopic \pKa{} values of $2.67 \pm 0.05$ ($\pm 0.05$: bootstrapped 95\% confidence interval) and $6.28 \pm 0.14$, respectively.
Our secondary force field \ffAmberUsed results in macroscopic \pKa values of $2.37 \pm 0.05$ and $5.65 \pm 0.13$ for these two residues.
Both \pKa value sets closely align with the NMR reference values of 2.21 and 6.54,
thus providing more accurate \pKa values than a simple H-H fit (\pKa 4.8 and 3.7).
Our constant pH implementation thus accurately reproduces the Asp~19 -- Asp~21 coupling in simulations, behaving remarkably close to the experimental data.

\subsubsection{Glu 75 -- His 121 Pair}

Given these encouraging results, we systematically applied NMI analysis to all residue pairs of the test system to identify other coupled titrations.
This analysis revealed an additional coupled pair, Glu 75 -- His 121, exhibiting maximum NMI values of 0.14 and 0.18 for the \Charmm and \Amber simulations, respectively --- comparable in strength to the above Asp~19 -- Asp~21 pair.
Consequently, we employed the same modeling approach, transitioning from individual titration curves (Figure~\ref{fig:SNaseCoupledExplainChart}D and \ref{fig:SNaseCoupledExplainChart}E) to a two-proton macroscopic titration (Figure~\ref{fig:SNaseCoupledExplainChart}F).

The macroscopic \pKa values for Glu 75 and His 121 are $3.18 \pm 0.09$ and $5.46 \pm 0.08$ for the \Charmm simulation, and $2.04 \pm 0.13$ and $6.90 \pm 0.09$ for the \Amber one.
Unlike our approach, the NMR titration study\cite{Castaneda2009} reported individual \pKa values (3.30 and 5.24) rather than using a macroscopic titration curve.
Therefore, we also used the H-H \pKa values for comparison with the NMR data.

Interestingly, the titration curve for Glu 75 is nearly sigmoid, despite the coupling (Figure~\ref{fig:SNaseCoupledExplainChart}D), which likely explains why this coupling was not detected by NMR.
In fact, it might also have been overlooked by a qualitative examination of simulation 
titration curve, underscoring the value of our NMI analysis.

Given that Asp 19, Asp 21, Glu 75 and His 121 were all found to be coupled, as well as being spread-replica titration, our idea of a link between these two characteristics seems to hold for the staphylococcal nuclease system too.

\subsubsection{Analysis of Conformation-Protonation Coupling}

Two residues with spread-replica titration remain, Asp 77 and Asp 83, which did not exhibit residue-residue coupling.
Thus, we investigated all residues for a second possible kind of coupling, conformation-protonation coupling, using FMA.

Indeed, the residue pairs exhibiting direct residue-residue coupling (Asp 19 -- Asp 21, Glu 75 -- His 121) also display conformation-protonation coupling.
These residues show minor conformational changes between low and high FMA value structures, primarily involving side chain motions without significant backbone rearrangements (Supplementary Information 3.2).
Although our approach cannot quantify the relative contributions of residue-residue and conformation-protonation coupling, the observed small motions suggest residue-residue coupling predominates.

Beyond these pairs, Asp 77 and Asp 83 also demonstrate conformation-protonation coupling in staphylococcal nuclease (Figure~\ref{fig:LysozymeSnasePkaChart} right, indicated by the four points with dashed x-axis error bars). 
The \pKa values of these residues are exceptionally low in both NMR and our simulations (1.35 and 0.50), falling below the global acid unfolding transition at pH 2 for this protein,\cite{Castaneda2009} thus precluding exact \pKa measurement.
We therefore speculate that the observed conformational coupling might also involve primary unfolding steps.

The conformational coupling affecting Asp 77 and Asp 83 further corroborates the system-wide relationship between coupling and spread-replica titrations.
This phenomenon, now observed across cardiotoxin V, HEWL, and staphylococcal nuclease, is robustly supported by our data.
Beyond identifying coupling, this pattern offers a foundation for exploring potential solutions, which we will address in subsequent sections.

\FloatBarrier
\subsection{Dynamic Barrier Optimization}
\label{dynbarwelladj}

We have seen above that convergence, e.g., for titration simulations, critically depends on the transition rates between the protonated and deprotonated state. 
To improve the accuracy of such calculations, it is therefore beneficial to artificially increase these rates. 
To this end, we have implemented Dynamic Barrier Optimization (DBO), which tunes the
double well potential \Vdw
to bring these transition rates to a desired value, for each residue during the simulation (Methods section~\ref{sec:dbo}). 

\begin{figure}[tb]
\includegraphics[width=\textwidth]{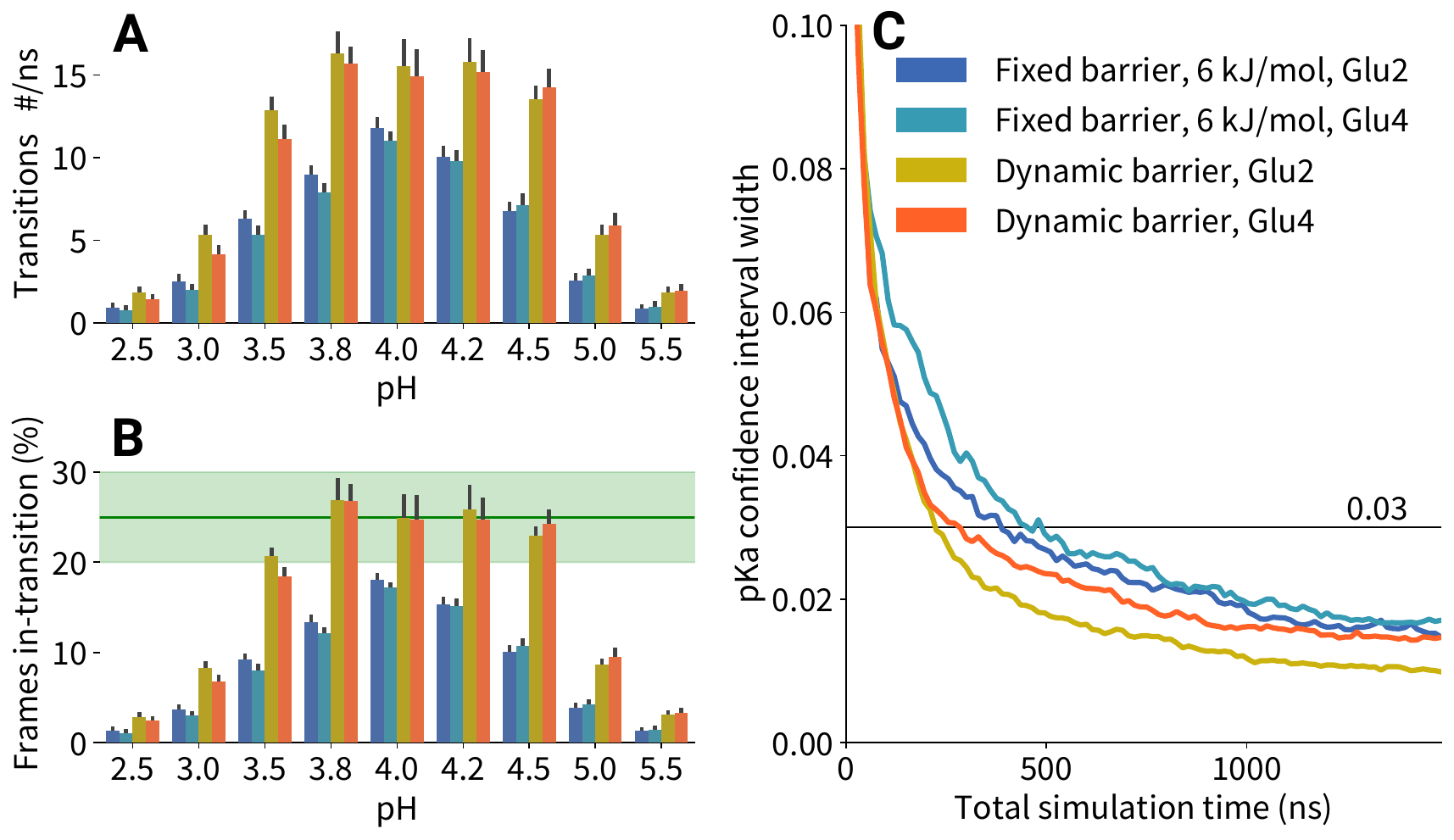} 
\caption{
\textbf{Effect of Dynamic Barrier Optimization (DBO) on protonation transition rates and \pKa convergence for the GEAEG pentapeptide.} Transition rates (A) and fraction of unphysical lambda values (B) observed in titration simulations for selected pH values, (C) precision in terms of \pKa confidence intervals widths. 
The colors indicate fixed barrier (blue, green) vs.\ DBO (yellow, red) for residues Glu2 (blue, yellow) and Glu4 (green, red), respectively.
The maximum tolerated unphysical fraction of frames (25\%) with 5\% threshold is indicated as a green bar in panel B; the 0.03 confidence interval (black line) in panel C indicates our convergence criterion. Black bars show standard deviations.}
\label{fig:pentapeptideDynBiasEffect}
\end{figure}

As an illustration, Figure~\ref{fig:pentapeptideDynBiasEffect} compares simulations of the pentapeptide GEAEG with and without DBO.
DBO results in markedly increased transition rates (panel A) as well as percentages of frame in-transition, defined by $0.2 < \lambda_p < 0.8$ (panel B).
The latter is directly monitored by DBO, which adjusts the height of the inter-well barrier to reach its target value, here 25\% with a 5\% tolerance.
While these results mean DBO is effective at adjusting rates, convergence remains the primary goal. 
In the absence of a direct observable reporting on convergence, we rely on proxy measurements, here the \pKa uncertainty and degree of homogeneity between replicas.
Panel C thus plots the width of the confidence interval (CI) on \pKa as a function of total simulation time, with lower CI width indicating less uncertainty, greater homogeneity, and thus more converged simulations.
For both Glu in GEAEG, DBO markedly accelerates the decrease in CI width over time, reaching our target width for fully converged simulations (0.03 pH points) 40\% faster --- a 200~nanosecond reduction in total simulation time. 
The improvement in transition rates brought about by DBO thus also translates into improved convergence.

Here, we test our DBO implementation to (a) verify that similar \pKa values are obtained, and (b) if and to what extent convergence is actually enhanced.
We conducted titration simulations for both cardiotoxin V and HEWL, with and without DBO, employing 40 replicas each. 
Each replica ran for 100~ns for cardiotoxin V and 75~ns for HEWL.
Consistent with our previous example (Figure~\ref{fig:pentapeptideDynBiasEffect}B), the target value for the percentage of frame in-transition ($0.2 < \lambda_p < 0.8$) was 25\% with a tolerance of 5\%.

\begin{figure}
\includegraphics[width=0.8\textwidth]{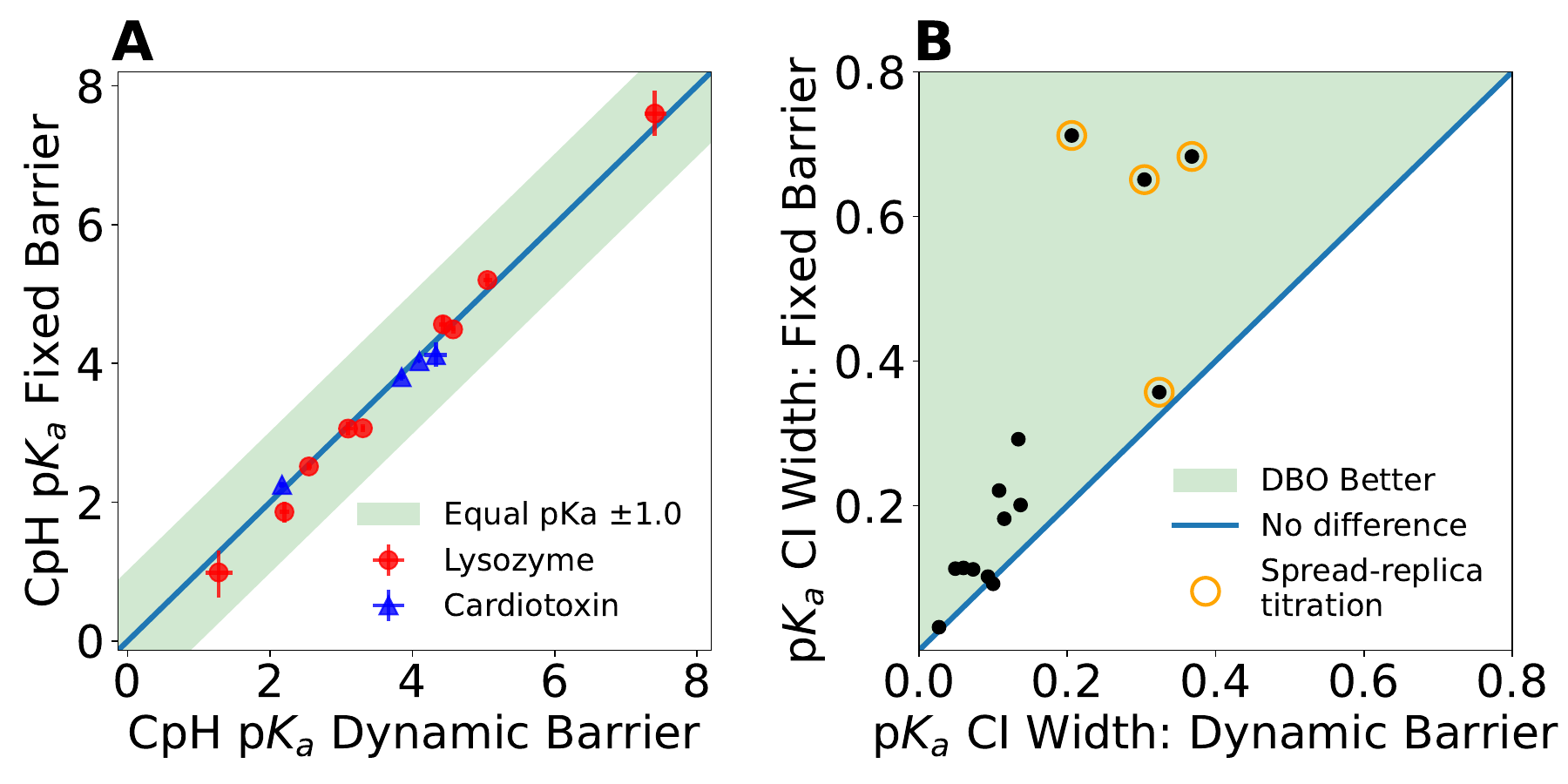} 
\caption{\textbf{Impact of DBO on the titration of cardiotoxin V (blue) and HEWL (red).}
(A): Comparison of \pKa values with and without DBO, with the green region indicating  difference $\le 1$ \pKa point.
Error bars are 95\% confidence intervals.
(B): Comparison of precision in terms of \pKa confidence interval width.
Residues for which the confidence interval is narrower when DBO is used lie in the green region, and residues with spread-replica titration are highlighted by orange circles.
}
\label{fig:DBpKaAndCIWidth}
\end{figure}

DBO introduces a time-dependence in the double well potential $\Vdw$ that could produce artifactual changes in residue \pKa, for instance if the protonation state of a residue affected its protonation transition rate.
We hypothesized this effect to be negligible, as DBO adjusts barrier height at most every nanosecond, much slower than changes in the $\lambda$ degree of freedom, and only rarely past the first five to ten nanoseconds of the simulation.
We nonetheless investigated this possibility by comparing the \pKa values of all residues in both titrations (Figure~\ref{fig:DBpKaAndCIWidth}A), revealing an average agreement within 0.15 \pKa units and a Pearson's correlation coefficient of 0.99, with no outliers.
We conclude that DBO does not introduce any such unwanted artifacts.

We next examined the impact of DBO on convergence, using the same proxy as in the illustrative example above (Figure~\ref{fig:pentapeptideDynBiasEffect}C), namely \pKa uncertainty as measured through CI width.
Figure~\ref{fig:DBpKaAndCIWidth}B compares the CI width with (x-axis) and without (y-axis) DBO for each residue, at the end of the simulation.
As can be seen, DBO reduces \pKa uncertainty for essentially all residues, demonstrating that the artificially lowered energy barrier indeed increases the protonation/deprotonation rates and, hence, the statistical precision of the resulting protonation fractions. 
As should be expected, the most substantial precision increases occur for those residues with the largest inter-replica spread (orange circles), resulting in \pKa CI width improving from $\pm 0.7$ to $\pm 0.3$ (95\% interval). 
Residues with already high transition rates showed no further improvement from DBO, but importantly, experienced no adverse effects.

Although DBO substantially improved sampling and reduced \pKa uncertainty for most residues, some (e.g., His 4 in cardiotoxin V, Glu 35 in HEWL) still exhibit slow convergence. 
DBO acts on the barrier of the double-well potential \Vdw, which is only one of the factors affecting protonation transitions.
Other barriers exist, such as (anti)cooperativity from residue-residue coupling and conformational rearrangements in pro\-tona\-tion-con\-for\-ma\-tion coupling. 
These barriers can only be partially compensated through DBO.
We thus analyze these additional barriers in detail in the following subsection.

\FloatBarrier
\subsection{Protonation Coupling and Convergence}
\label{ReplicaDurationAndConvergence}

We have already seen for our test systems above that the  protonation/deprotonation  dynamics of a given residue can be coupled to the protonation state of other residues and also, to conformational dynamics.
In such cases, the  protonation/deprotonation  dynamics may be markedly slowed down, e.g., by slow conformational dynamics.
This section outlines techniques to monitor and accelerate convergence in these situations.

\subsubsection{Conformation-Protonation Coupling}

In fact, conformational dynamics is very often far slower than DBO-enhanced protonation dynamics.
In such cases, the convergence of protonation fractions and, hence, \pKa values is limited by conformational sampling.
This is a challenge unrelated to constant pH simulations, of course, that has prompted the development of many methods to address it.
Among these techniques, some are untargeted, accelerating all protein degrees of freedom, such as parallel tempering,\cite{sugita1999replica,calvo2005all} or replica exchange with solute tempering (REST).\cite{liu2005replica}
Other techniques require the definition of a collective variable, like weighted ensemble path sampling,\cite{bhatt2010steady}
essential dynamics,\cite{amadei1994essential,escidoc:599910} flooding,\cite{escidoc:596066} or metadynamics.\cite{barducci2011metadynamics}
These latter methods require knowledge of the precise nature of the conformational motion coupled to the protonation dynamics.

As an example, Figure~\ref{fig:FmaTrajCompareHis4Asp42} shows the projection of the trajectory of the cardiotoxin V protein (simulated with CPH MD) onto the collective conformational motion that exhibits the largest correlation with the protonation states of His 4 and Asp 42, respectively, as determined from Functional Mode Analysis.
Strikingly different behavior is seen in these two cases.
His 4 (top row) shows very slow conformational dynamics, with long plateaus and infrequent transitions between them, such that each replica samples a different region of conformational space, resulting in slow convergence.
Conversely, Asp 42 (bottom row) displays frequent transitions, indicating well-overlapping conformational ensembles.
This results in well-converged protonation fractions and \pKa values.

\subsubsection{Residue-Residue Coupling}

We now turn to residue-residue coupling.
This coupling manifests either directly through electrostatic interactions between adjacent residues, or indirectly via conformational changes.
Both mechanisms can be analyzed in terms of protonation microstates, with the latter also amenable to the conformational dynamics methods described earlier.

For a pair of coupled residues, a microstate describes the joint protonation state of both residues, e.g., residue 1 protonated and residue 2 deprotonated (written $\mathrm{A}_1\mathrm{H}$/$\mathrm{A}_2^{-}$).
The fraction of time spent in each possible microstate is determined by 
individual \pKa values, pH, as well as by the energetics of the coupling. 
In the absence of any coupling, $P(\mathrm{A}_1  \land \mathrm{A}_2) = P(\mathrm{A}_1)\cdot P(\mathrm{A}_2)$ (where $\mathrm{A}_1 $ and $\mathrm{A}_2$ denote either protonation state), i.e., the protonation of each residue is independent of that of the other. 
Any deviation points to a coupling free energy of 

\begin{equation*}
k_B T \, \Big( \ln   P(\mathrm{A}_1 \land \mathrm{A}_2) \, -  \,\ln P(\mathrm{A}_1) -  \,\ln P(\mathrm{A}_2) \Big). 
\end{equation*}

\begin{figure}[tb]
\includegraphics[width=0.9\textwidth]{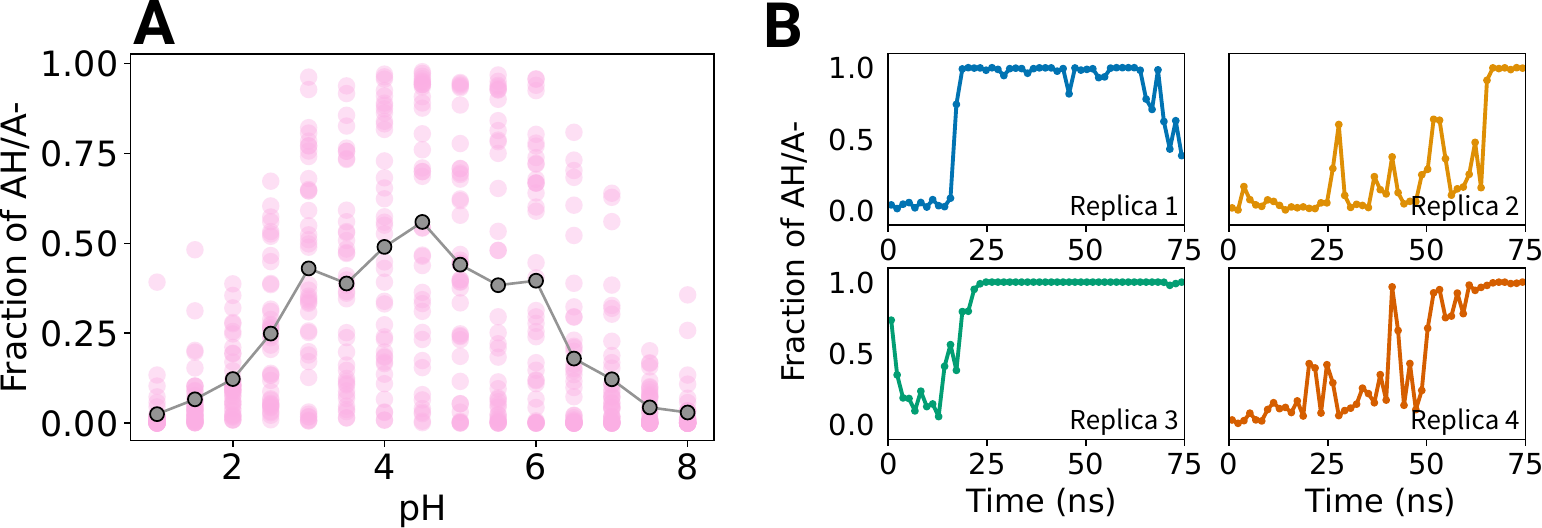} 
\caption{
\textbf{Residue-residue coupling and microstate fraction heterogeneity} in staphylococcal nuclease. Shown is the fraction of time during which Asp~19 is protonated and, simultaneously,  Asp~21 is deprotonated ($\mathrm{A}_1\mathrm{H}$/$\mathrm{A}_2^{-}$ microstate).
(A): as a function of \pH, for each replica (transparent circle) and averaged across replicas (black outlined circle).
(B): as a function of time for four replicas at \pH 4.0, computed in 1.5~ns windows.}
\label{fig:Asp19Asp21Microstate}
\end{figure}

Generalizing the single residue treatment above, the scatter of microstate fractions across replicas indicates the level of convergence.
To illustrate this convergence evaluation method, we examine the Asp 19 -- Asp 21 coupled residue pair. Figure~\ref{fig:Asp19Asp21Microstate}A depicts the fraction of the $\mathrm{A}_1\mathrm{H}$/$\mathrm{A}_2^{-}$ microstate (protonated Asp 19, deprotonated Asp 21) as a function of pH, revealing a substantial spread between replicas in the pH range of 2 to 7 and indicating incomplete convergence.

Figure~\ref{fig:Asp19Asp21Microstate}B shows examples of the time-dependent behavior of the $\mathrm{A}_1\mathrm{H}$/$\mathrm{A}_2^{-}$ microstate fraction, characterized by long plateaus with infrequent transitions.
These rare transitions indicate high barriers between microstates, accounting for the observed spread in microstate fractions across replicas. 
While our analysis readily detects this incomplete convergence, our current implementation does not provide automated solutions. 
In these exceptional cases, it may help to fix the protonation state of one of the coupled residues while allowing the other to titrate freely, and \textit{vice versa}.
Possible future implementations may extend DBO to multiple residues and/or incorporate biasing potentials akin to conformational flooding\cite{escidoc:596066} to accelerate microstate transitions.

To further enhance microstate sampling, several other methods could be employed: pH replica exchange, as implemented in the AMBER\cite{Harris2022} constant pH codes; adaptive landscape flattening, developed for the Brooks group's $\lambda$-dynamics code;\cite{hayes2017adaptive} or integration of $\lambda$ degrees of freedom into collective variable suites like Colvars\cite{Fiorin2013} or PLUMED.\cite{bonomi2009plumed}

\FloatBarrier
\subsection{Overall \pKa Accuracy}

\begin{figure}
\includegraphics[width=1.0\textwidth]{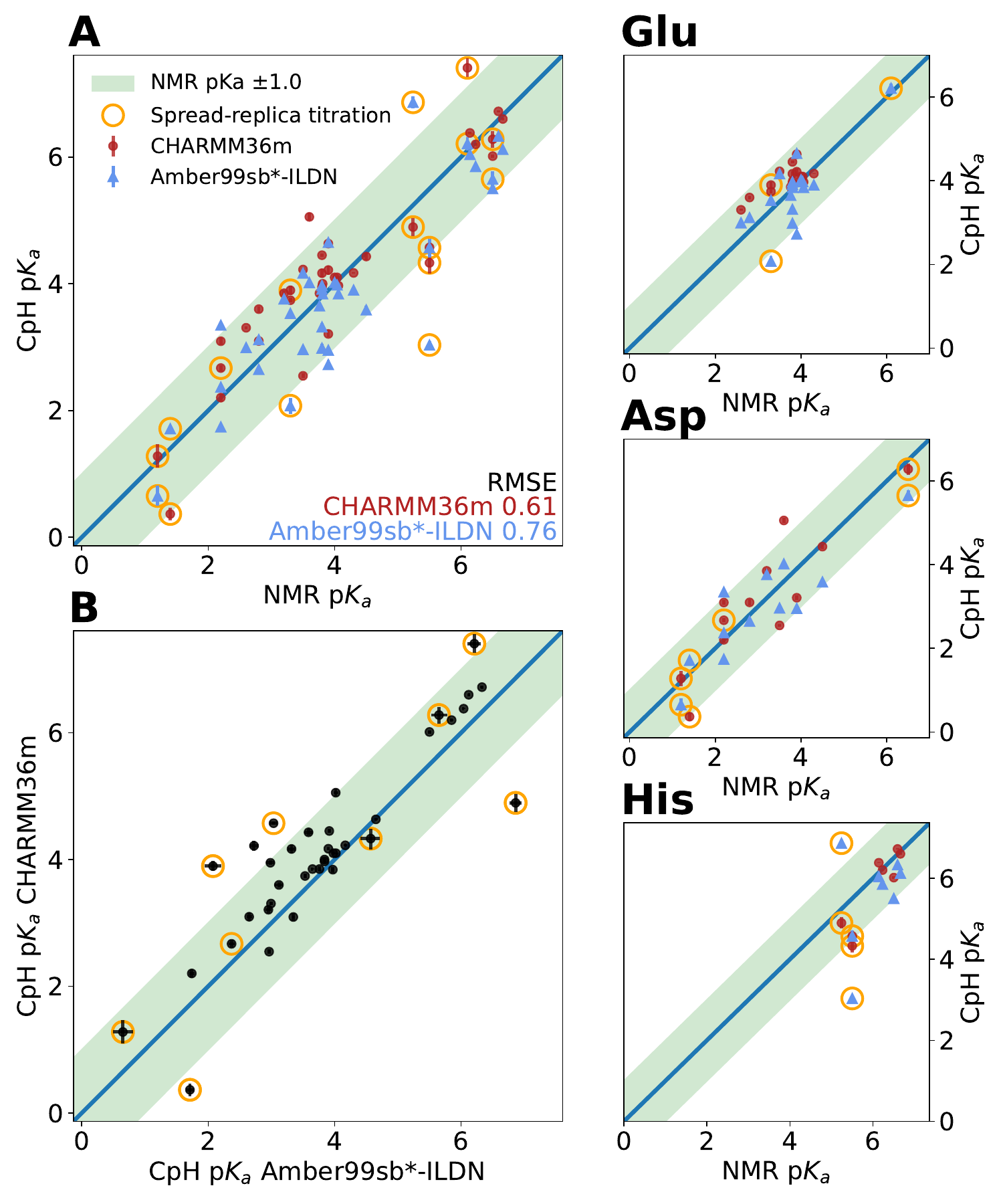} 
\caption{\textbf{Aggregate comparison between calculated and measured \pKa values} for 
 \ffCharmmUsed  (red circles) and
\ffAmberUsed (blue triangles), with bootstrapped 95\% confidence intervals as error bars.
The green region indicates $\le$ 1 \pKa point deviation.
Residues with spread-replica titration are highlighted by orange circles.
(A): Comparison between NMR and constant pH \pKa values for all residues.
(Glu), (Asp), (His): for specific residue types.
(B): between \ffCharmmUsed and \ffAmberUsed. }
\label{fig:AllPkaChartPanel}
\end{figure}

Figure~\ref{fig:AllPkaChartPanel} finally summarizes the overall accuracy of our CPH implementation across all benchmark systems (excluding single residues).
For the 38 titrated residues with measured \pKa values, our implementation achieves \pKa RMSE values of 0.61 and 0.76 with \ffCharmmUsed and \ffAmberUsed, respectively (Figure~\ref{fig:AllPkaChartPanel}A).
These RMSE values are similar to those from other explicit solvent constant 
pH methods,\cite{Harris2022,huang2016all,swails2014constant,aho2022constph,buslaev2022best} indicating that our FMM-based \lambdadyn approach achieves state-of-the-art \pKa accuracy.

When the \pKa values and their accuracy are grouped according to amino acid type, we find that the most abundant residue, Glu, with 18 occurrences (Figure~\ref{fig:AllPkaChartPanel}, Glu panel), exhibits  the best accuracy, with RMSE values of 0.54 and 0.55 for \Charmm and \Amber, respectively, significantly better than the overall \pKa RMSE of 0.61 and 0.76.
In our test systems, Glu typically exhibits \pKa shifts below 1.5 units, posing less challenge for constant pH methods.
More demanding are the 12 Asp residues (Figure~\ref{fig:AllPkaChartPanel}, Asp panel), with a markedly broader range of \pKa shifts up to $\pm 2.5$ \pKa units.
Nevertheless, very good agreement with experiments is achieved, with an RMSE of 0.72 and 0.66 for \Charmm and \Amber, respectively, and only one \pKa value deviating by more than one unit.
Interestingly, for the eight histidines (Figure~\ref{fig:AllPkaChartPanel}, His panel), \ffCharmmUsed predicts comparably accurate \pKa values (RMSE 0.58), while \ffAmberUsed achieves lower accuracy (RMSE 1.18).
Our implementation thus demonstrates a robust reproduction of the \pKa of Asp and His across the \pKa spectrum (the latter only with \ffCharmmUsed).

Regarding convergence, we noted a clear correlation between spread-replica titrations (Figure~\ref{fig:AllPkaChartPanel}, orange circles) and large \pKa shifts with respect to the solution \pKa of each respective residue type, most evidently for Asp and His.
As expected, the correlation between spread-replica titration and less accurate \pKa reproduction, observed for individual systems, also holds at the aggregate level. 
Indeed, all but two residues with $>1$ \pKa units difference between calculated and experimental \pKa exhibit this property (Figure~\ref{fig:AllPkaChartPanel}A).
Notably,  excluding these unconverged residues yields a \pKa RMSE of 0.55 for both force fields, 
 highlighting the potential gains from improving the titration of these residues.

To assess the force field dependence of our calculated \pKa values,  
Figure~\ref{fig:AllPkaChartPanel}B compares the \pKa values obtained with \ffCharmmUsed and \ffAmberUsed.
These values are in good agreement, with a Pearson correlation of 0.89.
A minority of residues show \pKa differences exceeding 1 \pH point between the two force fields. 
Across all test systems, the \pKa RMSE improves to an excellent 0.57 when averaging \pKa values per residue between both force fields, outperforming both our own individual force field simulations, and other constant pH studies\cite{Harris2022,huang2016all,swails2014constant,aho2022constph,buslaev2022best} (using a single force field).
This improved accuracy could stem from either cancellation of errors between force fields, or increased sampling from having effectively combined both titration series.
To differentiate these effects, we recalculated the averaged \pKa RMSE using half the sampling per titration series, matching the total sampling of single force field simulations.
This calculation yielded a \pKa RMSE of 0.61, identical to the \ffCharmmUsed simulations, suggesting that increased sampling was the primary factor for the improved accuracy.
Spread-replica titrations further support this hypothesis, demonstrating that unconverged titrations, which benefit from additional sampling, are present in our results.
Although force field biases likely contribute to \pKa error,\cite{Peeples2024} we identify insufficient sampling as the primary issue to address in future work, given that (1) the root issue -- protonation coupling -- remains a significant challenge in current CPH MD codes, and (2) well-converged titrations are a pre-requisite to force field tuning for \pKa reproduction.

\FloatBarrier
\subsection{Conclusions and Outlook}

This study introduced a new implementation of Hamiltonian interpolation \lambdadyn constant pH MD within the \gromacs MD suite,\cite{Abraham2015,Hess:2008tf,pall2020} supporting both \ffCharmmUsed and \ffAmberUsed force fields, which we tested on a variety of systems from simple pentapeptides to proteins.
We demonstrated the feasibility of using Hamiltonian interpolation for constant pH protein simulations, previously deemed impractical due to computational costs stemming from Particle Mesh Ewald (PME)\cite{Essmann:1995vj} electrostatics calculations. 
To circumvent this problem, our implementation therefore incorporated a Fast Multipole Method implementation  for efficient and accurate calculations of Coulomb interactions and their derivatives\cite{Kohnke2023} including the dynamically variable charges inherent in \lambdadyn.

We validated our approach through computational titrations of constant pH benchmark systems (cardiotoxin V, lysozyme, and staphylococcal nuclease), yielding \pKa values comparable to NMR measurements, with overall \pKa RMSEs of 0.61 and 0.76 for \ffCharmmUsed and \ffAmberUsed force fields, respectively.
Notably, averaging \pKa values on a per-residue basis across both force fields yielded an improved RMSE of 0.57.
However, we also observed marked correlations between \pKa errors, large inter-replica spread in titrations, and residues subject to protonation coupling.
Based on these findings, we posit that enhancing sampling for coupled residues offers the most promising avenue for improving CPH MD accuracy.

Across our test systems, our implementation achieved similar or better accuracy compared to recent constant pH studies,\cite{Harris2022,huang2016all,swails2014constant,aho2022constph,buslaev2022best} with the aforementioned averaged \pKa values across force fields reaching a remarkable 0.57 \pKa units RMSE. 
Notably, for the challenging staphylococcal nuclease system, known for its pronounced inter-residue and protonation-conformation coupling, we attained an RMSE of 0.53 using \ffCharmmUsed, surpassing previous reports.

To accelerate and control protonation/deprotonation transitions during CPH simulations, we implemented Dynamic Barrier Optimization (DBO), which dynamically adjusts the heights of barriers separating protonated from deprotonated states, thereby enhancing sampling and computational titration accuracy.
Our results demonstrated that DBO significantly improves convergence without compromising \pKa accuracy, particularly for notoriously slow-converging coupled residue pairs.

Furthermore, we developed post-simulation tools to analyze protonation couplings in proteins, encompassing both protonation-conformation coupling and inter-residue protonation state coupling, which provided considerable insight when applied to our benchmark systems. 
Our FMA-based analysis identified substantial \pKa shifts related to conformational changes, such as one exceeding 2.5 \pH units for His 4 in cardiotoxin V. 
Using mutual information analysis, we detected coupling  in both well-known interacting residue pairs (e.g., Asp 19 - Asp 21 in staphylococcal nuclease) and previously unidentified pairs (e.g., Glu 75 - His 121).
These tools also characterized the degree of convergence for individual residues, highlighting that coupled residues converge much more slowly than isolated ones. 
Consequently, our analysis tools not only detected couplings but also revealed causes of slow convergence, supporting the development of future convergence enhancement methods.
Future development of our code will thus focus on implementing sampling improvement techniques such as pH replica exchange.\cite{Harris2022}

Current force field-based fixed protonation simulations require pre-determination of protonation states for each titratable residue, posing challenges particularly for histidines and arginines in hydrophobic environments (e.g. active site, membrane or ion channel interior).
These methods often fail to account for \pKa changes during simulations due to conformational shifts, ligand binding, or nearby residue protonation changes. 
In contrast, the largely automated GROMACS constant pH implementation described here should provide accurate protonation states and ensembles, with control of protonation kinetics, and requiring little additional human and computer effort.
 This advancement paves the way for routine atomistic simulations at constant pH with protonation ensembles that dynamically adapt to structural and environmental changes --  eventually replacing the less accurate current fixed protonation simulations.

\subsection*{Appendix}

An enhanced version of \gromacs that enables constant pH simulations is available for download;
please follow the instructions at \url{https://www.mpinat.mpg.de/grubmueller/gromacs-fmm-constantph}.

Development can be followed at our git repository \url{https://gitlab.mpcdf.mpg.de/grubmueller/fmm}, 
and documentation can be found at 
\url{https://grubmueller.pages.mpcdf.de/docs-gromacs-fmm-constantph}

\subsection*{Acknowledgments}

This work was funded by the German Federal Ministry of Education and Research (BMBF)
as part of its SCALEXA (New Methods and Technologies for Exascale Computing) initiative
(BMBF project 16ME0713).
We thank Plamen Dobrev for fruitful discussions on many aspects of constant pH simulations and for beta-testing our implementation.
We are grateful to P. Buslaev, N. Aho, A. Jansen, P. Bauer, B. Hess and G. Groenhof
for sharing their cardiotoxin titration data \cite{buslaev2022best} for comparison with our results.
Special thanks to Gerrit Groenhof for providing valuable comments on the manuscript.

\hspace{10mm}
\bibliography{const-pH}

\end{document}


\tableofcontents

\section{Vmm Potential Calibration}

The set of $\lambda_p, \lambda_t$ used for force field-specific $\Vmm(\lambda)$ calibration is the Cartesian product of two of the following set of value: -0.1, -0.05, 0.0, 0.05, 0.1, 0.2, 0.4, 0.6, 0.8, 0.9, 0.95, 1.0, 1.05, 1.1. For example, the point ($\lambda_p$ = 0.1, $\lambda_t$ = 0.6) was used.

\section{Titration details}

\subsection{Pentapeptide titration}

Total period removed from analysis due to dynamic barrier/well adjustment less than 3 ns per simulation, typically less than 1 ns.

\begin{table}
\begin{tabular}{ll}
Peptide & pH points \\
GEAEG &  2.5, 3.0, 3.5, 3.75, 4.0, 4.25, 4.5, 5.0, 5.5 \\              
GHAHG &  5.0, 5.5, 6.0, 6.25, 6.5, 6.75, 7.0, 7.5, 8.0 \\         
GEAHG &   2.0, 2.5, 3.0, 3.25, 3.5, 3.75, 4.0, 4.5, 5.0, 5.5, 6.0, 6.25, 6.5, 6.75, 7.0, 7.5, 8.0            \\            
GHAEG & 2.5, 3.0, 3.5, 3.75, 4.0, 4.25, 4.5, 5.0, 5.5, 5.75, 6.0, 6.25, 6.5, 7.0, 7.5             \\         
\end{tabular}
\end{table}

Fitted H-H sigmoid to constant pH data as solid line. All replicas as point. H-H Sigmoid from NMR pKa value and pKa $\pm$ 0.5  as dashed lines

\begin{figure}[htbp]
\includegraphics[width=1.0\textwidth]{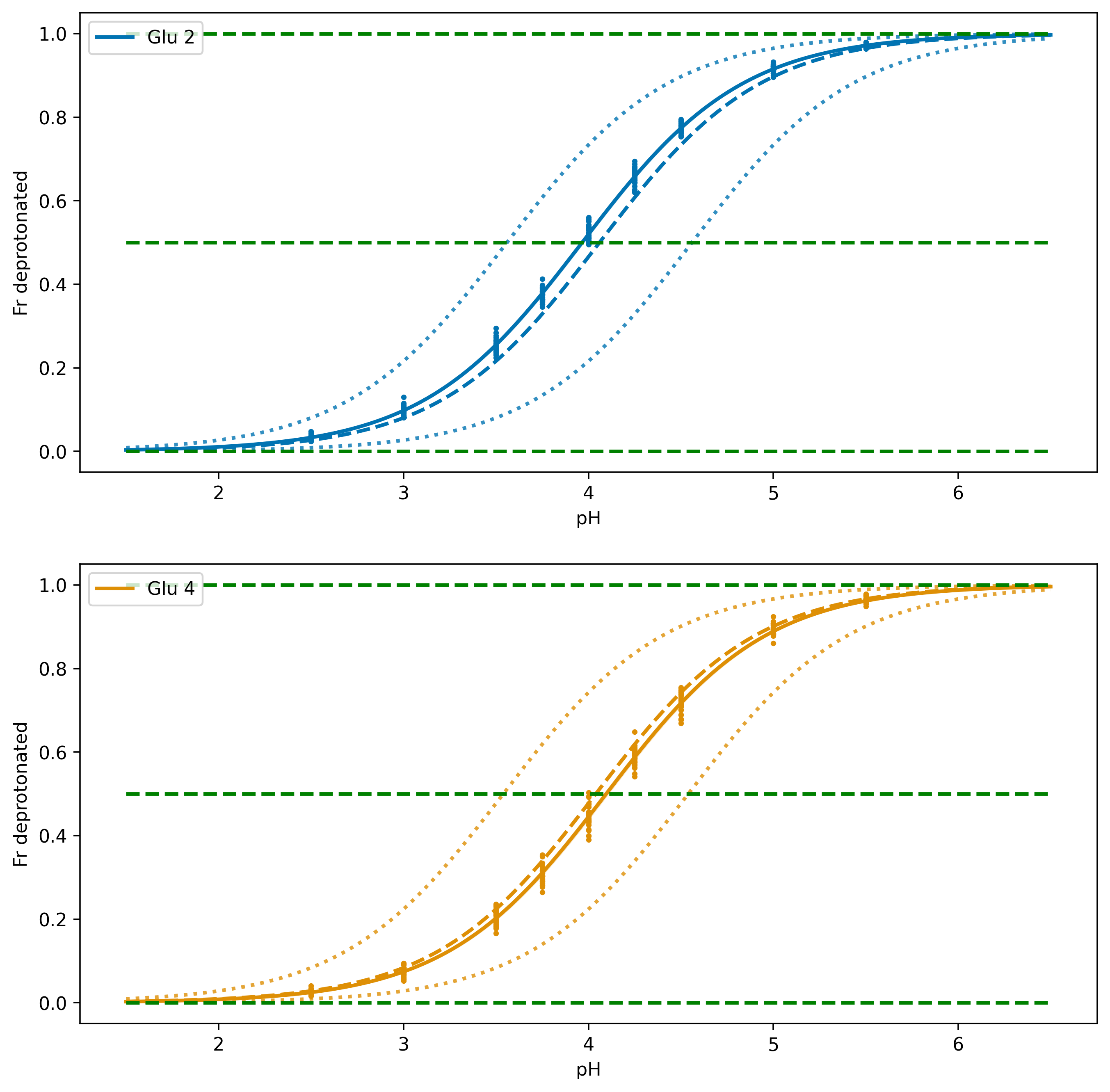} 
\caption{GEAEG titration, \ffCharmmUsed}
\end{figure}

\begin{figure}[htbp]
\includegraphics[width=1.0\textwidth]{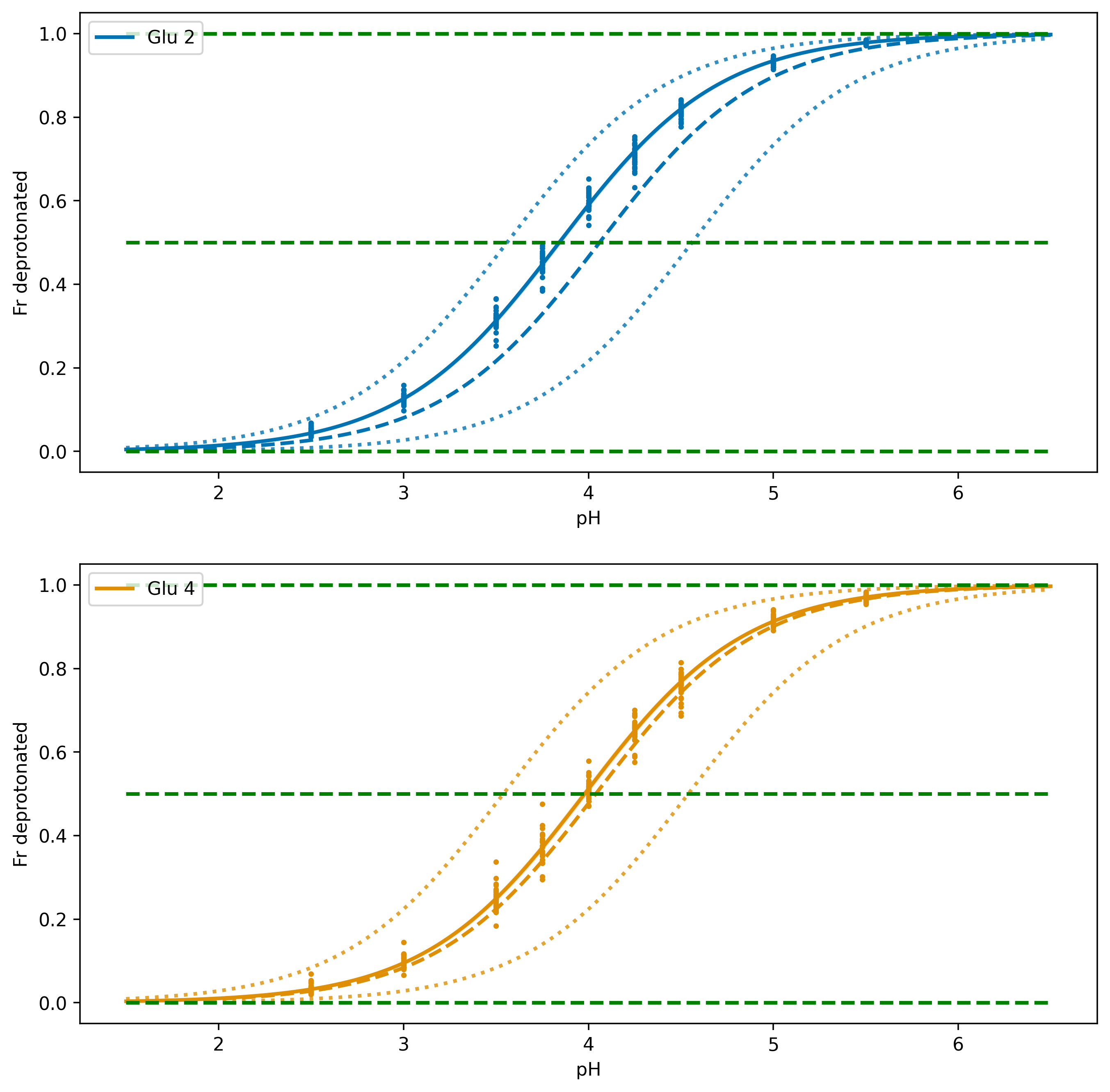} 
\caption{GEAEG titration, \ffAmberUsed}
\end{figure}

\begin{figure}[htbp]
\includegraphics[width=1.0\textwidth]{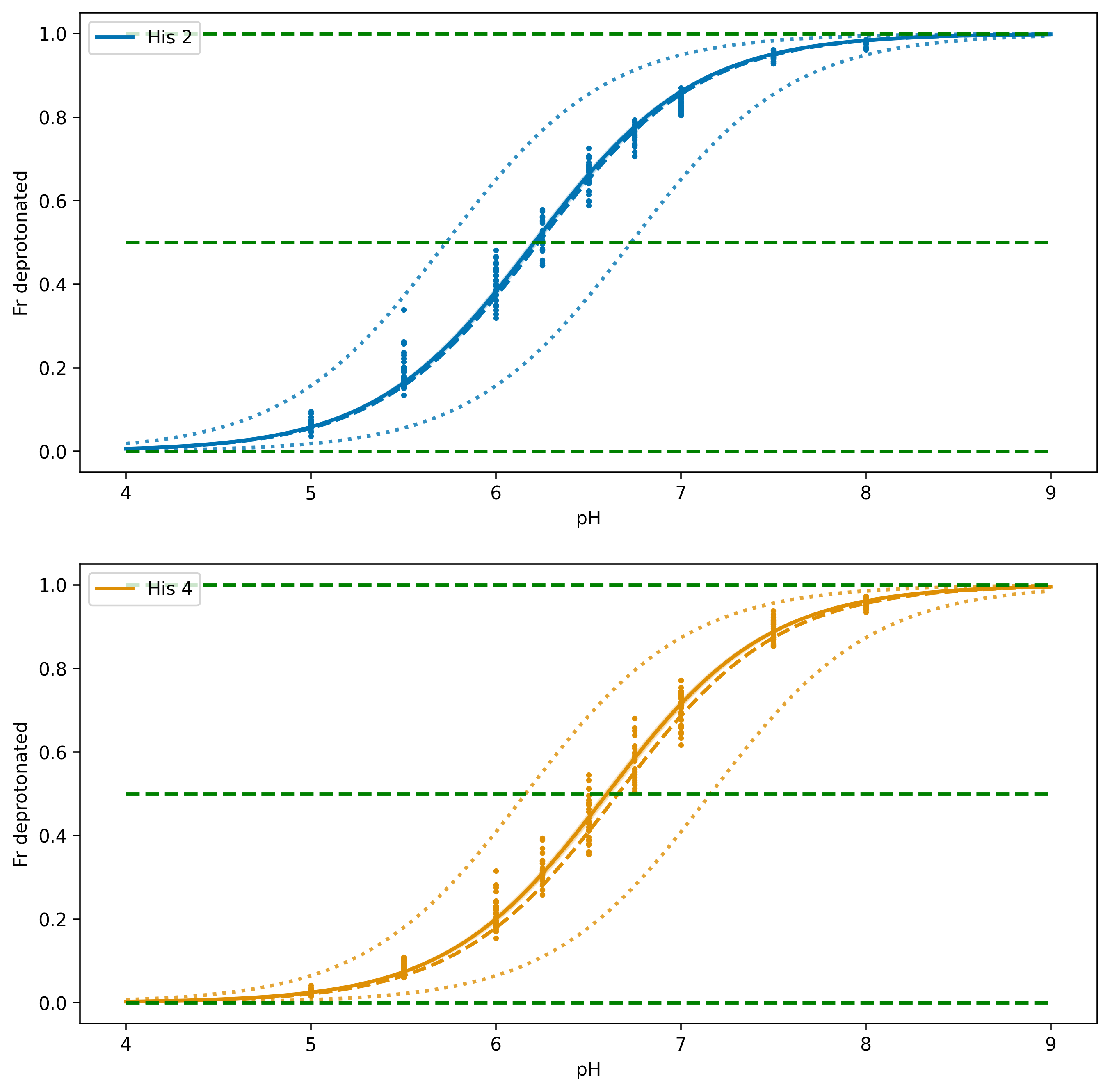} 
\caption{GHAHG titration,  \ffCharmmUsed}
\end{figure}

\begin{figure}[htbp]
\includegraphics[width=1.0\textwidth]{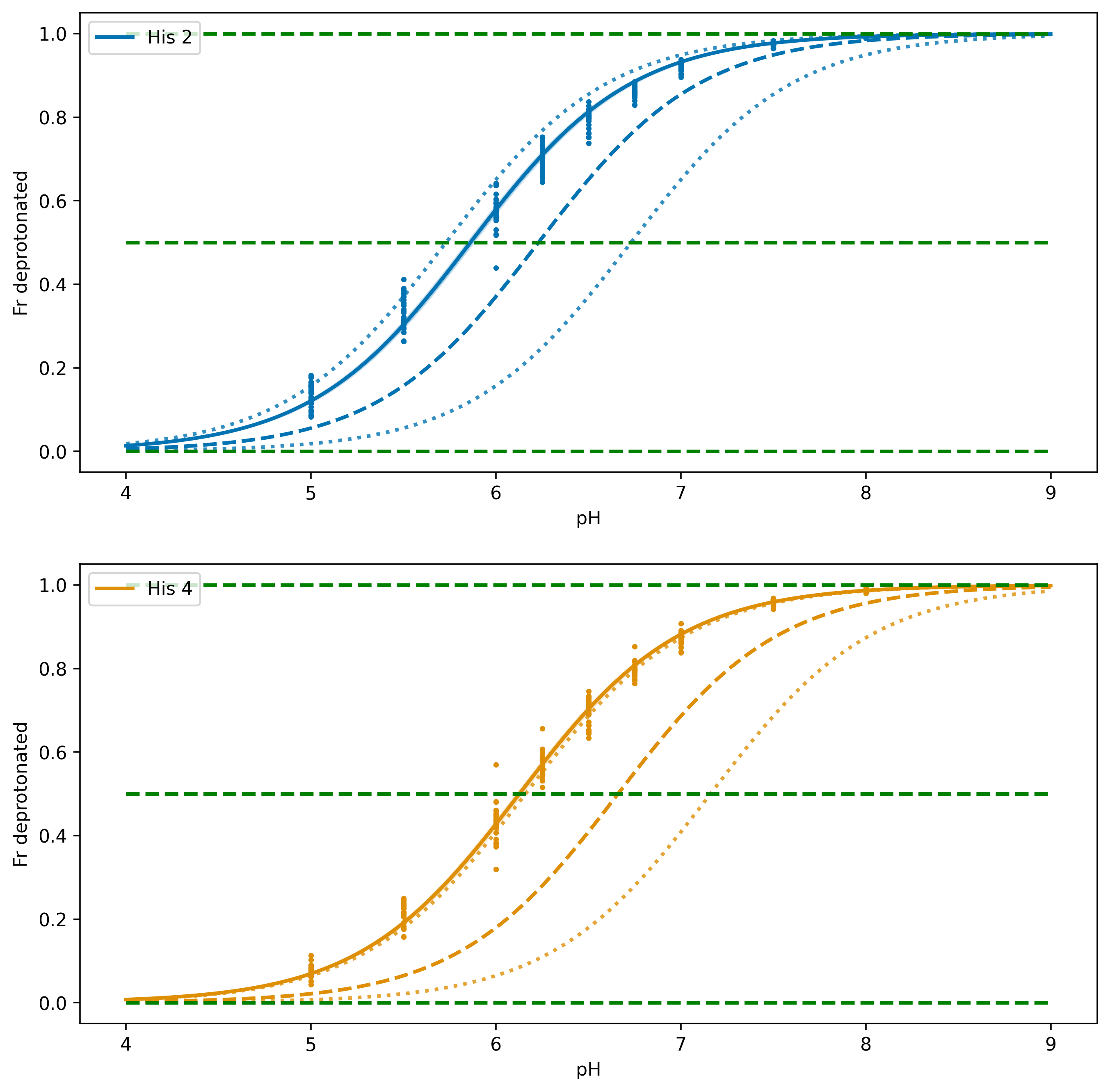} 
\caption{GHAHG titration,\ffAmberUsed}
\end{figure}

\begin{figure}[htbp]
\includegraphics[width=1.0\textwidth]{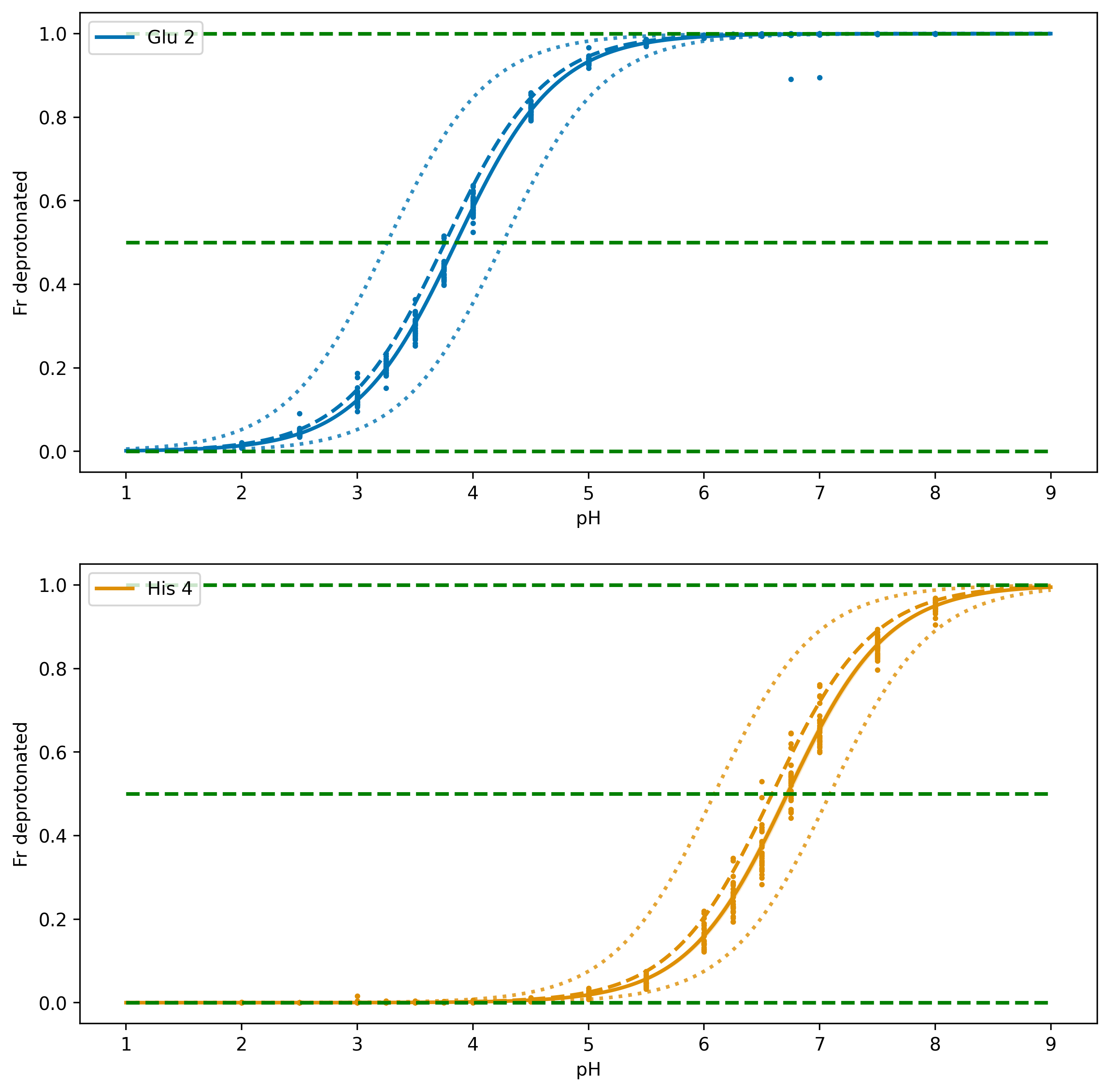} 
\caption{GEAHG titration,  \ffCharmmUsed}
\end{figure}

\begin{figure}[htbp]
\includegraphics[width=1.0\textwidth]{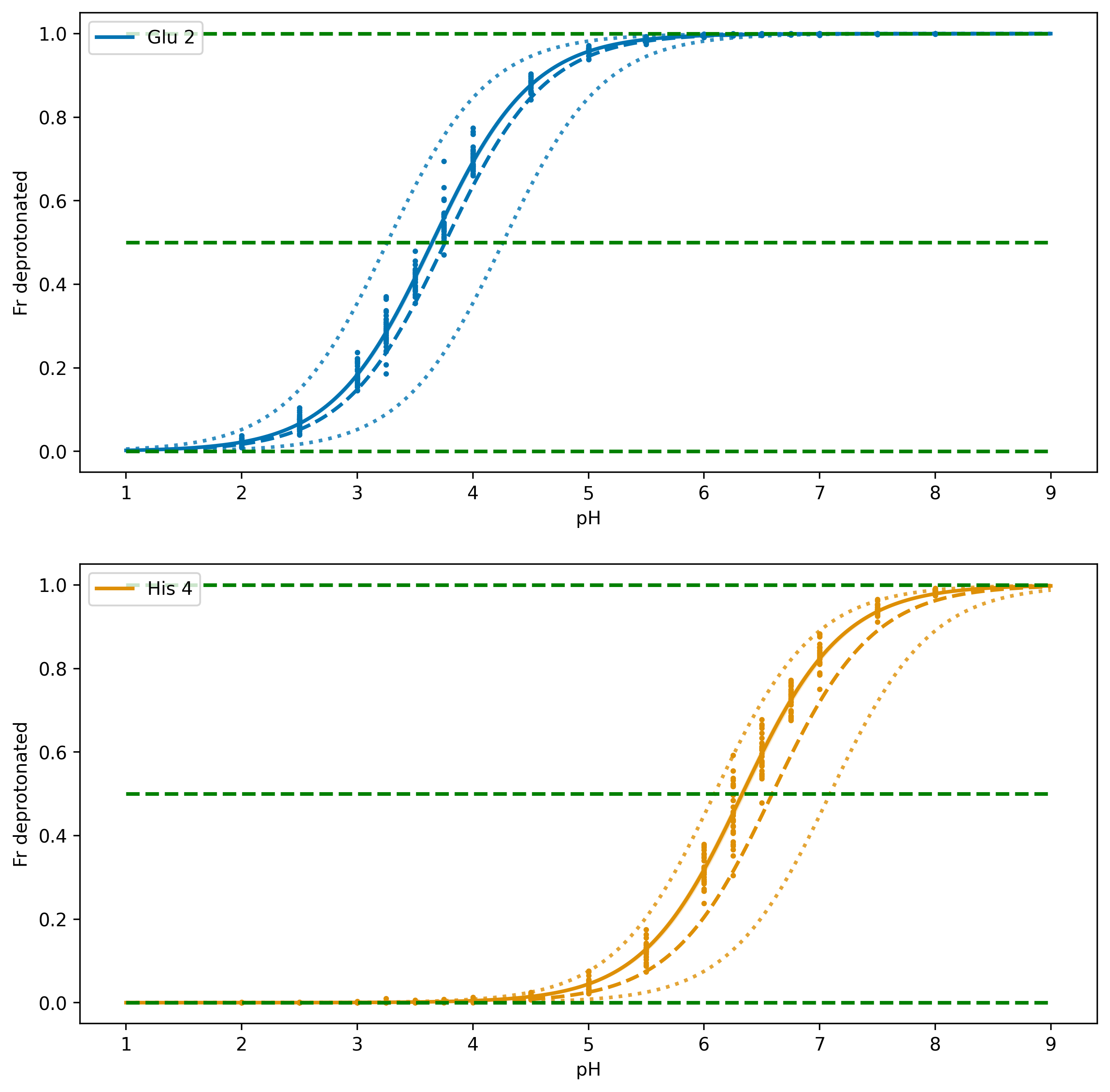} 
\caption{GEAHG titration, \ffAmberUsed}
\end{figure}

\begin{figure}[htbp]
\includegraphics[width=1.0\textwidth]{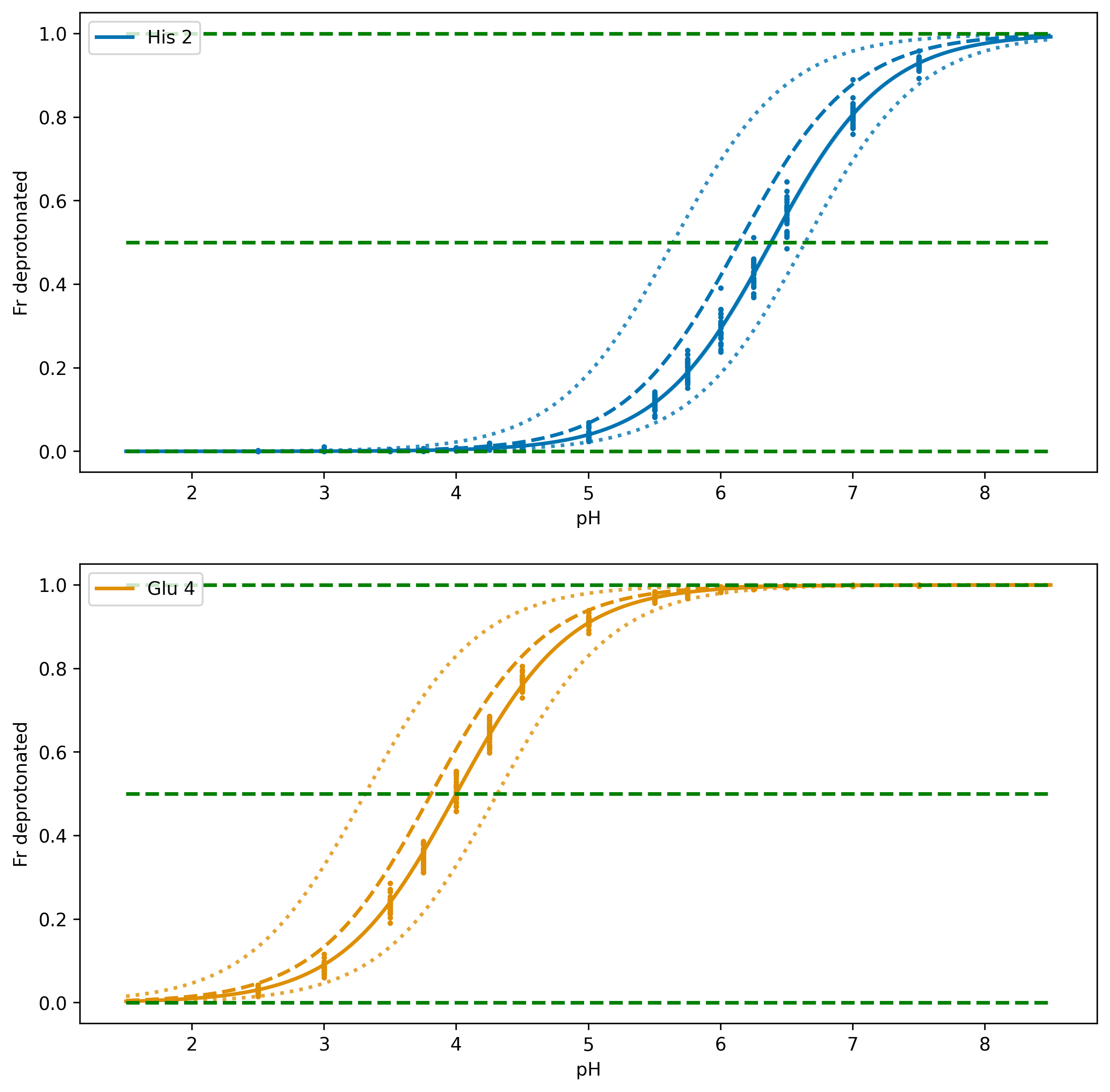} 
\caption{GHAEG titration,  \ffCharmmUsed}
\end{figure}

\begin{figure}[htbp]
\includegraphics[width=1.0\textwidth]{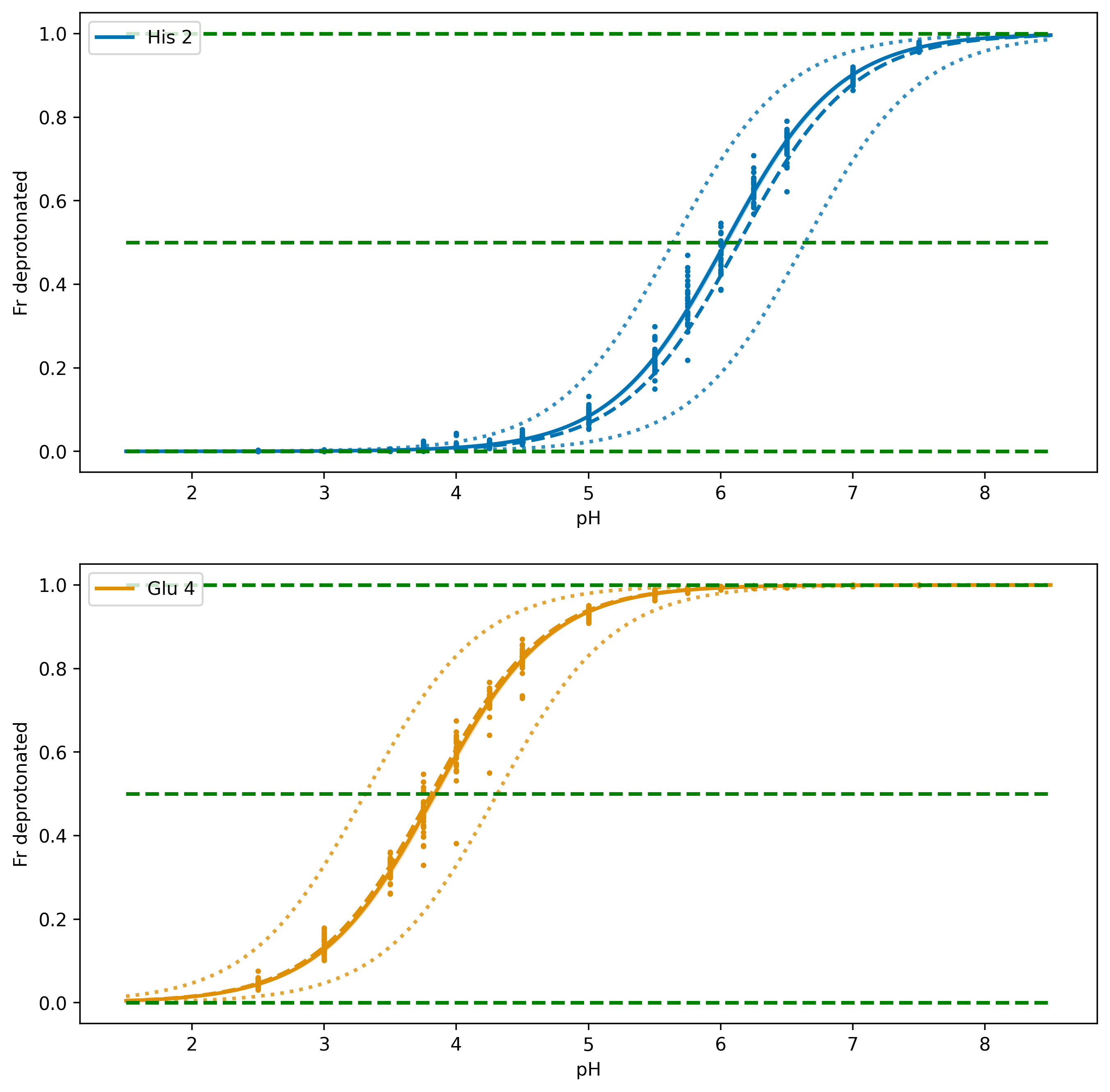} 
\caption{GHAEG titration, \ffAmberUsed}
\end{figure}

\begin{figure}[htbp]
\includegraphics[width=0.5\textwidth]{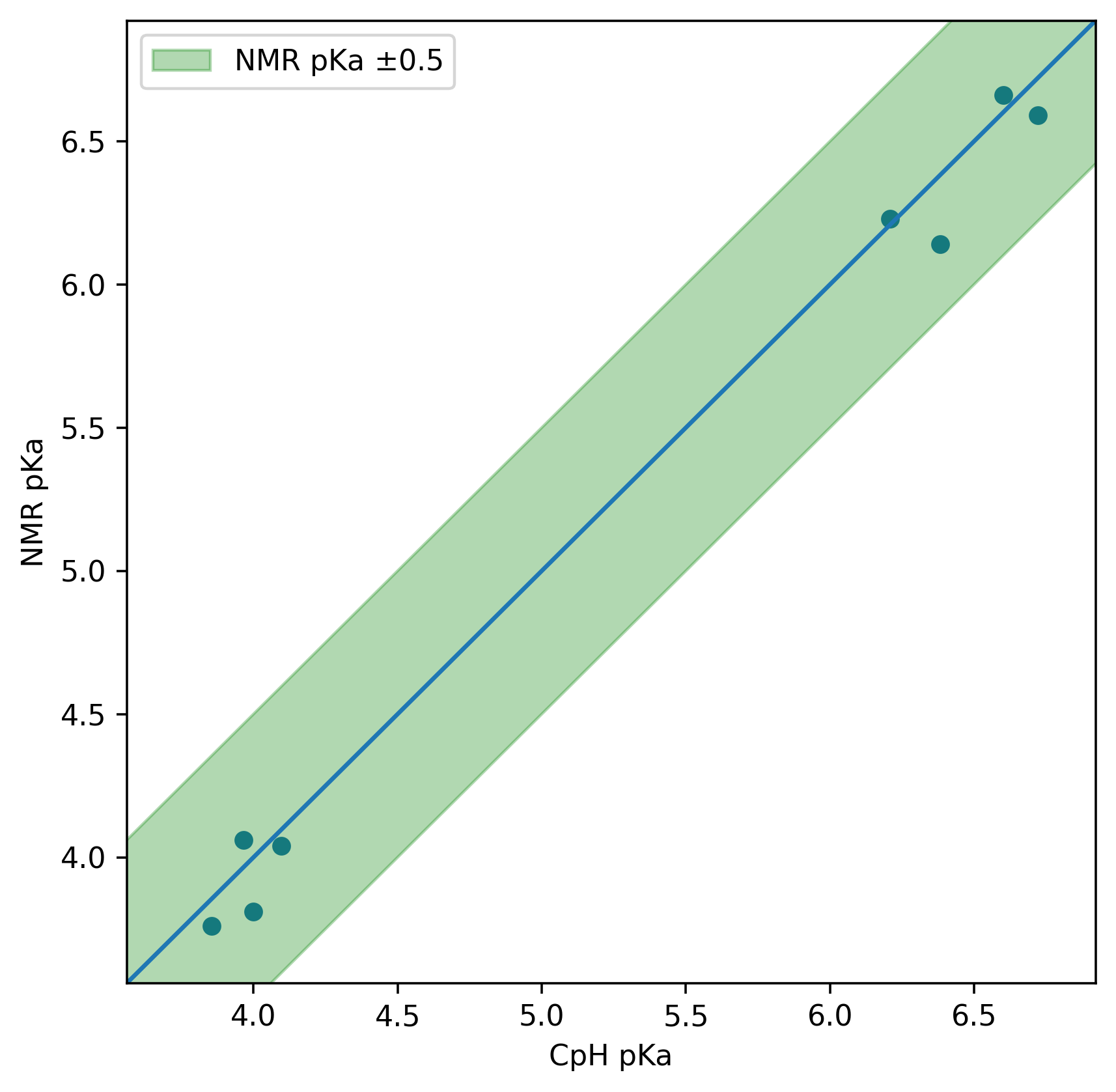} 
\caption{Overall accuracy for pentapeptide titration,  \ffCharmmUsed}
\end{figure}

\begin{figure}[htbp]
\includegraphics[width=0.5\textwidth]{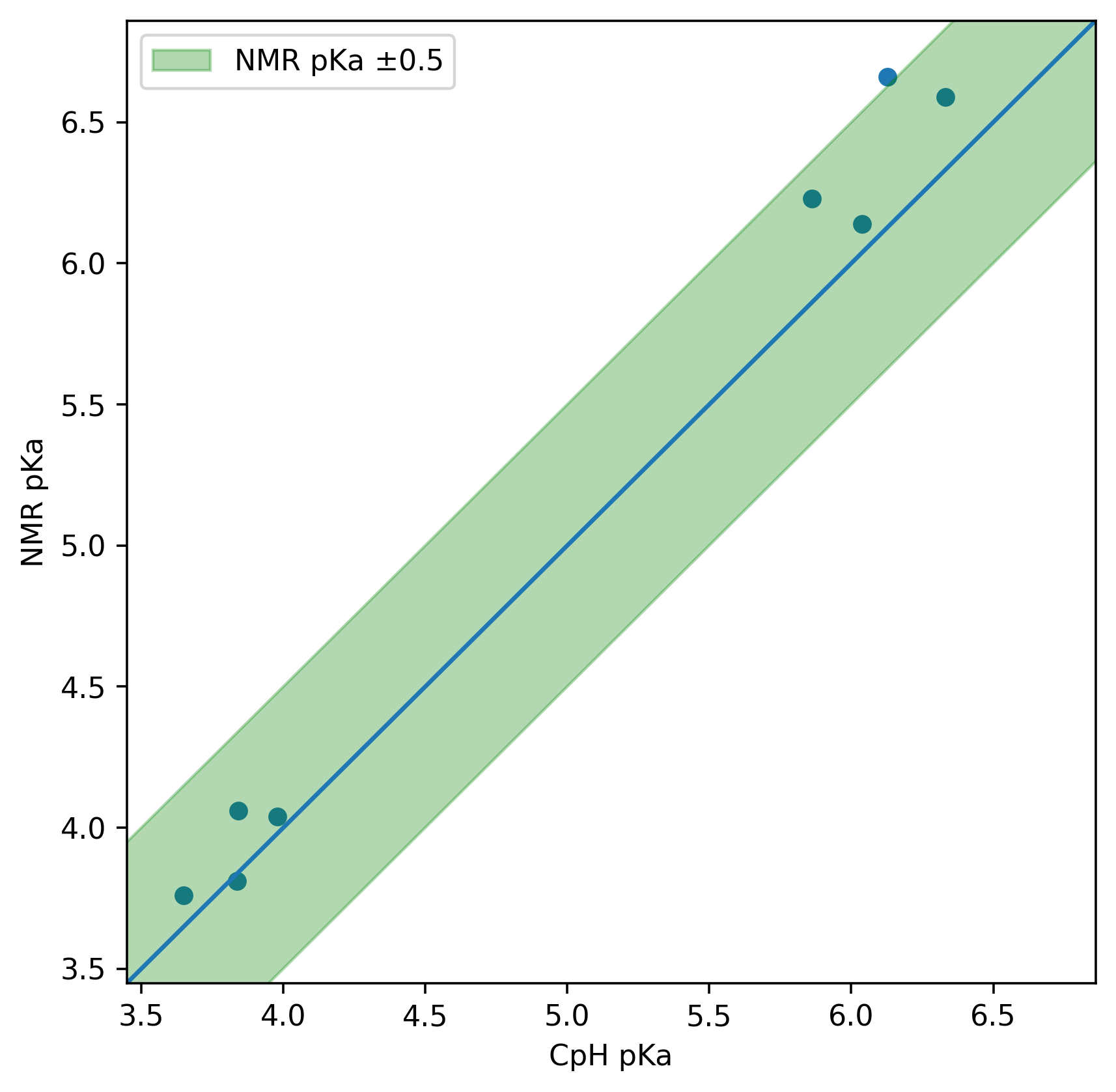} 
\caption{Overall accuracy for pentapeptide titration, \ffAmberUsed}
\end{figure}

\begin{table}[htbp]
\begin{tabular}{lllclcll}
\toprule
        &              & \multicolumn{4}{l}{Constant pH simulation}       & \multicolumn{2}{l}{Experiment\citep{dobrev2020}} \\
Peptide & Residue      & \pKa &  CI95           &  $n$   &  CI95          & \pKa       &          $n$            \\ \midrule
GEAEG   & Glu N-ter    & 3.97 &  3.96 -- 3.97   &  1.00  &  0.99 -- 1.02  & 4.06       &          0.90           \\   
        & Glu C-ter    & 4.10 &  4.09 -- 4.11   &  1.01  &  1.00 -- 1.03  & 4.04       &          0.79           \\              
GHAHG   & His N-ter    & 6.20 &  6.18 -- 6.22   &  0.90  &  0.88 -- 0.93  & 6.23       &          1.06           \\   
        & His C-ter    & 6.60 &  6.58 -- 6.62   &  0.96  &  0.93 -- 0.98  & 6.66       &          0.85           \\   
GEAHG   & Glu N-ter    & 3.85 &  3.84 -- 3.86   &  1.00  &  0.99 -- 1.03  & 3.76       &          0.94           \\   
        & His C-ter    & 6.72 &  6.70 -- 6.74   &  1.02  &  0.99 -- 1.05  & 6.59       &          0.91           \\   
GHAEG   & His N-ter    & 6.38 &  6.37 -- 6.40   &  0.99  &  0.97 -- 1.02  & 6.14       &          0.96           \\   
        & Glu C-ter    & 4.00 &  3.99 -- 4.01   &  1.03  &  1.02 -- 1.05  & 3.81       &          0.90           \\
\bottomrule
\end{tabular}
\caption{Computational titration of pentapeptides:  \pKa{}s and Hill coefficients $n$ 
for \ffCharmmUsed compared to NMR-measure \pKa values. CI95 = 95\% confidence interval obtained by bootstrapping.
NMR single residue \pKa: Glu 4.081, His 6.54. mean absolute error = 0.11, RMSE  = 0.13, Max abs. deviation = 0.24.}
\label{table:Pentapeptidec36m}
\end{table}

\begin{table}[htbp]
\begin{tabular}{lllclcll}
\toprule
        &              & \multicolumn{4}{l}{Constant pH simulation}       & \multicolumn{2}{l}{Experiment\citep{dobrev2020}} \\
Peptide & Residue      & \pKa &  CI95           &  $n$   &  CI95          & \pKa       &          $n$            \\ \midrule
GEAEG   & Glu N-ter    & 3.84 &  3.83 -- 3.85   &  0.98  &  0.96 -- 0.99  & 4.06       &          0.90           \\   
        & Glu C-ter    & 3.98 &  3.97 -- 3.99   &  0.98  &  0.96 -- 1.01  & 4.04       &          0.79           \\              
GHAHG   & His N-ter    & 5.85 &  5.83 -- 5.87   &  0.92  &  0.89 -- 0.94  & 6.23       &          1.06           \\   
        & His C-ter    & 6.12 &  6.11 -- 6.14   &  0.95  &  0.93 -- 0.97  & 6.66       &          0.85           \\   
GEAHG   & Glu N-ter    & 3.65 &  3.63 -- 3.66   &  1.00  &  0.98 -- 1.03  & 3.76       &          0.94           \\   
        & His C-ter    & 6.33 &  6.31 -- 3.35   &  1.02  &  0.99 -- 1.05  & 6.59       &          0.91           \\   
GHAEG   & His N-ter    & 6.04 &  6.02 -- 6.05   &  0.98  &  0.95 -- 1.01  & 6.14       &          0.96           \\   
        & Glu C-ter    & 3.84 &  3.82 -- 3.86   &  0.98  &  0.95 -- 1.01  & 3.81       &          0.90           \\
\bottomrule
\end{tabular}
\caption{Computational titration of pentapeptides:  \pKa{}s and Hill coefficients $n$ 
for \ffAmberUsed compared to NMR-measure \pKa values. CI95 = 95\% confidence interval obtained by bootstrapping.
NMR single residue \pKa: Glu 4.081, His 6.54. mean absolute error = 0.21, RMSE  = 0.26, Max abs. deviation = 0.53.}
\label{table:PentapeptideA99}
\end{table}

\FloatBarrier
\subsection{Cardiotoxin V}

The pH point used for the titration were: 1.0, 1.5, 2.0, 2.5, 3.0, 3.5, 4.0, 4.5, 5.0, 5.5, 6.0, 6.5, 7.0, 7.5, 8.0.



\begin{table}[tbp]
\begin{tabular}{llll}
Residue & \ffCharmmUsed & \ffAmberUsed & Exp.\citep{chiang1996,chiang1996role} \\
\hline
His 4   & 4.33 (4.16 - 4.49) \textbf{!}
        & 4.57 (4.41 - 4.71) \textbf{!}
        & 5.5 \\   
Glu 17  & 4.10 (4.08 - 4.11) 
        & 4.02 (4.02 - 4.03) 
        & 4.0 \\
Asp 42  & 3.85 (3.81 - 3.90) 
        & 3.76 (3.75 - 3.78) 
        & 3.2 \\   
Asp 59  & 2.17 (2.12 - 2.22) 
        & 2.63 (2.58 - 2.68) 
        & $ < $2.0 \\  
\hline  
RMSE   & 0.77 & 0.63 &  \\ 
\end{tabular}
\caption{Computational titration results for cardiotoxin V simulated using \ffCharmmUsed and \ffAmberUsed, 
with measured NMR \pKa as reference. 
(x.xx - x.xx): bootstrapped 95th percentile confidence interval .
RMSE excluding Asp59 for which only a ceiling value is known. \textbf{!}: spread-replica titration}
\label{table:CardiotoxinC36mA99}
\end{table}

\FloatBarrier
\subsection{Hen Egg White Lysozyme (HEWL)}

The following pH value were used for the titration of this protein: -1.0, -0.5, 0.0, 0.5, 1.0, 1.5, 2.0, 2.5, 3.0, 3.5, 4.0, 4.5, 5.0, 5.5, 6.0, 6.5, 7.0, 7.5, 8.0, 8.5, 9.0.

\begin{table}
\begin{tabular}{llll}
Residue 
& CPH \ffCharmmUsed 
& CPH \ffAmberUsed
& Exp. NMR\citep{webb2011}\\
\hline	
Glu 7   & 3.31 (3.28 - 3.33) 
		& 3.00 (2.97 - 3.02) 
        & 2.6 $\pm 0.2$ \\   
His 15  & 4.57 (4.54 - 4.60)
		& 3.03 (2.97 - 3.10) \textbf{!}
        & 5.5 $\pm 0.2$ \\  
Asp 18  & 3.10 (3.03 - 3.17)
		& 2.65 (2.61 - 2.69) 
        & 2.8 $\pm 0.3$ \\  
Glu 35  & 7.41 (7.26 - 7.56) \textbf{!}
		& 6.21 (6.10 - 6.32) \textbf{!}
        & 6.1 $\pm 0.4$ \\  
Asp 48  & 0.37 (0.27 - 0.48) \textbf{!}
		& 1.71 (1.65 - 1.77) \textbf{!}
        & 1.4 $\pm 0.2$ \\  
Asp 52  & 5.05 (5.00 - 5.12)
		& 4.02 (3.98 - 4.06) 
        & 3.6 $\pm 0.3$ \\ 
Asp 66  & 1.28 (1.10 - 1.46) \textbf{!}
		& 0.65 (0.50 - 0.81) \textbf{!}
        & 1.2 $\pm 0.2$ \\  
Asp 87  & 2.20 (2.14 - 2.27) 
		& 1.74 (1.70 - 1.79) 
        & 2.2 $\pm 0.1$ \\ 
Asp 101 & 4.42 (4.38 - 4.48) 
		& 3.59 (3.52 - 3.66) 
        & 4.5 $\pm 0.1$ \\ 
Asp 119 & 2.55 (2.51 - 2.59) 
		& 2.96 (2.94 - 2.99) 
        & 3.5 $\pm 0.3$ \\ 
 \hline	     
       
RMSE                    & 0.85  & 0.90 & \\ 
\end{tabular}
\caption{Computational titration results for lysozyme (40 replicas per pH point, 75 ns per replica) (x.xx - x.xx): Bootstrapped 95th percentile confidence intervals. \textbf{!}: spread-replica titration }
\label{table:LysozymeTitrResTable}
\end{table}

\FloatBarrier
\subsection{Staphilococcal Nuclease}

The following pH value were used for the titration of this protein: 1.0, 1.5, 2.0, 2.5, 3.0, 3.5, 4.0, 4.5, 5.0, 5.5, 6.0, 6.5, 7.0, 7.5, 8.0.

\begin{table}
\begin{tabular}{lll}
Residue & HendHass \pKa  & Exp.\citep{Castaneda2009}\\
\hline	
His 8   & 6.01    (5.99 -  6.04)
        & 6.5 \\
Glu 10  & 3.60    (3.55 -  3.64) 
        & 2.8 \\         
Asp 19  & 2.67    (4.49 -  5.12) \textbf{!} 
        & 2.2, 6.5 \\         
Asp 21  & 6.28    (3.44 -  3.89)\textbf{!} 
        & 3.0, 6.5 \\         
Asp 40  & 3.21    (3.16 -  3.26)
        & 3.9 \\  
Glu 43  & 4.17    (4.16 -  4.19)
        & 4.3 \\  
Glu 52  & 4.63    (4.62 -  4.65) 
        & 3.9 \\          
Glu 57  & 4.22    (4.21 -  4.24)
        & 3.5 \\  
Glu 67  & 4.45    (4.42 -  4.48)
        & 3.58 \\         
Glu 73  & 3.74    (3.71 -  3.77)
        & 3.3 \\   
Glu 75  & 3.90    (3.82 -  3.98) \textbf{!} 
        & 3.3 \\                   
Asp 77  & 1.45    (1.24 -  1.65) \textbf{!} 
        & $<$2.2 \\                   
Asp 83  & 0.60    (0.36 -  0.78) \textbf{!} 
        & $<$2.2 \\                   
Asp 95  & 3.09    (3.04 -  3.15)  
        & 2.2 \\  
Glu 101 & 4.17    (4.10 -  4.23) 
        & 3.8 \\ 
His 121 & 4.89    (4.75 -  5.03) \textbf{!} 
        & 5.24 \\          
Glu 122 & 4.22    (4.14 -  4.29)  
        & 3.9 \\  
Glu 129 & 3.84    (3.79 -  3.89)
        & 3.8 \\
Glu 135 & 3.95    (3.91 -  3.99)
        & 3.8 \\          
                 
 \hline	     
       
RMSE                      & 0.53   &\\ 
\end{tabular}
\caption{Computational titration results for staphylococcal nuclease $\Delta\mathrm{PHS}$ 
(40 replicas per pH point, 75 ns per replica) using the \ffCharmmUsed force field. 
(x.xx - x.xx): Bootstrapped 95th percentile confidence intervals. 
\textbf{!}: spread-replica  titration.
Asp 19, Asp 21: the pH of the two chemical shift transition observed in NMR are reported in the Exp. column }
\label{table:StaphNucleaseTitrResTableC36m}
\end{table}

\begin{table}
\begin{tabular}{lll}
Residue & HendHass \pKa & Exp.\citep{Castaneda2009}\\
\hline	
His 8   & 5.50 (5.48 -   5.52) 
        & 6.5 \\
Glu 10  & 3.12 (3.05 -   3.18)  
        & 2.8 \\         
Asp 19  & 2.37 (2.32 -   2.42) 
        & 2.2, 6.5 \\         
Asp 21  & 5.65 (5.53 -   5.78) \textbf{!} 
        & 3.0, 6.5 \\         
Asp 40  & 2.95 (2.91 -   3.00) 
        & 3.9 \\  
Glu 43  & 3.90 (3.88 -   3.92) 
        & 4.3 \\  
Glu 52  & 4.65 (4.63 -   4.67) 
        & 3.9 \\          
Glu 57  & 4.17 (4.15 -   4.18) 
        & 3.5 \\  
Glu 67  & 3.92 (3.90 -   3.93) 
        & 3.58 \\         
Glu 73  & 3.53 (3.51 -   3.55) 
        & 3.3 \\   
Glu 75  & 2.08 (1.95 -   2.20) \textbf{!} 
        & 3.3 \\                   
Asp 77  & 1.82 (1.68 -   1.96) \textbf{!} 
        & $<$2.2 \\                   
Asp 83  & 0.07 (-0.21 -   0.28) \textbf{!} 
        & $<$2.2 \\                   
Asp 95  & 3.34 (3.33 -   3.37)  
        & 2.2 \\  
Glu 101 & 3.31 (3.25 -   3.37) 
        & 3.8 \\ 
His 121 & 6.86 (6.77 -   6.95) \textbf{!} 
        & 5.24 \\          
Glu 122 & 2.72 (2.65 -   2.80)  
        & 3.9 \\  
Glu 129 & 3.97 (3.94 -   4.00) 
        & 3.8 \\
Glu 135 & 2.98 (2.92 -   3.05) 
        & 3.8 \\          
                 
 \hline	     

RMSE                      & 0.83  &\\ 
\end{tabular}
\caption{Computational titration results for staphylococcal nuclease $\Delta\mathrm{PHS}$ 
(40 replicas per pH point, 75 ns per replica) using the \ffAmberUsed force field. 
(x.xx - x.xx): Bootstrapped 95th percentile confidence intervals.
\textbf{!}: spread-replica titration.
Asp 19, Asp 21: the pH of the two chemical shift transition observed in NMR are reported in the Exp. column }
\label{table:StaphNucleaseTitrResTableA99}
\end{table}

\begin{figure}[H]
\includegraphics[scale=0.4]{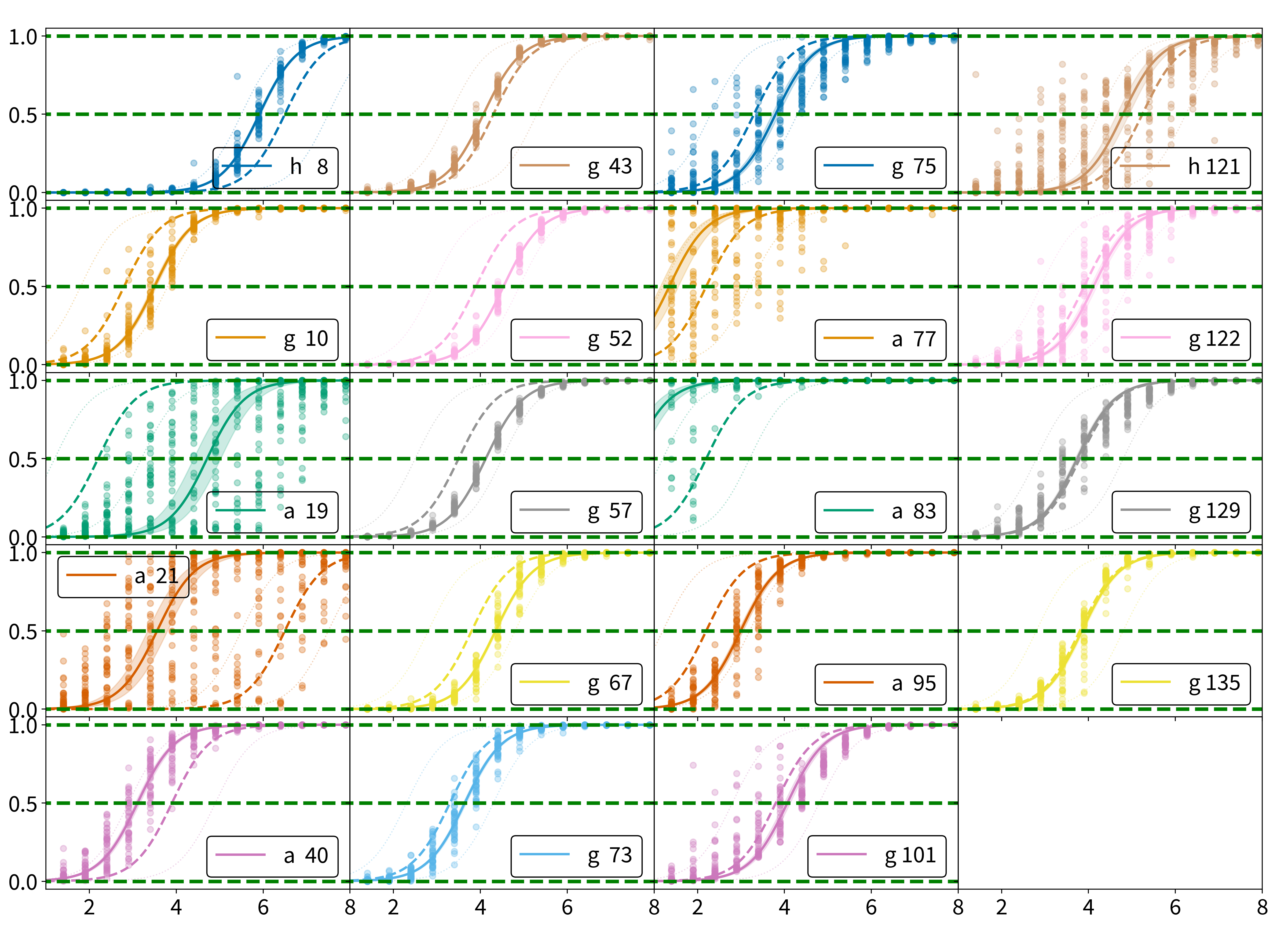} 
\caption{Individual computational titration curves for all residues in Staph. Nuclease $\Delta$ PHS, with all replicas shown (circles), for \ffCharmmUsed. Labelled as first letter of amino acid type and sequence number of the residue. Solid line: fitted Henderson-Hasselbach curve to constant pH data. Dashed line: Henderson-Hasselbach curve from NMR-determined pKa}
\end{figure}

\begin{figure}[H]
\includegraphics[scale=0.5]{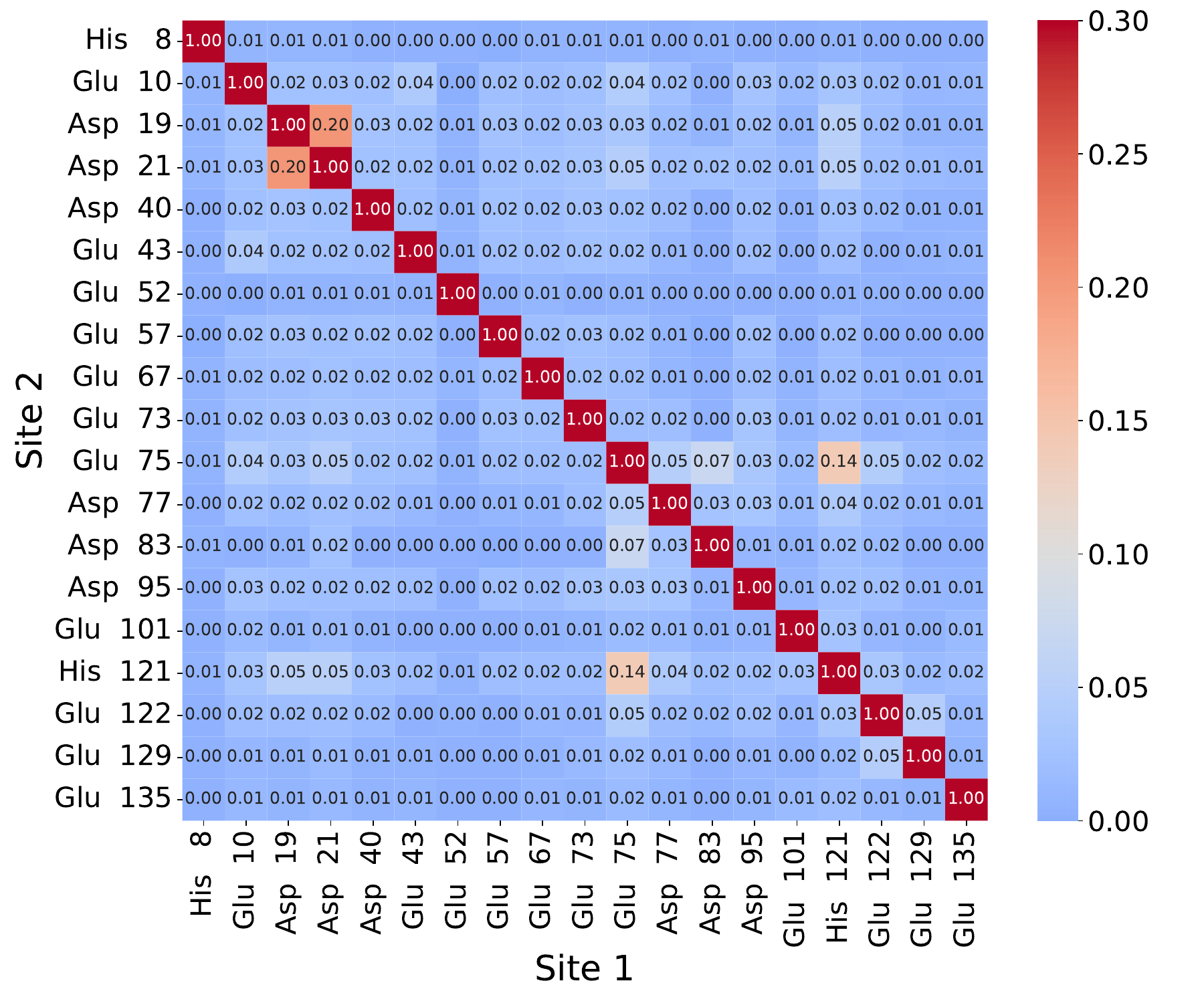} 
\caption{Matrix of the mean Normalized Mutual Information for each pairs of residue (using the \ffCharmmUsed force field), at the pH point where is is maximum for that pair. Non-diagonal red squares correspond to a positive criteria (NMI $> 0.1$) under our criteria. }
\label{fig:SNasePkaChart}
\end{figure}

\FloatBarrier
\section{FMA projections}

\subsection{Cardiotoxin Asp 59}

\begin{figure}[H]
\centering
\includegraphics[scale=0.2]{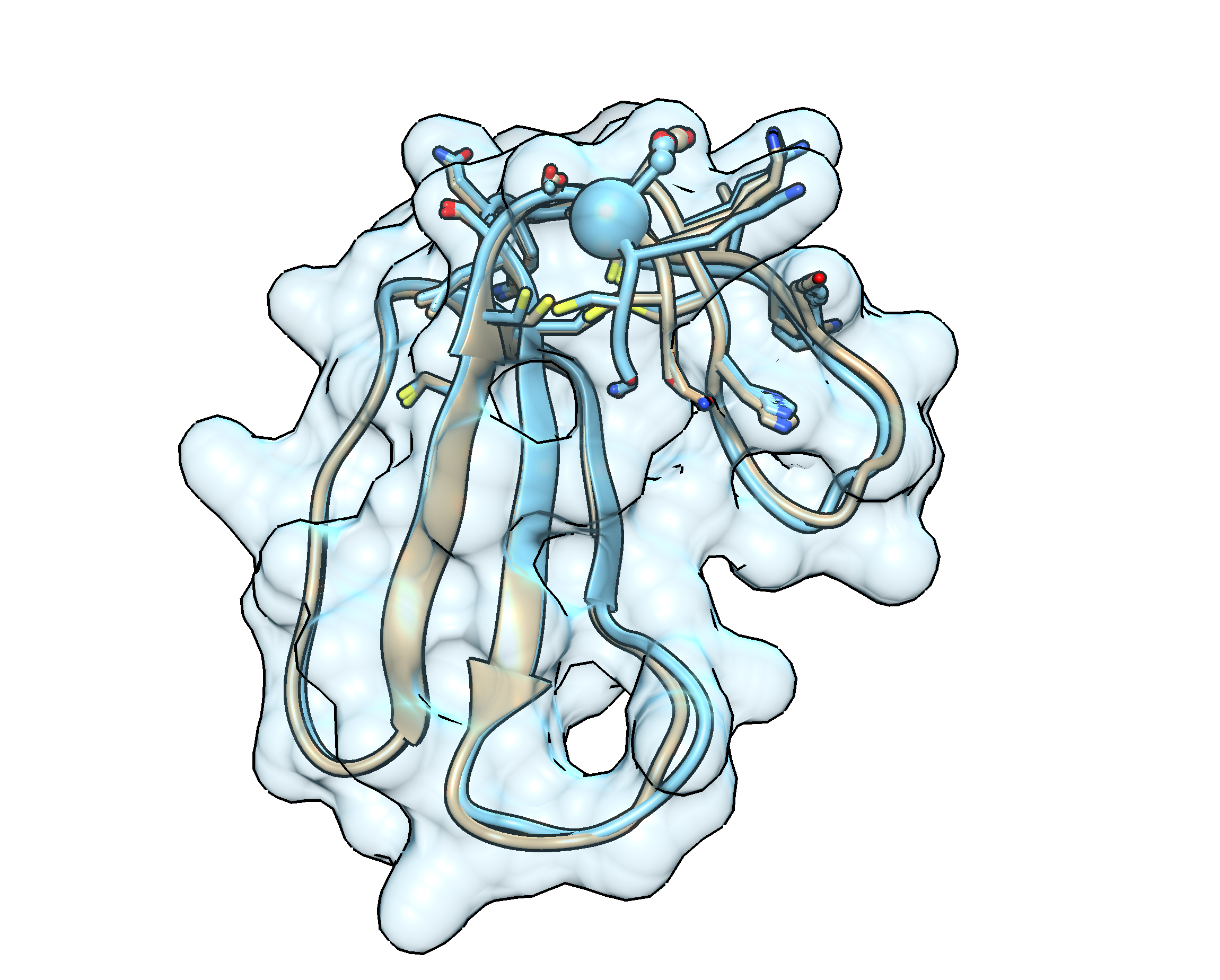} 
\caption{Low (tan) and high (blue) FMA value structure for Asp 59 in Cardiotoxin. Solvent-excluded surface for the high FMA value structure shown, with sphere locating the Asp 59 residue.}
\end{figure}

\FloatBarrier
\subsection{Staphiloccocal Nuclease}

\begin{figure}[H]
\centering
\includegraphics[scale=0.1]{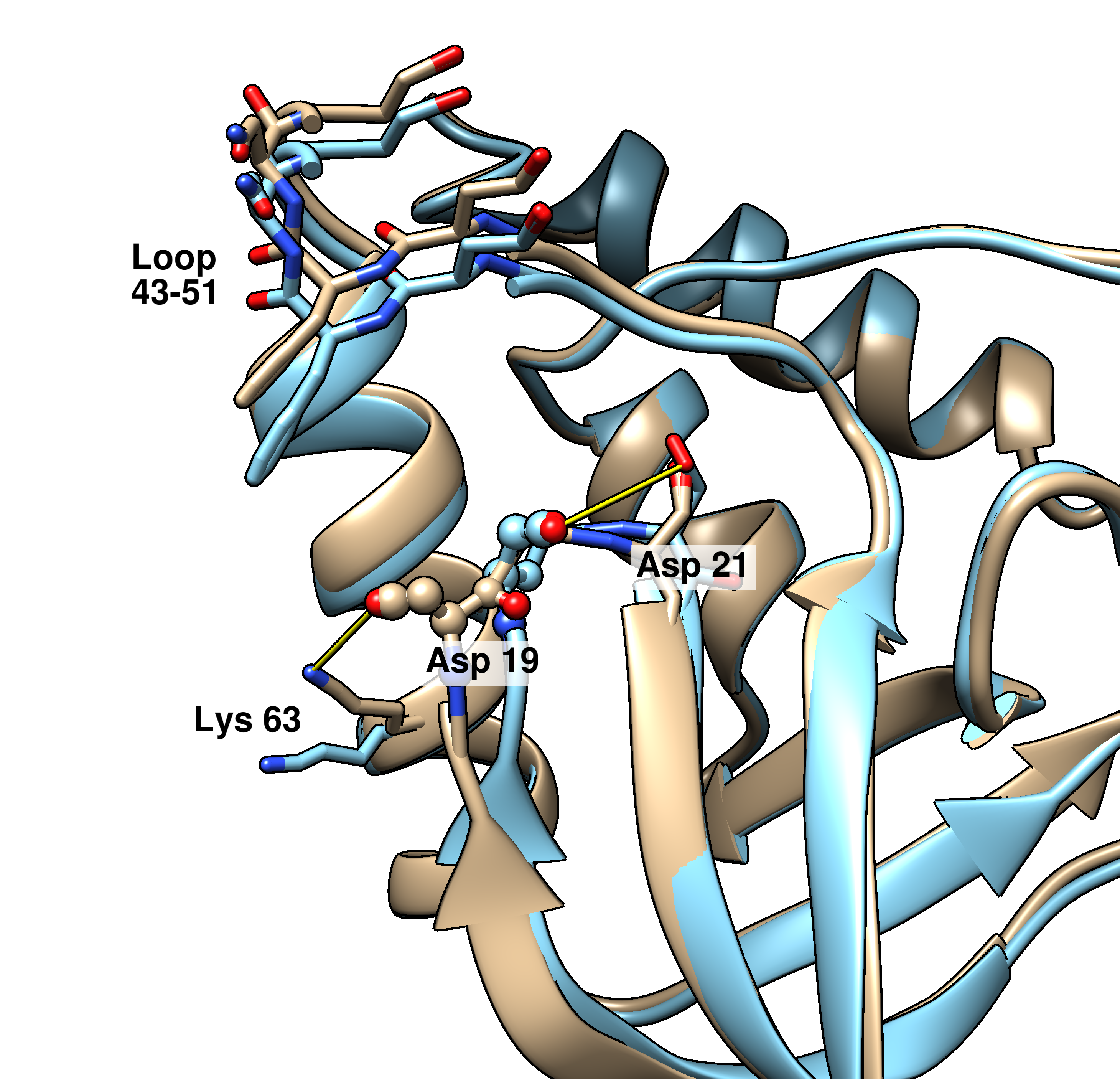} 
\caption{Low (tan) and high (blue) FMA value structure for Asp 19 in SNase. Relevant interaction of Asp 19 depicted with yellow pseudobonds.}
\end{figure}

\begin{figure}[H]
\centering
\includegraphics[scale=0.8]{supplementary/illustrations/snase_asp21_fma_extremestruct_cropped.png} 
\caption{Low (tan) and high (blue) FMA value structure for Asp 21 in SNase. Relevant interaction of Asp 19 depicted with yellow pseudobonds.}
\end{figure}

\begin{figure}[H]
\centering
\includegraphics[scale=0.8]{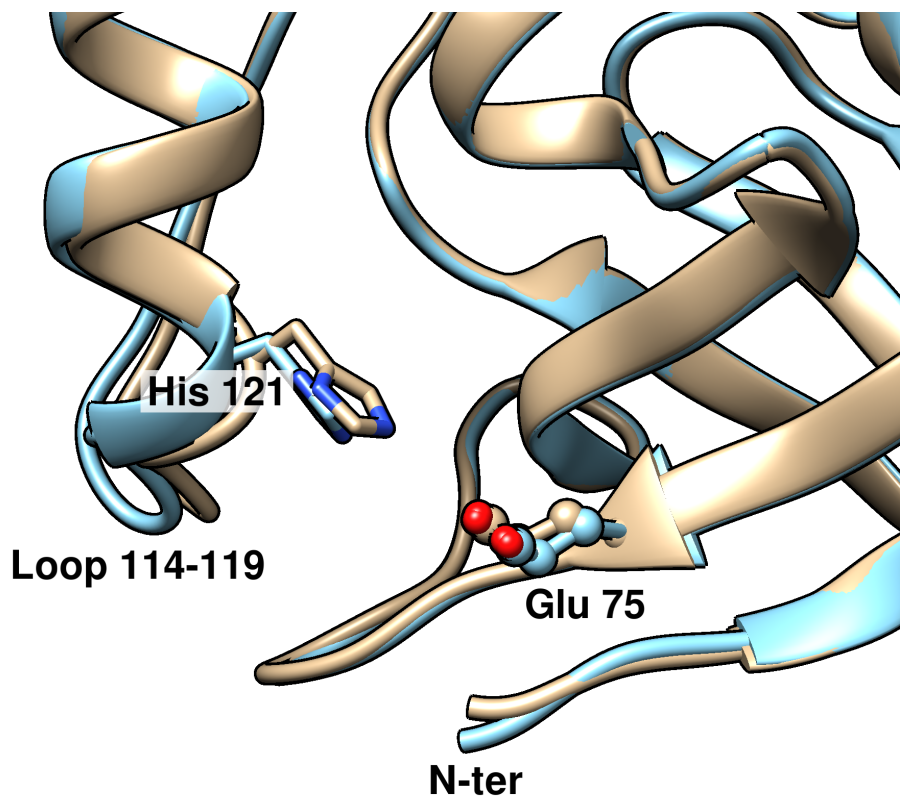} 
\caption{Low (tan) and high (blue) FMA value structure for Glu 275 in SNase.}
\end{figure}

\section{Simulations Setup Details}

\subsection{NVT simulations}

The constant pH simulations in this work were performed in the NVT ensemble, as our GPU-accelerated FMM codes does not yet support pressure coupling.
System solvation was performed using the \gromacs \texttt{solvate} tool, which yields solvent density slightly lower in our subsequent constant pH NVT simulation than an equivalent NPT simulations (around 1004 versus 1027 $\mathrm{kg}/\mathrm{m}^3$).
To correct the density, the user may choose to run a pre-equilibration NPT simulation (without constant pH) prior to the NVT constant pH MD simulations.
The subsequent constant pH simulations will requires a different set of calibration data, which can be enabled by setting the \lstinline{CPH_PRESSUREEQUIL_FIT} define in the \mdp file -- see Listing~\ref{fig:mdpPressEquil}.
We suspect that this is necessary due to the slightly different dielectric permitivity $\varepsilon$ of water owing to the increased density, which affects the residue -- buffer site contribution to $\nicefrac{\partial \mathcal{H}}{\partial \lambda}$.

\begin{figure}[H]
\centering
\includegraphics[scale=0.8]{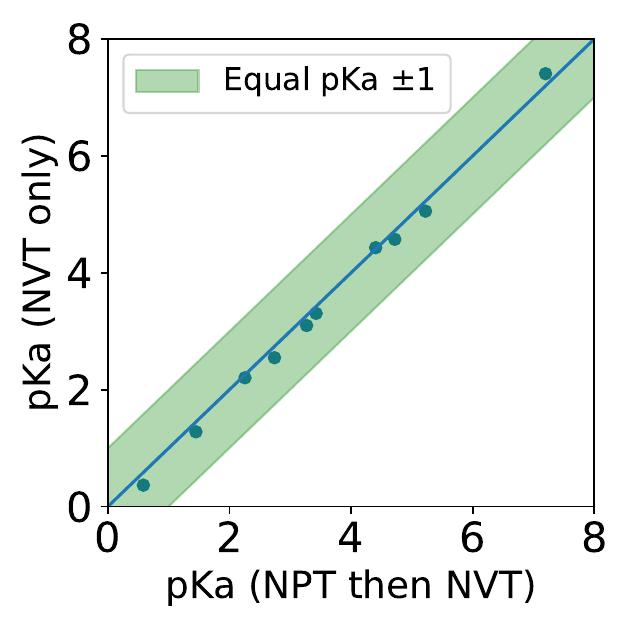} 
\caption{\pKa values for constant pH titration of HEWL (\ffCharmmUsed) where all simulations took place in the NVT ensemble (y-axis) compared to a protocol with a non-constant pH NPT equilibration preceding the titration itself (also in the NVT ensemble). Maximum \pKa deviation is 0.2, Pearson correlation coefficient is 0.99}
\label{fig:nvtVersusNptNvt}
\end{figure}

However, we found empirically that such pre-equilibration did not affect \pKa values for our test system lysozyme (Figure~\ref{fig:nvtVersusNptNvt}), a globular protein.
We therefore did not use this equilibration procedure in this work.
We speculate that, in contrast, this result would not necessarily hold for non-homogeneous systems such as non-globular proteins or lipid membranes.

Future development of our FMM implementation will provide a definitive solution to this issue by incorporating pressure coupling.

\begin{figure}[h!]
\begin{lstlisting}
define                  =  -DCPH_PRESSUREEQUIL_FIT
\end{lstlisting}
\caption{Enabling the set of calibration data for NPT-equilibrated box}
\label{fig:mdpPressEquil}
\end{figure}

